\documentclass[11pt]{article}
\usepackage[utf8]{inputenc}

\usepackage{geometry}
\usepackage[utf8]{inputenc}
\usepackage{babel}
\usepackage{hyperref}
\usepackage{bm}
\usepackage{mathtools}

\usepackage{amsmath,amsfonts,amssymb,amsthm}
\usepackage{mathrsfs}
\usepackage{mathenv}
\usepackage{xspace}
\usepackage{graphicx}
\usepackage{dsfont}
\usepackage{appendix}
\usepackage{enumitem}
\usepackage{xcolor}
\usepackage{subfig}
\usepackage{authblk}

\newcommand{\Tr}[1]{\ensuremath{{\rm Tr}\,#1}}

\newcommand{\Diag}[1]{\ensuremath{{\rm Diag}#1}}
\newcommand{\mat}[1]{\ensuremath{\mathbf{#1}}}
\newcommand{\vect}[1]{\ensuremath{\bm{#1}}}

\title{Right large deviation principle for the top eigenvalue of the sum or product of invariant random matrices}
\author[,1,2]{Pierre Mergny \thanks{\texttt{mergny.pierre@gmail.com}}}
\affil[1]{LPTMS,  CNRS,  Univ.   Paris-Sud,  Université  Paris-Saclay,  91405  Orsay,  France}
\affil[2]{Chair of Econophysics $\&$ Complex Systems, Ecole Polytechnique, 91128 Palaiseau Cedex, France}
\author[3]{Marc Potters}
\affil[3]{Capital Fund Management, 23 rue de l’Université, 75007 Paris, France}

\date{\today}

\begin{document}
\maketitle

\abstract{In this note we study the right large deviation of the top eigenvalue (or singular value) of the sum or product of two  random matrices $\mathbf{A}$ and $\mathbf{B}$ as their dimensions goes to infinity. The matrices $\mathbf{A}$ and $\mathbf{B}$ are each assumed to be taken from an invariant (or bi-invariant) ensemble with a confining potential with a possible \emph{wall} beyond which no eigenvalues/singular values are allowed. The introduction of this wall puts different models in a very generic framework. In particular, the case where the wall is exactly at the right edge of the limiting spectral density is equivalent, from the point of view of large deviations, to considering a fixed diagonal matrices, as studied previously in Ref. \cite{GuionnetMaida20}. We show that that the tilting method introduced in  Ref. \cite{GuionnetMaida20} can be extended  to our general setting and is equivalent to the study of a  spherical spin glass model specific to the operation - sum of symmetric matrices /  product of symmetric matrices / sum of rectangular matrices - we are considering. }
\tableofcontents

\section{Introduction}
\label{sec:Intro}
Since the pioneer work of Wishart \cite{Wishart} and Wigner \cite{Wigner} in the 60s, Random Matrix Theory (RMT) has found applications in many domains of research, ranging from the theory of disordered system \cite{Edwards_1975,sherrington1975solvable} to telecommunications \cite{tulino2004random} and finance \cite{bouchaud2000theory} and  even more recently to statistical learning theory \cite{Pennington,Benigni}, to cite few examples.  In particular the top eigenvalue/singular value of a random matrix plays a fundamental role in many fields: for example in statistics, Principal Component Analysis (PCA) \cite{Jolliffe2005} is often used to reduce the dimensions of a raw matrix based on the values of first top eigenvalues; in the study of the stability of a complex system \cite{may72,allesina2015stability,sompolinsky1988chaos,wainrib2013topological,moran2019may} the top eigenvalue of the (opposite of the) stability matrix of a randomly linear system indicates whether the system is locally stable or not; in the theory of disordered system, the law of the top eigenvalue is related to the so-called complexity of the system, see Refs. \cite{Fyodorov04,Ros19,BenArousFyodorov21}. 
 
Therefore, a natural question in RMT and related fields is the following:  given a random symmetric (respectively rectangular) matrix $\mathbf{A}$, as the dimensions of this matrix grows, what is the typical value of the top eigenvalue (resp. singular value) and what is the probability to find it at a position, say $x$,  far from this typical value? When this probability is exponentially small, one says that the top eigenvalue satisfies a \emph{large deviation principle} and the goal is to calculate both the \emph{speed of convergence} (what is the power in $N$ in the exponential decay?) and the \emph{rate function} (what is the leading prefactor and how does it depend on the position $x$?). The first natural case is to consider, $\mathbf{A}$ as taken from an \emph{invariant ensemble}, a family containing both the famous \emph{Gaussian orthogonal ensemble} (GOE)
 and the \emph{Wishart} ensemble of RMT. Based on a direct analogy between the ensemble of eigenvalues and a Coulomb gas of particles restricted on the real line, the large deviation can be computed explicitly, see Refs. \cite{Arous2001,Dean06,Dean08,Majumdar09,Vivo07,Borot11} and \cite{Majumdar14} for a review. In particular, both the speed of convergence and the rate functions depend on whether $x$ is above or below the typical value. The former is known as the right large deviation principle and the latter is known as the left large deviation principle. Several studies have gone beyond the invariant ensemble case by looking for example at generalized Wigner matrices, see Refs. \cite{Augeri16,Augeri2021,husson2020,GuionnetHusson20} or looking at small-rank deformation of an invariant ensemble, see Refs. \cite{Maida07,Bianchi11,BenaychGeorges11}. Recently, there has been an interest in the so-called full-rank deformation: 
 \begin{enumerate}
     \item In Ref. \cite{GuionnetMaida20}, the authors studied the right large deviation for the sum $\mathbf{C} = \tilde{\mathbf{A}}+\mathbf{O}\tilde{\mathbf{B}}\mathbf{O}^{\mathsf{T}}$, where $\mathbf{O}$ is a random (uniform) orthogonal matrix, and $\tilde{\mathbf{A}}$ and $\tilde{\mathbf{B}}$ are two 'fixed diagonal' matrices, based on a tilting method with the additive spherical integral.
     \item In Ref. \cite{McKenna21}, based on similar ideas, the author studied the (right) large deviation of the top eigenvalue of the sum $\mathbf{C} = \tilde{\mathbf{A}}+\mathbf{B}$ where $\mathbf{B}$ is a (slight generalization of a) GOE matrix. 
     \item In Ref. \cite{Maillard21} and based again on similar methods, the case of the large deviation of the top eigenvalue of the product $\mathbf{C}  = \sqrt{\tilde{\mathbf{A}}} \mathbf{B} \sqrt{\tilde{\mathbf{A}}}$, where $\tilde{\mathbf{A}}$ is a diagonal matrix and $\mathbf{B}$ is a Wishart matrix, was studied. 
 \end{enumerate}
 
In this paper, we obtain \emph{explicitly} the right large deviation principle for the top eigenvalue (or singular value if the matrices are rectangular) of the sum or the product of two \emph{arbitrary} matrices; by which we mean that each matrix can be either taken from an invariant ensemble or can be a (randomly rotated) diagonal matrix. In particular, our results allows one to recover the  three specific cases considered previously in Refs. \cite{GuionnetMaida20,McKenna21,Maillard21} and to obtain new results for cases that have not been previously considered. To obtain our result, we first  introduce  an \emph{invariant ensemble with a wall}, a natural generalization of the classical invariant ensemble. From the point of view of large deviations, we show that for a proper choice of the position of the wall (namely the wall is exactly at the edge of the spectrum), we recover the cases of deterministic matrices. Second, we extend the tilting method  introduced in Ref. \cite{GuionnetMaida20}. We argue that when one is considering respectively the product of two symmetric matrices or the sum of two rectangular matrices, one should replace the additive spherical integral with respectively the multiplicative spherical integral and the  rectangular spherical integral. In each case, we give a natural interpretation of those spherical integrals as the partition function of a disordered system. Based on ideas develop in Ref. \cite{foini2021annealed}, we give the precise asymptotic behavior for the \emph{annealed free energies} of any invariant random matrix, going beyond the case of GOE and Wishart matrices which can tackled by direct Gaussian integration. Combining this result with  the asymptotic behavior of the \emph{quenched free energy} of those three (additive, multiplicative, rectangular) spherical integrals derived in Refs. \cite{GuionnetMaida05,Collins07,MergnyPotters20,husson2021asymptotic,benaych2011rectangular} allows us to get the rate function in each case. 
 
The rest of the paper is organized as follows: in Sec.\ \ref{sec:LDP_1RM}, we recall the main results concerning large deviations of the top eigenvalue of \emph{one} random matrix from a (classical) invariant ensemble. This allows us to introduce the main concepts and notations used in this paper. In particular, we define the so-called invariant ensemble with a wall and also described the case of bi-invariant rectangular random matrices. In Sec.\ \ref{Sec:Tilt}, we introduce the main tool to compute the rate function: the tilting method with spherical integrals. Our description of the tilting method has been made in a general framework in order to describe the idea of the computation for each of the three cases (sum of symmetric matrices, product of symmetric matrices, sum of rectangular matrices) simultaneously. We then go into detail for each case separately. We consider the case of the sum of two symmetric matrices in Sec.\ \ref{sec:LDP_sum}, the product of two symmetric matrices in Sec.\ \ref{sec:LDP_prod} and  the sum of rectangular matrices in Sec.\ \ref{sec:LDP_Rect}. In each Section, we give explicitly the expression for the rate function together with concrete examples. This rate function admits up to three different regimes. Based on the similarity with the rate function of the simpler model of a rank-one deformation of one invariant random matrix, described in App.\ \ref{sec:Ap:rk1_deformation} and the rate function of the toy model of the sum of two rank-one matrix, described in App.\ \ref{sec:Ap:RateFunction.rk1rk1}, we give a natural interpretation for each regime. In App.\ \ref{sec:Ap:prop}, we recall classical properties of RMT and \emph{free probability} that are used in the main text and  App.\ \ref{sec:Ap:As_AFE_QFE} contains derivation of the quenched and annealed free energies.

\section{Reminder on right large deviation for one random matrix: The pulled Coulomb gas approach}
\label{sec:LDP_1RM}
Before considering the addition or multiplication of two random matrices, let's first briefly recall the simpler case of \emph{one} random matrix in an invariant ensemble. We first start with \emph{symmetric} random matrices.
\subsection{Classical Rotationally Invariant Ensemble}
\label{sec:RotInvEns}
 For an analytic confining potential $V(.)$, we say that a matrix is drawn from a (classical rotationally) invariant ensemble if the probability to observe the  matrix $\mathbf{A} \equiv \mathbf{A}_N$ in a region $R$ in the space of $(N \times N)$ symmetric matrices is defined as:
\begin{align}
\label{eq:InvEns.1}
    \mathbb{P}_{V} \left[ \mathbf{A} \in R \right] &= \frac{1}{Z_{N,V}} \int_{R} \mathrm{e}^{ - \frac{N}{2} \Tr V(\mathbf{A}) } \mathrm{d}\mathbf{A} \, ,&&
\end{align}
where $\mathrm{d}\mathbf{A}$ is the Lebesgue measure over the space of $(N \times N)$ symmetric matrices and $Z_{N,V}$ is a constant ensuring that this probability is normalized to one. When considering positive semi-definite matrices, we will implicitly assume the potential to be defined on $\mathbb{R}_+$.  For any orthogonal matrix $\mathbf{O}$ we have $V(\mathbf{O} \mathbf{A} \mathbf{O}^{\mathsf{T}}) =  \mathbf{O} V(\mathbf{A})  \mathbf{O}^{\mathsf{T}}$ which together with the cyclical property of the trace gives:
\begin{align}
\label{eq:propInvEns}
     \mathbb{P}_{V} \left[ \mathbf{O} \mathbf{A} \mathbf{O}^{\mathsf{T}}  \in R \right] &= \mathbb{P}_{V} \left[ \mathbf{A} \in R \right] & \quad  \left(\forall  \mathbf{O} \in \mathsf{O}(N) \right) \, , &&
\end{align}
where $\mathsf{O}(N)$ is the group of $(N \times N)$ orthogonal matrix; hence the name (rotationally) invariant ensemble.

\vskip 0.3cm 
\noindent \textit{Example (GOE matrices):} If one considers  the elements $A_{ij}$ of the matrix $\mathbf{A}$ to be independent (up to the symmetry) Gaussian random variables with mean zero and variance $\sigma^2/N$ for the off-diagonal elements and $2\sigma^2/N$ for the diagonal elements, then this ensemble corresponds to the famous \emph{Gaussian Orthogonal Ensemble}  (GOE) for which the potential is equal to:
    \begin{align}
    \label{eq:pot_GOE}
        V(\lambda) &= \frac{\lambda^2}{2\sigma^2} \, . &&
    \end{align}
\vskip 0.3cm 
\noindent \textit{Example (Wishart matrices):} Consider a $(N \times M )$ rectangular matrix $\mathbf{X}$ with $M \geq N$, where the elements of $\mathbf{X}$ are Gaussian independent random variables with mean zero and variance one, from which we construct the $(N \times N)$ square matrix $\mathbf{A} := \frac{1}{M}\mathbf{X} \mathbf{X}^{\mathsf{T}}$.  The matrix $\mathbf{A}$ is a (Gaussian White) \emph{Wishart} matrix. It is rotationally invariant with potential:
    \begin{align}
    \label{eq:pot_LOE}
        V_{N}(\lambda) &=  \frac{M}{N} \lambda + \left(1 - \frac{M}{N} + \frac{1}{N}\right) \log \lambda \to V(x) = \frac{\lambda}{q} + \left( 1 - \frac{1}{q} \right) \log \lambda \, , &&
    \end{align}
    where the asymptotic behavior in Eq.\ \eqref{eq:pot_LOE} corresponds to the double scaling limit $N \to \infty$ and $M \to \infty$ with $\frac{N}{M} \to q \in (0,1)$. The invariant ensemble with the potential $V(.)$ in  Eq.\ \eqref{eq:pot_LOE} is sometimes known as the \emph{Laguerre Orthogonal Ensemble}.

\vskip 0.3cm 
\noindent From there, it is a standard result of RMT that $\mathbf{A}$ admits the following spectral decomposition $\mathbf{A} = \mathbf{V} \mathrm{Diag}( \lambda_1(\mathbf{A}), \dots, \lambda_N(\mathbf{A}))  \mathbf{V}^{\mathsf{T}}$, where the matrix of eigenvectors $\mathbf{V}$ is taken uniformly over $\mathsf{O}(N)$. To ease notation when it is needed, we  write $\lambda_i \equiv \lambda_i(\mathbf{A})$ for the eigenvalues of the matrix $\mathbf{A}$. The joint density of the eigenvalues is given by:
\begin{align}
\label{eq:JointDensityEigvals}
    \mathcal{P}_N(\lambda_1, \dots, \lambda_N) &= \frac{1}{Z_N} \mathrm{exp} \left[ - \frac{N}{2} \sum_{i=1}^N V(\lambda_i) + \frac{1}{2} \sum_{ij | i \neq j} \log | \lambda_i - \lambda_j| \right] \, . &&
\end{align}
The term $ \sum_{ij | i \neq j} \log | \lambda_i - \lambda_j|$ is the Jacobian of the change of variable $\mathbf{A} \to ( \mathbf{V}, \{\lambda_i\})$ and is exactly the pairwise repulsive interaction of the $2d$-Coulomb gas. In the large $N$ limit, the empirical (random) spectral density converges to a non-random smooth density $\mu_A(.)$:
\begin{align}
\label{eq:conv_ESD}
    \mu_{\mathbf{A}}(\lambda) &:= \frac{1}{N} \sum_{i=1}^N \delta(\lambda - \lambda_i(\mathbf{A})) \underset{N \to \infty}{\to} \mu_A(\lambda) \, . && 
\end{align}
where $\mu_A(.)$ is the solution of the Tricomi problem:
\begin{align}
\label{eq:Tricomi}
  \mathrm{P.V.} \int \frac{\mu_A(\lambda')}{\lambda-\lambda'} \mathrm{d}\lambda' &= \frac{V'(\lambda)}{2} \, , &&
\end{align}
where $\mathrm{P.V.}$ stands for Principal Value. 
\vskip 0.3cm 
\noindent  \textit{Example (GOE matrices and the semi-circle distribution): } For GOE matrices with  a quadratic potential given by Eq.\ \eqref{eq:pot_GOE}, the solution of this Tricomi equation is given by the famous \emph{semi-circle distribution}:
        \begin{align}
        \label{eq:SC_dist}
        \mu_{\mathrm{sc}}(\lambda) &= \frac{\sqrt{4\sigma^2 - \lambda^2}}{2\pi \sigma^2} \mathbb{I}_{[-2 \sigma, 2 \sigma]} \, .&& 
    \end{align}
    where $\mathbb{I}_{[a,b]}$ is the indicator function, it is equal to $1$ if $\lambda \in [a,b]$ and zero otherwise.
\vskip 0.3cm 
\noindent  \textit{Example (Wishart matrices and the Mar\v{c}enko-Pastur distribution):}  For  Wishart matrices with potential given by Eq.\ \eqref{eq:pot_LOE}, one obtains the  \emph{Mar\v{c}enko-Pastur distribution}:
    \begin{align}
    \label{eq:MP_dist}
        \mu_{\mathrm{MP}_q}(\lambda) &= \frac{\sqrt{( \mathrm{a}_+ -\lambda)(\lambda - \mathrm{a}_-)}}{2\pi q \lambda} \mathbb{I}_{[\mathrm{a}_-, \mathrm{a}_+]} \, . && 
    \end{align}
    where the edges are given by $\mathrm{a}_{\pm} = (1  \pm \sqrt{q})^2$. 

\subsection{Invariant Ensemble with a wall}
\label{sec:InvEns_w_wall}
\noindent Importantly, Eq.\ \eqref{eq:JointDensityEigvals} for the joint density of eigenvalues still makes sense if we introduce a \emph{wall} at  a position $w_A\ge\mathrm{a}_+$ beyond which the potential is infinite. We say that a matrix is taken from an invariant ensemble with a wall at $w_A$, which we denote by $\mathbf{A} \sim \mathbb{P}_{V,w_A}$, if $\mathbf{A} = \mathbf{O} \mathrm{Diag} \left(a_1, \dots, a_N \right) \mathbf{O}^{\mathsf{T}}$ with $\mathbf{O}$ uniform over $\mathsf{O}(N)$ and the $\{ \lambda_i \}$ follow the joint law of Eq.\ \eqref{eq:JointDensityEigvals} with $V(.)$ a confining potential such that $V(x>w_A) = \infty$. By construction, we still have the property \eqref{eq:propInvEns} for random matrices taken from this ensemble. It is important to notice that the introduction of this wall does not change the limiting equilibrium density $\mu_A(.)$ since $w_A \ge \mathrm{a}_+$ and the solution of the Tricomi problem of Eq.\ \eqref{eq:Tricomi} only depends on the values of the potential between the two edges $\mathrm{a}_{\pm}$.  The introduction of these invariant ensembles with a wall might seem odd at first, but as we will see later on, this construction allows the study of the sum of two matrices where one (or both\footnote{When considering two diagonal  matrices,  we will be considering the sum $\mathbf{C} = \tilde{\mathbf{A}} + \mathbf{O} \tilde{\mathbf{B}} \mathbf{O}^{\mathsf{T}}$  with $\mathbf{O}$  uniform over $\mathsf{O}(N)$, such that $\tilde{\mathbf{A}}$ and $\mathbf{O} \tilde{\mathbf{B}} \mathbf{O}^{\mathsf{T}}$ are asymptotically free and the spectrum of the  eigenvalues of the sum (respectively of the product) is given asymptotically by the \emph{free convolution}  described in Sec.\ \ref{sec:add_free_conv}  (resp. the \emph{multiplicative free convolution} described in Sec.\ \ref{sec:mult_free_conv}).   })  matrix is a fixed diagonal matrix: 
\begin{align}
\label{eq:DeterministicMat}
    \tilde{\mathbf{A}} = \begin{pmatrix}
\tilde{a}_1 & 0 & \dots & 0 \\
0  & \ddots & & \vdots \\
\vdots & & \ddots &0 \\  
0 & \dots & 0& \tilde{a}_N
\end{pmatrix} \, . &&
\end{align}
The $\{\tilde{a}_i\}$ are 'frozen' sequences of numbers such that as we increase the size of the matrix, their empirical distribution converges to the same $\mu_A(.)$:
\begin{align}
\label{eq:conv_ESD.2}
      \mu_{\tilde{\mathbf{A}}}(\lambda) &:= \frac{1}{N} \sum_{i=1}^N \delta(\lambda - \tilde{a}_i) \underset{N \to \infty}{\to} \mu_A(\lambda) \, . && 
\end{align}
and importantly the minimum and the maximum of the $\{\tilde{a}_i\}$ converge to the edges $\mathrm{a}_-$ and $\mathrm{a}_+$ of $\mu_A(.)$ (that is, in the large $N$ limit, no outlier survives). To be more concrete, a typical example for the choice of the $\{ \tilde{a}_i \}$ is to take them as the quantiles of the distribution $\mu_A(.)$, sometimes called the "classical positions" of the particles. For a given $N$ and for each $i$ ranging from $1$ to $N$, this means that  $\tilde{a}_i$ is solution of the integral equation:
\begin{align}
    \int_{\mathrm{a}_-}^{\tilde{a}_i} \mu_A(\lambda) \mathrm{d}\lambda &= \frac{i}{N+1} \, , &&
\end{align}
and by construction their empirical distribution converge to $\mu_A(.)$. Another natural choice is to draw independently each $\tilde{a}_i$ from  the distribution $\mu_A(.)$.  Now in general, diagonal matrices of the form of Eq.\ \eqref{eq:DeterministicMat} are very different from random matrices taken from an invariant ensemble since their eigenvalues are fixed and don't have a true repulsion as in Eq.\ \eqref{eq:JointDensityEigvals}. However, when one consider the problem of large deviation of the top eigenvalue, we will see that they behave as an invariant ensemble with a wall exactly at the edge ($w_A = \mathrm{a}_+$). This will be made more precise later on, but one can already see that if there is a wall exactly at the edge, then at finite $N$ the top eigenvalue cannot fluctuate outside the support of the bulk density $\mu_A(.)$ and hence it is in a sense fixed. This construction is  superfluous when considering just one matrix (in this case the question of large deviation for the top eigenvalue of a deterministic matrix as in Eq.\ \eqref{eq:DeterministicMat} is trivial), but turns out to be very convenient when considering the sum or product of two matrices. 

\vskip 0.3cm 
\noindent \textit{Remark (wall at infinity and classical invariant ensemble):} One may note that we recover the case of classical invariant ensemble of Sec.\ \ref{sec:RotInvEns} by sending $w_A \to \infty$. 

\begin{figure}
     \centering
         \includegraphics[width= 0.55\textwidth]{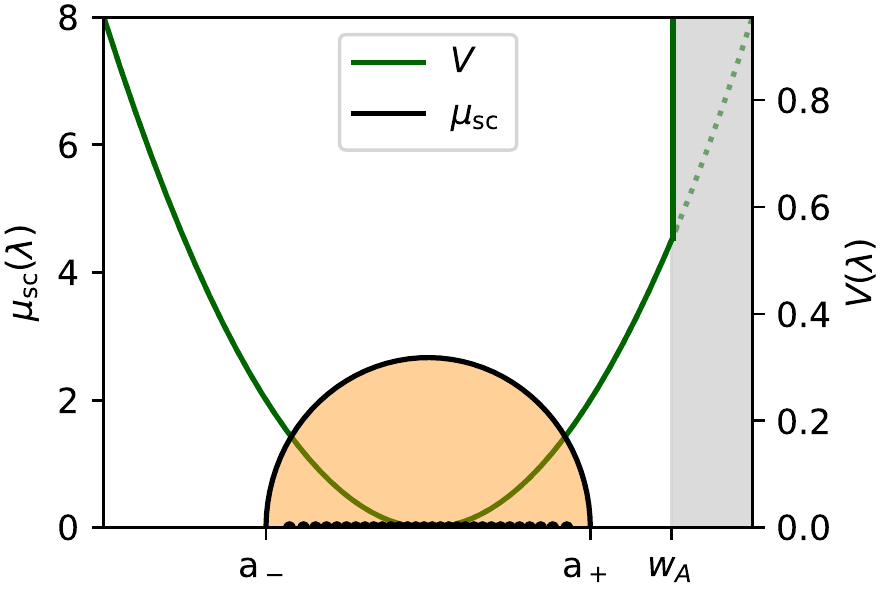}
\caption{Potential and limiting density for a random matrix taken from an invariant ensemble with a wall. Beyond the wall, the potential is infinite. The limiting density is the same as if there were no wall $(w_A \to \infty)$ since $w_A \geq \mathrm{a}_+$. The black dots represent a 'typical' configurations  of the eigenvalues at finite $N$.     } 
\label{fig:Pot_w_wall}
\end{figure}

\subsection{Coulomb gas approach}
\label{sec:LDP_Cg}
\noindent Coming back to random symmetric matrices from an invariant ensemble, if we order the eigenvalues in decreasing order, $\lambda_1(\mathbf{A}) \geq \dots \geq \lambda_N(\mathbf{A})$ then we have: 
\begin{align}
\label{conv_TopToEdge}
    \lambda_1(\mathbf{A}) &\underset{N \to \infty}{\to} \mathrm{a}_+ \, . && 
\end{align}
This result is valid in the formal $N \to \infty$ and a natural question is to estimate the probability at large but finite $N$, $\mathbb{P} \left[ \lambda_1(\mathbf{A}) \simeq x \right]$, of having the top eigenvalue $\lambda_1(\mathbf{A})$  at a position $x$ different from its typical value $\mathrm{a}_+$. If  the density $\mu_A(.)$ is 'non-critical', by which we mean that it behaves as a square-root near the edge $\mathrm{a}_{+}$,
\begin{align}
\label{eq:nondegen_density}
    \mu_A(x) & \underset{x \nearrow \mathrm{a}_+}{\sim} \frac{\gamma_0^{3/2}}{\pi}\sqrt{\mathrm{a}_+-x} \quad \, ,  && 
\end{align}
then the small deviations of $\lambda_1(\mathbf{A})$ around the limiting value $\mathrm{a}_+$ are given by the Tracy-Widom law for fluctuations of order $ O(N^{-\frac{2}{3}})$, see Refs. \cite{Tracy1994,Tracy1996}. For a value of $x$ far from the edge $\mathrm{a}_+$, one is outside the scope of the Tracy-Widom regime describing typical fluctuations and one needs to estimate a very rare event dictated by a \emph{large deviation principle}. This can be summarized (see Ref. \cite{Majumdar14}) by the following set of equations:
\begin{align}
    \label{eq:summary}
     \mathbb{P} \left[ \lambda_1(\mathbf{A}) \simeq x \right]&\approx \left\{
    \begin{array}{lll}
        \mathrm{exp} \left[ - N^2 \Psi_-(x) +o(N^2)  \right]  
& \mbox{for } x < \mathrm{a}_+  \mbox{ and } |x-\mathrm{a}_+| \sim O(1)  \, ,\\
\\
      \gamma_0 N^{2/3}  (\mathcal{F}^{(1)})'\left( \gamma_0 N^{2/3}(x-\mathrm{a}_+) \right) & 
\mbox{for } |x -\mathrm{a}_+| \sim O(N^{-\frac{2}{3}}) \, , \\
\\
        \mathrm{exp} \left[ - N \Psi(x) +o(N) \right] & 
\mbox{for } x>\mathrm{a}_+ \mbox{ and } |x-\mathrm{a}_+| \sim O(1) \, .
    \end{array}
\right. &&
\end{align}
The function $\mathcal{F}^{(1)}$ in Eq.\ \eqref{eq:summary} is the $\beta=1$ Tracy-Widom function. The scaling (or speed of convergence) of the large deviation principle is different if $x$ is either above or below the edge $\mathrm{a}_+$, and can be naturally interpreted thanks to the $2d$-Coulomb gas picture.

\subsubsection{The pulled Coulomb gas ($x>\mathrm{a}_+$)}
\label{sec:PullCoulombgas}
Let's consider the case $x > \mathrm{a}_+$ which is the main subject of this note. Integrating the joint density in Eq.\ \eqref{eq:JointDensityEigvals}, one has that computing the (logarithm of the)  probability $\mathbb{P}[\lambda_1(\mathbf{A}) \simeq x]$ is equivalent to computing the difference of energy between the configuration of a $2d$-Coulomb gas where the top particle is \emph{pulled} at the position $x$ and the configuration of the unperturbed $2d$-Coulomb gas. For the perturbed gas,  we are just moving one particle away from the bulk, and thus we expect that this perturbation does not change the equilibrium density $\mu_A$ of the $N-1$ other particles inside the bulk. This induces the scaling in Eq.\ \eqref{eq:summary} for $x>\mathrm{a}_+$, and the right rate function is given by:
\begin{align}
\label{eq:Psi_1RM}
    \Psi(x) &= \frac{1}{2} \left[ V(x) - V(\mathrm{a}_+) - 2 \int \log( x - \lambda) \mu_A(\lambda) \mathrm{d}\lambda + 2  \int \log( \mathrm{a}_+ - \lambda) \mu_A(y) \mathrm{d}\lambda  \right] \, , &&
\end{align}
which can be written in integral form as: 
\begin{align}
\label{eq:Psi_1RM.1}
     \Psi(x) &= \int_{\mathrm{a}_+}^{x} \left( \frac{V'(t)}{2} - g_A(t) \right) \mathrm{d}t \, , &&
\end{align}
where $g_A(.)$ is the \emph{Stieltjes transform} of $\mu_A$:
\begin{align}
    \label{eq:def:Stieltjes}
    g_A(z) &:= \int_{\mathrm{a}_-}^{\mathrm{a}_+}  \frac{\mu_A(\lambda)}{z-\lambda} \mathrm{d}\lambda \, . &&
\end{align}
It will be convenient to introduce the \emph{second branch of the Stieltjes transform}, (see App.\ \ref{sec:Ap:Stieltjes}), which, in the case of invariant ensemble, is given for $x>\mathrm{a}_+$ by:
\begin{align}
\label{eq:def:2ndBranchStieltjes}
    \bar{g}_A(x) &:= V'(x) - g_A(x) \, , && 
\end{align}
so that from Eq.\ \eqref{eq:Psi_1RM.1} we can interpret the rate function as (half) the area between the two branches of the Stieltjes transform up to the position $x$:
\begin{align}
\label{eq:Psi_1RM.2}
     \Psi(x) &= \frac{1}{2} \int_{\mathrm{a}_+}^{x} \left( \bar{g}_A(t) - g_A(t) \right)  \mathrm{d}t \, . &&
\end{align}
Note that for $g_A(\mathrm{a}_+) = \infty$, the rate  function is finite. 
\vskip 0.3cm
\noindent \textit{Remark (Recovering the potential):} From the eigenvalue density $\mu_A(\lambda)$ and the rate function $\Psi(x)$, one can recover the potential $V(x)$ (up to an arbitrary constant) over the whole range of possible eigenvalues. Indeed, from the density one can compute the Stieltjes transform using Eq.\ \eqref{eq:def:Stieltjes} and rewrite Eqs.\ \eqref{eq:Tricomi} and \eqref{eq:Psi_1RM.1} as\footnote{The potential for $x<\mathrm{a}_-$ can be recovered similarly using the rate function for $-A$.}
\begin{align}
\label{eq:effective_potential}
    \frac{V'(x)}{2}=\left\{
    \begin{array}{lll}
    \mathfrak{Re}\, g_A(x - \mathrm{i} 0^+ ) &  \mbox{for } \mathrm{a}_-\leq x\leq \mathrm{a}_+\\
    \\
    \Psi'(x)+g_A(x) & \mbox{for } x\geq \mathrm{a}_+\, .\\
    \end{array}\right.
\end{align}
 For random matrices that are not necessarily drawn from an invariant ensemble, we can use this formula to define an effective potential, i.e. the potential of the invariant ensemble that has the same density and rate functions. In particular, for a fixed diagonal matrix with eigenvalue density $\mu_A(x)$ the rate function is infinite, and the effective potential is infinite beyond $w=\mathrm{a}_+$ exactly as described in Sec.\ \ref{sec:InvEns_w_wall}. 

\begin{figure}[h]
\centering
    \subfloat{\includegraphics[width=0.45\linewidth]{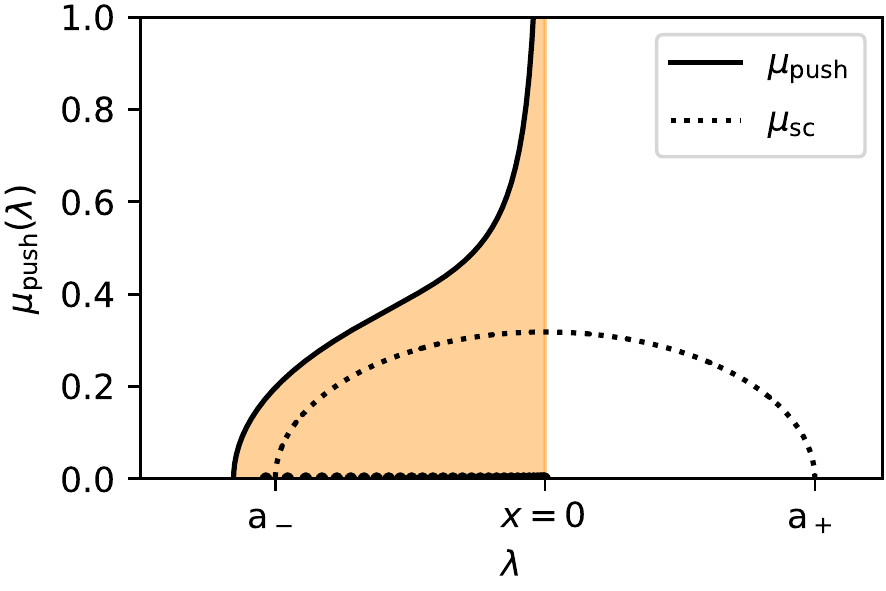}}
    \qquad
    \subfloat{\includegraphics[width=0.45\linewidth]{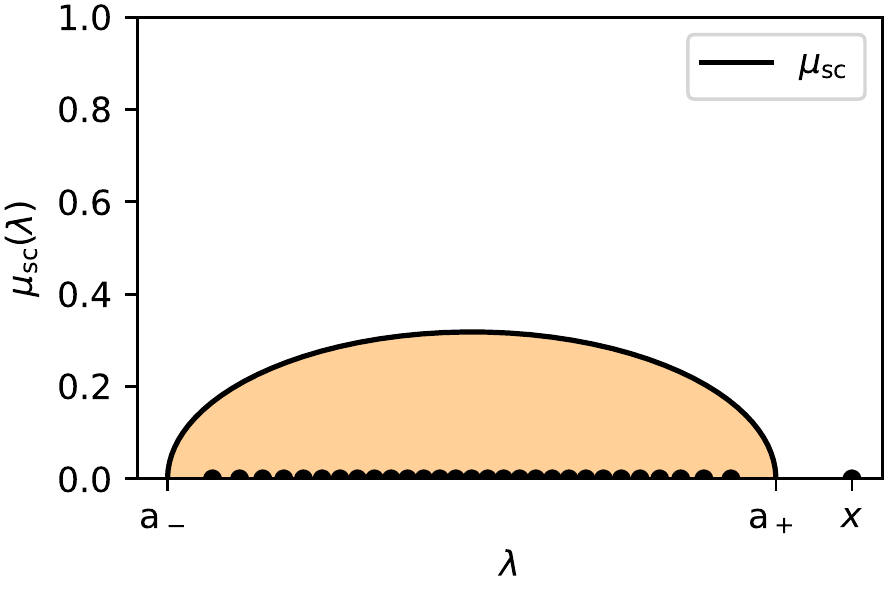}}
\caption{On the left, a representation of a 'typical' configuration of a pushed-to-the-origin Coulomb gas in a harmonic potential. To have the top eigenvalue at the position $x=0$, one needs to also push all the other eigenvalues to the right and in the large $N$ limit, this creates a different limiting density (with black solid line) compared to the unperturbed semi-circle distribution (with dotted line). On the right, a 'typical' configuration  corresponding to a Coulomb gas in a harmonic potential pulled at the position $x=2.5 > \mathrm{a}_+ =2$. Only the top eigenvalue pops out of the limiting distribution. }
    \label{fig:push_pull}
\end{figure}

\begin{figure}[h]
\centering
    \subfloat{\includegraphics[width=0.45\linewidth]{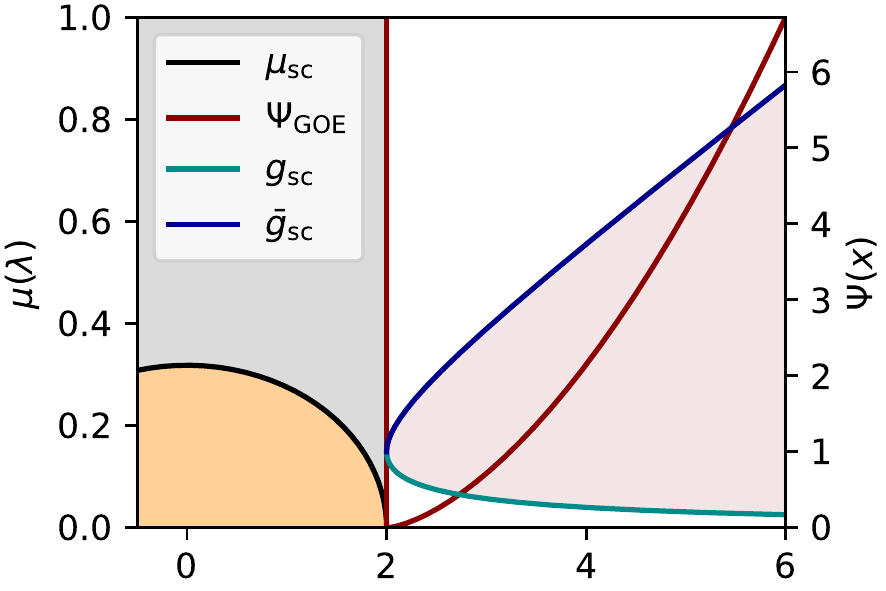}}
    \qquad
    \subfloat{\includegraphics[width=0.45\linewidth]{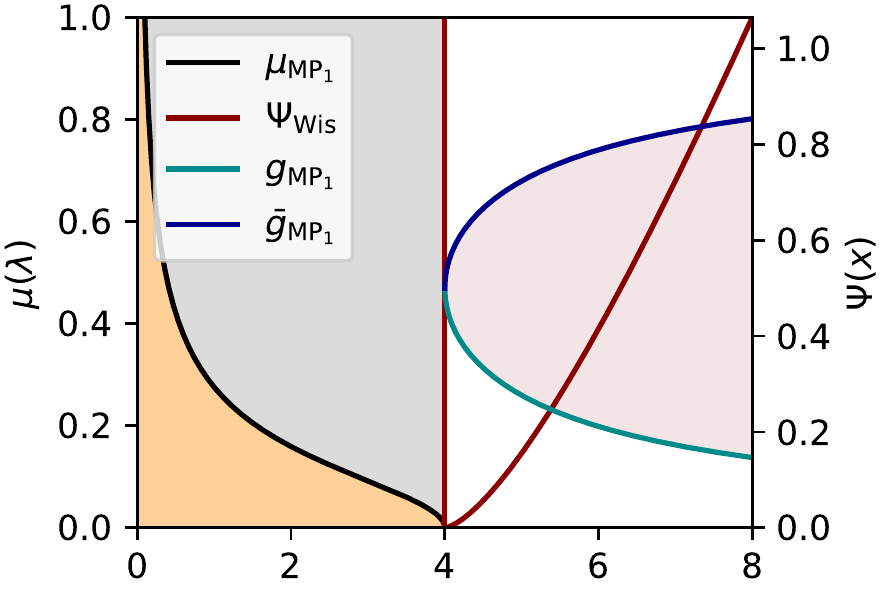}}
\caption{On the left, the rate function (in red) for the largest eigenvalue of a GOE matrix with $\sigma=1$, whose expression is given by Eq.\ \eqref{eq:RateFunction_GOE}. On the right, the rate function (in red)  for the largest eigenvalue of a Wishart matrix with $q=1$, given by Eq.\ \eqref{eq:RateFunction_LOE.q1}. In each case, the rate function is infinite for values below the edge of the limiting distribution and is otherwise  given as half the area between the curve of the second branch of the Stieltjes transform (in blue) and the curve of the Stieltjes transform (in cyan), see Eq.\ \eqref{eq:Psi_1RM.2}.  }
    \label{fig:RF_GOE_Wis}
\end{figure}

\vskip 0.3cm
\noindent \textit{Example (Rate function for GOE matrices):} for a  GOE matrix, whose limiting spectrum is the semi-circle distribution of Eq.\  \eqref{eq:SC_dist}, the Stieltjes transform is given by:
\begin{align}
    \label{eq:Stieltjes_SC}
    g_{\mathrm{sc}}(z) &= \frac{z - \sqrt{z^2 - 4\sigma^2}}{2 \sigma^2} &&  (\mbox{for } z\geq 2\sigma) \, ,
\end{align}
and the second branch of the Stieltjes is given by: 
\begin{align}
    \label{eq:2BStieltjes_SC}
    \bar{g}_{\mathrm{sc}}(z) &= \frac{z + \sqrt{z^2 - 4\sigma^2}}{2 \sigma^2} &&  (\mbox{for } z\geq 2\sigma) \, ,
\end{align}
and therefore integrating according to Eq.\ \eqref{eq:Psi_1RM.1}, the  right large deviation of the top eigenvalue is given by the rate function:
\begin{align}
    \label{eq:RateFunction_GOE}
    \Psi_{\mathsf{GOE}}(x) &=  \frac{x \sqrt{x^2 - 4 \sigma^2}}{4\sigma^2} +  \log \left( \frac{2 \sigma}{\sqrt{x^2 - 4\sigma^2}+x} \right)  \, . &&
\end{align}
The two branches of the Stieltjes transform and the rate function are given in Fig.\ \ref{fig:RF_GOE_Wis} (Left).
\vskip 0.3cm
\noindent \textit{Example (Rate function for Wishart matrices):} For a (Gaussian White) Wishart  matrix, whose limiting spectrum is the Mar\v{c}enko-Pastur distribution of Eq.\  \eqref{eq:MP_dist}, the Stieltjes transform is given by:
\begin{align}
    \label{eq:Stieltjes_MP}
    g_{\mathrm{MP}_q}(z) &= \frac{z -(1-q) - \sqrt{z - \mathrm{a}_-} \sqrt{z - \mathrm{a}_+}}{2qz} &&  (\mbox{for } z\geq \mathrm{a}_+) \, ,
\end{align}
with $\mathrm{a}_{\pm} = (1\pm \sqrt{q})^2$, and the second branch is given by: 
\begin{align}
    \label{eq:2BStieltjes_MP}
    \bar{g}_{\mathrm{MP}_q}(z) &= \frac{z -(1-q) + \sqrt{z - \mathrm{a}_-} \sqrt{z - \mathrm{a}_+}}{2qz} &&  (\mbox{for } z\geq \mathrm{a}_+) \, .
\end{align}
The two branches of the Stieltjes transform are illustrated in Fig.\ \ref{fig:RF_GOE_Wis} (Right). The right rate function is given by:
\begin{align}
    \label{eq:RateFunction_LOE}
    \Psi_{\mathsf{Wish}}(x) &= \int_{\mathrm{a}_+}^x \frac{\sqrt{(t-\mathrm{a}_+) ( t - \mathrm{a}_-)}}{2qt} \mathrm{d}t  \, . &&
\end{align}
and is also represented in  Fig.\ \ref{fig:RF_GOE_Wis} (Right). Note that in integral in Eq.\ \eqref{eq:RateFunction_LOE} can be computed analytically, but the result is not very enlightening. For $q=1$ the rate function simplifies considerably, and we have:
\begin{align}
    \label{eq:RateFunction_LOE.q1}
    \Psi_{\mathsf{Wish}}(x) &= \frac{\sqrt{x(x-4)}}{2} + \log \left( \frac{x-2 -\sqrt{x(x-4)} }{2}\right)   & (\mbox{for } q=1) \, . &&
\end{align}
In this case, one may notice the following identity: 
\begin{align}
    \label{eq:RateFunction_LOE.q1.2}
    \Psi_{\mathsf{Wish}}(x^2) &=  2 \Psi_{\mathsf{GOE}}(x)  & (\mbox{for } q=1) \, , &&
\end{align}
which will be discussed in more detail in Sec.\ \ref{sec:BiInvEns}  concerning bi-invariant rectangular matrices. 

\vskip 0.3cm
\noindent \textit{Remark (Behavior near the edge and the Tracy-Widom '3/2' scaling):}
If one is looking at a non-critical density satisfying the condition of Eq.\ \eqref{eq:nondegen_density} near the edge, then both the Stieltjes and its second branch behaves near the top edge as:
\begin{align}
    \label{eq:3o2scaling_Stieltjes}
    \begin{array}{ll}
          g_A(\mathrm{a}_+ + \epsilon) &=    g_A(\mathrm{a}_+) - \gamma_0^{\frac{3}{2}} \sqrt{\mathrm{a}_+ + \epsilon} +o(\sqrt{\epsilon})  \, , \\
          \bar{g}_A(\mathrm{a}_+ + \epsilon) &=    g_A(\mathrm{a}_+) + \gamma_0^{\frac{3}{2}} \sqrt{\mathrm{a}_+ + \epsilon} +o(\sqrt{\epsilon}) \, . 
    \end{array}&&
\end{align}
so approximating the integral of Eq.\ \eqref{eq:Psi_1RM.2} by Euler method, at first order one has for the rate function:
\begin{align}
    \label{eq:3o2scaling_RateFunction}
    \Psi(\mathrm{a}_+ + \epsilon) &= \frac{2\gamma^{3/2}}{3}\epsilon^{3/2} + o\left(\epsilon^{\frac{3}{2}}\right) \, ,&&
\end{align}
and hence the probability behaves as
\begin{align}
    \label{eq:3o2scaling_Proba}
\mathbb{P} \left[ \lambda_1(\mathbf{C}) \simeq  x \right] \approx \mathrm{exp} \left[- \frac{2 }{3} u^{3/2} +o(N) \right] \quad \quad  \text{where } \, u=\gamma N^{2/3}(x-\mathrm{c}_+) \, .
\end{align}
The scaling of the reduced variable $u$ and the asymptotic behavior matches the large argument behavior of the Tracy-Widom regime which describes the probability of finding an eigenvalue near the edge, see Ref. \cite{Majumdar14}.

\subsubsection{A word on the pushed Coulomb gas ($x < \mathrm{a}_+$)}
\label{sec:PushCoulombgas}
For $x <\mathrm{a}_+$, one can still use the Coulomb gas analogy, but now the perturbed $2d$-Coulomb gas is compressed such that its top particle is at the position $x$. Unlike the case $x >\mathrm{a}_+$, the equilibrium measure in the bulk is modified, since one needs to \emph{push} a large fraction of the particles to satisfy this constraint. This explains the different scaling in Eq.\ \eqref{eq:summary} in this scenario, and we refer to Ref. \cite{Majumdar14} for a description of the rate function in this case. This left large deviation  will not be discussed in the rest of this note.

\subsection{The case of one rectangular random matrices}
\label{sec:Rect}
In this paragraph, we describe the case of rectangular random matrices. The reader only interested in symmetric random matrices may skip this section and jump directly to Sec.\ \ref{Sec:Tilt}.  

\subsubsection{Singular value decomposition}
\label{sec:svd}

\vskip 0.3cm
\noindent Let $\mathbf{A} \equiv  \mathbf{A}_{N,M}$ be a $(N \times M)$ real matrix. Without loss of generality, we consider $M \geq N$. We will be interested in the double scaling limit where $N \to \infty$ and $M \to \infty$   but the ratio stays finite:\footnote{In the rest of this section and when we are considering rectangular matrices, all limits are assumed to be taken under this setting.}
\begin{align}
\label{Ap:double_scaling_limit}
    \frac{N}{M} &\to q \in (0,1) \, . && 
\end{align}
The singular value decomposition (SVD) of  $\mathbf{A}$ is given by:
\begin{align}
\label{Ap:SVD_decomposition}
    \mathbf{A} &= \mathbf{U}_1 \mathbf{\Sigma} \mathbf{V}_1^{\mathsf{T}} \, , &&
\end{align}
where $\mathbf{U}_1$ (resp. $\mathbf{V}_1$) is a $(N \times N)$ (resp. $(M \times M)$) orthogonal matrix, $\mathbf{\Sigma}$ is a diagonal rectangular matrix: 
\begin{align}
\label{Ap:Rect_Diag_mat}
\mathbf{\Sigma} &=
\begin{pmatrix}
    s_1(\mathbf{A}) & 0  & 0\\
    0 & \ddots & 0 & \mathbf{0}_{M - N} \\
    0 & 0 & s_{N}(\mathbf{A})
\end{pmatrix} \, ,&&
\end{align}
and the $s_i(\mathbf{A})$ are the \emph{singular values} of $\mathbf{A}$. They are related to the eigenvalues of the matrix $\mathbf{A} \mathbf{A}^{\mathsf{T}}$ by the relation:
\begin{align}
\label{eq:prop_sv_eigv}
    s_i(\mathbf{A}) &= \sqrt{\lambda_i(\mathbf{A} \mathbf{A}^{\mathsf{T}}) } \, . &&
\end{align}

\subsubsection{Bi-invariant ensemble}
\label{sec:BiInvEns}
\noindent We say that a rectangular matrix is taken from a bi-invariant ensemble if its law can be written as: 
\begin{align}
    \label{eq:proba_Birot}
    \mathbb{P}_{V} \left[ \mathbf{A} \in R \right] &= \frac{1}{Z_{N,V}} \int_{R} \mathrm{e}^{ - \frac{N}{2} \Tr V(\mathbf{A} \mathbf{A}^{\mathsf{T}}) } \mathrm{d}\mathbf{A} \, ,&&
\end{align}
for an analytic potential $V(.)$, since it satisfies the property:
\begin{align}
\label{eq:prop_Birot}
     \mathbb{P}_{V} \left[ \mathbf{U} \mathbf{A} \mathbf{V}^{\mathsf{T}}  \in R \right] &= \mathbb{P}_{V} \left[ \mathbf{A} \in R \right] \, ,&&
\end{align}
for any $\mathbf{U} \in  \mathsf{O}(N)$ and any  $\mathbf{V} \in  \mathsf{O}(M)$. The change of variable $\mathbf{A} \to \left(\mathbf{U}, \{ s_i \equiv s_i \left( \mathbf{A} \right) \}, \mathbf{V} \right)$ introduces a Jacobian \cite{hua1963harmonic} which is equal to:
\begin{align}
    \mathrm{Jac} \left[ \mathbf{A} \to \left(\mathbf{U}, \{ s_i \} , \mathbf{V} \right) \right] &\propto  \mathrm{e}^{ \frac{1}{2} \sum_{i,j | i \neq j} \log |s_i^2 - s_j^2|} \prod_{i=1}^{N} s_i^{M -N} \, . &&
\end{align}
In this case, the matrix $\mathbf{A}$ admits the SVD of Eq.\ \eqref{Ap:SVD_decomposition} where $ \mathbf{U}_1$ (resp. $ \mathbf{V}_1$) is taken uniformly over  $\mathsf{O}(N)$ (resp. over  $\mathsf{O}(M)$) and the singular values $s_i$  admit the following joint density:
\begin{align}
\label{eq:JointDensitySv.1}
    \mathcal{P}_{N}(s_1, \dots, s_{N}) &= \frac{1}{Z_{N}} \mathrm{exp} \left[ - \frac{N}{2} \sum_{i=1}^{N} \left( V(s_i^2) + \left(1- \frac{M}{N} \right) \log (s_i^2)  \right) + \frac{1}{2} \sum_{i,j | i \neq j} \log | s_i^2 - s_j^2| \right] \, . &&
\end{align}
If we introduce the modified potential 
\begin{align}
    \label{eq:modpot_Birot}
    \tilde{V}_q(x) &:= V(x) + \left( 1 - \frac{1}{q} \right) \log x \, , && 
\end{align}
with $q = N/M$, this can be  written as:
\begin{align}
\label{eq:JointDensitySv.2}
    \mathcal{P}_{N}(s_1, \dots, s_N) &= \frac{1}{Z_{N}} \mathrm{exp} \left[ - \frac{N}{2} \sum_{i=1}^{N} \tilde{V}_q(s_i^2) + \frac{1}{2} \sum_{i,j | i \neq j} \log | s_i^2 - s_j^2| \right] \, . &&
\end{align}

\vskip 0.3cm 
\noindent Note that if we do the change of variable $ \{ s_i(\mathbf{A}) \} \to \{ \lambda_i \left( \mathbf{A} \mathbf{A}^{\mathsf{T}} \right) \}$ given by Eq.\ \eqref{eq:prop_Birot} in the joint law of Eq.\ \eqref{eq:JointDensitySv.2}, we have that the matrix $\mathbf{A} \mathbf{A}^{\mathsf{T}}$ is taken from an invariant ensemble with the modified potential $\tilde{V}_q(.)$ (plus a vanishing term coming from the change of variable).\footnote{Taking into account the Jacobian amounts to do the change $\tilde{V}_q(x) \to \tilde{V}_q(x) + \frac{1}{N} \log x $. In particular, this  vanishing term prevents  the eigenvalues of the matrix $\mathbf{A} \mathbf{A}^{\mathsf{T}}$ to be negative for $q=1$ where the coefficient in front of the logarithm in Eq.\ \eqref{eq:modpot_Birot} is null in this case. For $q=1$, one has to consider $\tilde{V}_1(.)$ as being infinite for negative values.} As a consequence, in the large double scaling limit of Eq.\ \eqref{Ap:double_scaling_limit}, the empirical singular value distribution converges to a smooth limit:
\begin{align*}
\label{Ap:ESVD_conv}
    \rho_{\mathbf{A}}(s) &:= \frac{1}{N} \sum_{i=1}^{N} \delta( s - s_i(\mathbf{A}) ) \to \rho_A(s) \, , &&
\end{align*}
where the limiting singular value distribution (LSVD)  $\rho_A$ is  given as: 
\begin{align}
    \label{eq:LSVD.1}
    \rho_A(s) &= 2s\mu_{AA^{\mathsf{T}}}(s^2)  \, , && 
\end{align}
where $\mu_{AA^{\mathsf{T}}}$ is solution of the Tricomi problem of Eq.\ \eqref{eq:Tricomi} with the potential $V(.)$ replaced by $\tilde{V}_q(.)$ of Eq.\ \eqref{eq:modpot_Birot}:
\begin{align}
\label{eq:Tricomi_Rect.0}
   \mathrm{P. V.} \int \frac{\mu_{AA^{\mathsf{T}}}(\lambda')}{\lambda - \lambda'} \mathrm{d}\lambda' &=  \frac{\tilde{V}'_q(\lambda)}{2} \, . &&
\end{align}
The edges $\mathrm{a}_{\pm}$ of the LSVD $\rho_A$ are the square root of the edges of the distribution $\mu_{AA^{\mathsf{T}}}$. Equivalently, the LSVD $\rho_A$ is the solution of the following equation: 
\begin{align}
\label{eq:Tricomi_Rect}
   \mathrm{P. V.} \int \frac{\rho_A(s')}{s^2 - {s'}^2} \mathrm{d}s' &=  \frac{\tilde{V}'_q(s^2)}{2} \, , &&
\end{align}
which can be directly seen from the joint law of Eq \eqref{eq:JointDensitySv.2}.

\vskip 0.3cm
\noindent \textit{Example (Gaussian rectangular random matrices and Ginibre matrices):} Let's consider a $(N \times M)$  matrix $\mathbf{A}$ with Gaussian i.i.d entries with mean zero and variance $1/M$. This corresponds to $V(x) = \frac{x}{q}$.  The matrix $ \frac{1}{\sigma^2} \mathbf{A}\mathbf{A}^{\mathsf{T}}$ is a Gaussian white Wishart matrix with density given by Eq.\ \eqref{eq:MP_dist}. Since The LSVD of the matrix $\mathbf{A}$ is related to the spectrum of Mar\v{c}enko-Pastur distribution  by $\rho_A(x)=2 \, \frac{x}{\sigma^2} \,  \mu_{\mathrm{MP}_q}(\frac{x^2}{\sigma^2})$, one gets:
\begin{align}
    \label{eq:LSVD_gauss}
    \rho_A(s) &= \frac{\sqrt{4q \sigma^4 - (s^2 - \sigma^2(1+q))^2}}{\pi \sigma^2 q s} \mathbb{I}_{[ \sigma(1-\sqrt{q}),  \sigma (1+\sqrt{q})]} \, . &&
\end{align}
In particular the case $q=1$ corresponds to \emph{Ginibre random matrices} and Eq.\ \eqref{eq:LSVD_gauss} becomes the so-called \emph{quarter-circle distribution:}
\begin{align}
    \label{eq:LSVD_Ginibre}
    \rho_A(s) &= \frac{\sqrt{4 \sigma^2 - s^2}}{\pi \sigma^2} \mathbb{I}_{[0, 2\sigma]} \, . &&
\end{align}

\subsubsection{Bi-invariant ensemble with a wall and fixed diagonal rectangular matrices}
\label{sec:BiInvEns_wall}
\vskip 0.3cm 
\noindent Similarly to the symmetric case, we say that a rectangular random matrix is taken from a bi-invariant ensemble \emph{with a wall} at the position $w_A$ if its SVD is given by Eq.\ \eqref{Ap:SVD_decomposition} with $\mathbf{U}$ (resp. $\mathbf{V}$) taken uniformly over $\mathsf{O}(N)$ (resp.  over $\mathsf{O}(M)$) and the singular values follow the law of Eq.\ \eqref{eq:JointDensitySv.2} with now $\tilde{V}(x > w_A^2)=\infty$, such that no singular value can be higher than the position $w_A$.  The introduction of this wall will allow us to study the sum of rectangular matrices where one (or both\footnote{When considering two diagonal rectangular matrices,  we will be considering the sum $\mathbf{C} = \tilde{\mathbf{A}} + \mathbf{U} \tilde{\mathbf{B}} \mathbf{V}^{\mathsf{T}}$  with $\mathbf{U}$ and $\mathbf{V}$ uniform over $\mathsf{O}(N)$ and $\mathsf{O}(M)$ respectively, such that $\tilde{\mathbf{A}}$ and $\mathbf{U} \tilde{\mathbf{B}} \mathbf{V}^{\mathsf{T}}$ are asymptotically bi-free, and the spectrum of the  singular values of the sum  is given asymptotically by the \emph{rectangular free convolution}  described in Sec.\ \ref{sec:Rect_conv}.  }) of them is a \emph{fixed diagonal rectangular matrix} of the form:
\begin{align}
\label{Ap:fix_Diag_Rect_mat}
\tilde{\mathbf{A}} &=
\begin{pmatrix}
    \tilde{a}_1 & 0  & 0\\
    0 & \ddots & 0 & \mathbf{0}_{M - N} \\
    0 & 0 & \tilde{a}_N
\end{pmatrix} \, ,&&
\end{align}
where the $\{ \tilde{a}_i \}$ are frozen sequence of number such that at large $N$, the empirical distribution of the $\{ \tilde{a}_i \}$ converges to the same $\rho_A$:
\begin{align*}
\label{Ap:ESVD_conv_fixdiag}
    \rho_{\tilde{\mathbf{A}}}(s) &:= \frac{1}{N} \sum_{i=1}^{N} \delta( s -  \tilde{a}_i ) \to \rho_A(s) \, .&&
\end{align*}

\subsubsection{Right large deviation of the top singular value}
\label{sec:LDP_1rect}
As in the symmetric case, the goal is to estimate for large $N$, the probability of finding the top singular value at a position $x$ above its typical value given by the edge $\mathrm{a}_+$ of $\rho_A$:
\begin{align}
    \label{eq:LDP_rect}
    \mathbb{P} \left[ s_1 \left( \mathbf{A} \right) \simeq x \right] &= \mathrm{exp} \left[ - N \Phi(x) \right] \, . &&
\end{align}
From the relation \eqref{eq:prop_Birot}, we have:
\begin{align}
    \label{eq:LDP_rect.2}
    \mathbb{P} \left[ s_1 \left( \mathbf{A} \right) \simeq x \right] &= \mathbb{P} \left[ \lambda_1(\mathbf{A} \mathbf{A}^{\mathsf{T}}) \simeq x^2 \right]  \, , &&
\end{align}
and hence 
\begin{align}
    \label{eq:LDP_rect.3}
   \Phi(x) &= \Psi_{AA^{\mathsf{T}}}(x^2) \, , &&
\end{align}
where $\Psi_{AA^{\mathsf{T}}}(.)$ is the rate function associated with the invariant matrix $\mathbf{A} \mathbf{A}^{\mathsf{T}}$ in the potential $\tilde{V}_q(.)$. Namely, using Eq.\ \eqref{eq:Psi_1RM.2}, we have:
\begin{align}
    \label{eq:RateFunction_1Rec}
    \Phi(x) &=  \frac{1}{2} \int_{\mathrm{a}_+^2}^{x^2} \left( \bar{g}_{AA^{\mathsf{T}}}(t) - g_{AA^{\mathsf{T}}}(t) \right) \mathrm{d}t \, , &&
\end{align}
where $g_{AA^{\mathsf{T}}}$ and $\bar{g}_{AA^{\mathsf{T}}}$ are respectively the Stieltjes and second branch of the Stieltjes of $\mu_{AA^{\mathsf{T}}}$. Equivalently, we can write the rate function $\Phi$ as: 
\begin{align}
      \label{eq:RateFunction_1Rec.2}
    \Phi(x) &= \int_{\mathrm{a}_+}^{x}  t \left( \tilde{V}'_q(t^2) - \frac{1}{2} \int_{\mathrm{a}_-}^{\mathrm{a}_+} \frac{\rho_A(s)}{t^2 - s^2} \mathrm{d}s  \right) \mathrm{d}t \, . &&  
\end{align}
\vskip 0.3 cm
\noindent \textit{Remark (square matrix and symmetrized density):} In the case where $q=1$, corresponding to the case where $\mathbf{A}$ is an (asymptotic) square matrix but not necessarily symmetric, there exist a nice relation with the symmetric case of Sec.\ \ref{sec:RotInvEns}. Let's consider a \emph{symmetric} matrix $\hat{\mathbf{A}} \sim \mathbb{P}_{\hat{V}}(.)$, where its potential $\hat{V}(.)$ is related to the potential $V(.)$ of Eq.\ \eqref{eq:proba_Birot} of the square (but non-symmetric) matrix $\mathbf{A}$ by:
\begin{align}
    \label{eq:q1_pot}
    \hat{V}(\lambda) &:= \frac{V(\lambda^2)}{2} \, .&&
\end{align}
In the large $N$ limit, the limiting density $\mu_{\hat{A}}(.)$ of $\hat{\mathbf{A}}$ satisfies Eq.\ \eqref{eq:Tricomi} with $V(.)$ replaced by $ \hat{V}(.)$, that is using Eq.\ \eqref{eq:q1_pot}:
\begin{align}
\label{eq:Tricomi_q1}
   \mathrm{P. V.} \int \frac{\mu_{\hat{A}}(\lambda')}{\lambda - \lambda'} \mathrm{d}\lambda' &=  \lambda V'(\lambda^2) \, . &&
\end{align}
On the other hand, for $q=1$ since $V_1(\lambda^2)=V(\lambda^2)$ (see Eq.\ \eqref{eq:modpot_Birot}), Eq.\ \eqref{eq:Tricomi_Rect} reads: 
\begin{align}
\label{eq:Tricomi_q1.1}
   \mathrm{P. V.} \int \frac{\rho_{A}(\lambda')}{\lambda^2 - {\lambda'}^2} \mathrm{d}\lambda' &=  \frac{V'(\lambda^2)}{2} \, , &&
\end{align}
so using the identity:
\begin{align}
       \label{eq:Apart.q1}
     \frac{1}{\lambda^2-{\lambda'}^2} &=  \frac{1}{2 \lambda} \left( \frac{1}{\lambda-{\lambda'}} + \frac{1}{\lambda+{\lambda'}} \right)   \, , &&
\end{align}
we have: 
\begin{align}
\label{eq:Tricomi_q1.2}
   \mathrm{P. V.} \int \frac{1}{\lambda - \lambda'} \left( \frac{\rho_A(\lambda') + \rho_A(-\lambda')}{2} \right) \mathrm{d}\lambda' &=  \lambda V'(\lambda^2) \, , &&
\end{align}
and hence by comparing Eq.\ \eqref{eq:Tricomi_q1} and \eqref{eq:Tricomi_q1.2}, we have that the distribution $\mu_{\hat{A}}(.)$ is the \emph{symmetrized} distribution of $\rho_A(.)$: 
\begin{align}
    \label{eq:symm_density}
    \mu_{\hat{A}}(.) &= \frac{\rho_A(.) + \rho_A(-.)}{2} \, . &&
\end{align}
Furthermore, using again the identity Eq.\ \eqref{eq:Apart.q1} in Eq.\ \eqref{eq:RateFunction_1Rec.2} for $q=1$ with the definition of $\hat{V}(.)$ given by Eq.\ \eqref{eq:q1_pot}, we can write the rate function as:
\begin{align}
\label{eq:RateFunction_1Rec.3}
    \Phi(x) &= \int_{\mathrm{a}_+}^{x} \left( \hat{V}'(t) - 2 g_{\hat{A}}(t) \right) \mathrm{d}t = 2 \Psi_{\hat{A}}(x) \, . &&  
\end{align}
where $g_{\hat{A}}$ is the Stieltjes transform of $\mu_{\hat{A}}$, and $\Psi_{\hat{A}}$ is the rate function associated to the large deviation of the largest eigenvalue of $\hat{\mathbf{A}}$. In other words, for a bi-invariant square matrix, the rate function associated to the largest singular value is \emph{twice} the one associated to the largest eigenvalue of the invariant symmetric matrix whose limiting eigenvalue density is equal to the symmetrized distribution of $\rho_A$. This can be heuristically guessed by remarking that for $q=1$, we can express Eq.\ \eqref{eq:JointDensitySv.2} as: 
\begin{align}
\label{eq:JointDensitySv.3}
    \mathcal{P}_{N}(s_1, \dots, s_N) &= \frac{1}{Z_{N}} \mathrm{exp} \left[ - \frac{N}{2} \sum_{i=1}^{N} \left[ \hat{V}(s_i) +  \hat{V} (-s_i)   \right]+ \frac{1}{2} \sum_{i,j | i \neq j} \log | s_i - s_j| + \frac{1}{2} \log | s_i + s_j| \right] \, , &&
\end{align}
where the potential $\hat{V}(x)$ is symmetric by construction. The joint law of the $N$ (positive) singular values can be interpreted as the law of $2N$ eigenvalues following the usual Eq.\ \eqref{eq:JointDensityEigvals} where the first $N$ variables are constraint to be positive and each of the last $N$ is constraint to equal minus its positive counterpart. In the large $N$ limit, these constraints are irrelevant: the two problems have the same density (which does not depend on $N$) and differ by a factor of two for the rate function (which as an explicit $N$ factor).

\vskip 0.3 cm
\noindent \textit{Example (Rate function for Ginibre matrices):} If $\mathbf{A}$ is a Ginibre matrix, the LSVD is the quarter-circle law of Eq.\ \eqref{eq:LSVD_Ginibre} and its symmetrized density is the semi-circle law of Eq.\ \eqref{eq:SC_dist}. As a consequence, the rate function $\Phi_{\mathsf{Gin}}$ of the largest singular value of a Ginibre matrix is given by:
\begin{align}
    \label{eq:RateFunction_Gin}
    \Phi_{\mathsf{Gin}}(x) = 2 \Psi_{\mathsf{GOE}}(x) &=  \frac{x \sqrt{x^2 - 4 \sigma^2}}{2\sigma^2} + 2 \log \left( \frac{2 \sigma}{\sqrt{x^2 - 4\sigma^2}+x} \right) \, . && 
\end{align}
For $\sigma=1$, $\mathbf{A}\mathbf{A}^{\mathsf{T}}$ is a Wishart with shape parameter $q=1$, so  Eq.\ \eqref{eq:RateFunction_Gin} and Eq.\ \eqref{eq:LDP_rect.3} give the relation \eqref{eq:RateFunction_LOE.q1.2}.

\begin{figure}
     \centering
         \includegraphics[width= 0.55\textwidth]{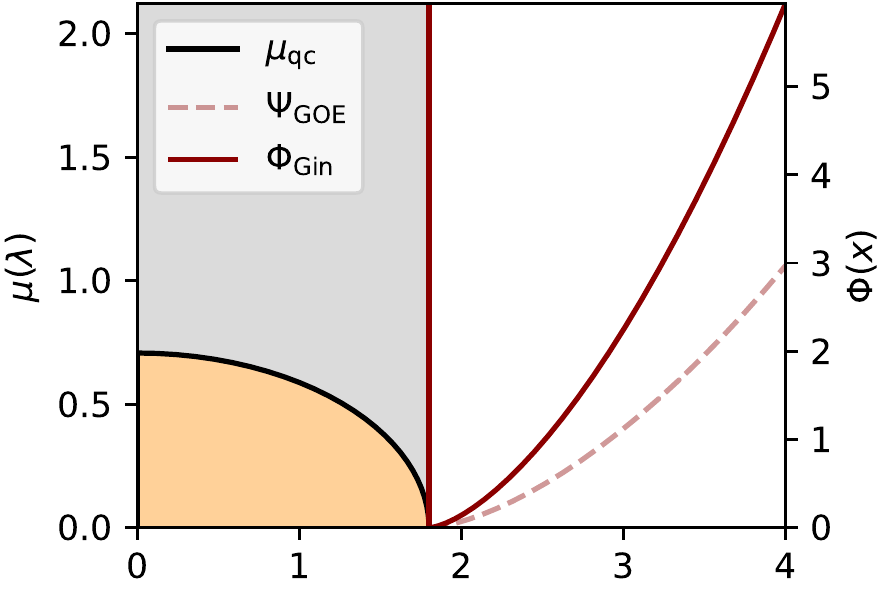}
    \caption{ The Rate function  (in red solid line) for the largest singular value of a Ginibre random matrix with $\sigma=1$. This rate function is twice the rate function of the largest eigenvalue of a GOE matrix with also $\sigma =1$, see Eq.\ \eqref{eq:RateFunction_Gin}.  } 
\label{fig:RF_Gin}
\end{figure}

\section{The tilting method}
\label{Sec:Tilt} 
\subsection{Notation}
\label{sec:notation_tilt}
In the previous section, we have reviewed the results concerning the right large deviation of the top eigenvalue (or singular value) of \emph{one} random matrix. The goal of this section is to describe the general framework to tackle the case of the large deviation of the largest eigenvalue (resp. singular value) of a matrix $\mathbf{C}$ given as  the (free) sum or product of two symmetric (resp. rectangular) random matrices $\mathbf{A}$ and $\mathbf{B}$. In the large $N$ limit, the limiting density $\mu_C$ (resp. $\rho_C$) of eigenvalues  (resp. singular values) of the matrix $\mathbf{C}$ is given by  \emph{free probability theory} and depends precisely on the operation (sum of symmetric matrices, product of symmetric matrices or sum of rectangular matrices\footnote{As we will see, the case of the product of two rectangular matrices can be deduced from the symmetric case.})  one is considering. This will be discussed in each dedicated section, see Secs. \ref{sec:LDP_sum}, Sec.\  \ref{sec:LDP_prod} and Sec.\ \ref{sec:LDP_Rect} respectively, but we argue that the strategy to get the right large deviation is the same. To put everything in the same framework we denote by
\begin{align}
\label{eq:zeta}
   \zeta_i \left( \mathbf{C} \right) :=\left\{
    \begin{array}{ll}
  \lambda_i(\mathbf{C}) & \mbox{if } \mathbf{C} \mbox{ is the sum/product of symmetric matrices} \, ,\\
\\
   s_i(\mathbf{C}) & \mbox{if } \mathbf{C} \mbox{ is the sum of rectangular matrices} \, .
    \end{array}
\right. &&
\end{align}
In any case, if we denote by $\mathrm{c}_+$ the edge of the spectrum of $\mu_C$ (resp. $\rho_C$), we have: 
\begin{align}
\label{eq:conv_TopToEdge.2}
    \zeta_1 \left( \mathbf{C} \right)  &\underset{N \to \infty}{\to} \mathrm{c}_+ \, , &&
\end{align}
and similar to the one-random matrix model, the natural question is to estimate the probability $\mathbb{P}\left[\zeta_1 \left( \mathbf{C} \right) \simeq x \right]$  at large but finite $N$. Since one recovers the original setting by taking the limit $\mathbf{B} \to \mathbf{0}$ (the null matrix) in the additive case or $\mathbf{B} \to \mathbf{I}$ (the identity matrix) in the multiplicative case, one should expect again to have the same different scaling as before for $x > \mathrm{c}_{+}$. As before, to tackle both the symmetric and rectangular cases at the same time, let's denote by:
\begin{align}
\label{eq:Pi_C}
   \Pi_C(x) :=\left\{
    \begin{array}{ll}
  \Psi_C(x) & \mbox{if } \mathbf{C} \mbox{ is the sum/product of symmetric matrices} \, ,\\
\\
   \Phi_C(x) & \mbox{if } \mathbf{C} \mbox{ is the sum of rectangular matrices} \, ,
    \end{array}
\right. &&
\end{align}
such that $\Pi_C$ is the rate function we want to compute: 
\begin{align}
\label{eq:Pi_C.2}
\mathbb{P}\left[\zeta_1(\mathbf{C}) \simeq x \right] &\approx \mathrm{exp} \left[-N \Pi_C(x) +o(N) \right] \quad \quad  \text{for} \, x> \mathrm{c}_+ \, ,&&
\end{align}
from the knowledge of the laws of the matrices $\mathbf{A}$ and $\mathbf{B}$. 

\subsection{Idea of the tilting method and general expression for the rate function}
\label{sec:idea_tilt}
The starting point for the derivation of the large deviation of \emph{one} random matrix is the joint law density of the eigenvalues/singular values $\zeta_i(\mathbf{C})$, from which we can use the Coulomb gas approach. For the sum or the product of matrices, one does not have a simple expression for the joint  density $\mathcal{P}_N \left(\zeta_1, \dots, \zeta_N  \right)$ of the eigenvalues/singular values, except in some specific cases, see Ref. \cite{MergnyMajumdar21}.  Instead of directly looking at the matrix $\mathbf{C}$,  the idea introduced in Ref. \cite{GuionnetMaida20} in the context of RMT, is to look at a \emph{weighted} realization of this matrix.  If we denote by $\mathcal{P}(.)$ the probability density of the random matrix $\mathbf{C}$ in the space of symmetric/rectangular matrix, let's consider \emph{another} random matrix $\mathbf{C}'$ whose probability density is given by:
\begin{align}
\label{eq:Tilt}
     \mathcal{P}^{(\theta)}(\mathbf{C}') &:= \frac{\mathcal{Z}_{\mathbf{C}'}(\theta)}{\mathbb{E} \left[ \mathcal{Z}_{\mathbf{C}}(\theta) \right] } \mathcal{P}(\mathbf{C}') \, . && 
\end{align}
where $\mathcal{Z}_{\mathbf{C}'}(\theta) \equiv \mathcal{Z}(\zeta_1(\mathbf{C}'), \dots, \zeta_N(\mathbf{C}'), \theta) $ is a function whose precise expression will be given later on. For now, let us note that this function  depends on $\mathbf{C}'$ only through its eigenvalues/singular values in addition to the free parameter $\theta$. The expectation in the denominator of the RHS of Eq.\ \eqref{eq:Tilt} is an average over  $ \mathbf{C} \sim \mathcal{P}(.)$ (and hence an average over both $\mathbf{A}$ and $\mathbf{B}$): $\mathbb{E} \left[ \mathcal{Z}_{\mathbf{C}}(\theta) \right] := \int \mathcal{Z}_{\mathbf{C}}(\theta) \, \mathcal{P}(\mathbf{C}) \mathrm{d} \mathbf{C} $ such that $\mathcal{P}^{(\theta)}(.)$ is well normalized. Because  $\mathcal{Z}_{\mathbf{C}'}(\theta)$ only depends on the $\zeta_i(\mathbf{C}')$ we can relate the joint density $\mathcal{P}_N^{(\theta)}(\zeta_1, \dots, \zeta_N)$ of the eigenvalues/singular values of $\mathbf{C}'$ to the (unknown) joint density $\mathcal{P}_N \left(\zeta_1, \dots, \zeta_N  \right)$ of eigenvalues/singular values of $\mathbf{C}$: 
\begin{align}
\label{eq:Tilt.2}
    \mathcal{P}_N^{(\theta)}(\zeta_1, \dots, \zeta_N) &= \frac{\mathcal{Z}_{\mathbf{C}'}(\theta)}{\mathbb{E} \left[ \mathcal{Z}_{\mathbf{C}}(\theta) \right] } \mathcal{P}_N \left(\zeta_1, \dots, \zeta_N  \right) \, . && 
\end{align}
by integrating Eq.\ \eqref{eq:Tilt.2} we have:
    \begin{align}
    \label{eq:Tilt.zeta1}
       \mathrm{Prob.}\left[\zeta_1(\mathbf{C}') \simeq x \right] &= \frac{1}{\mathbb{E} \left[ \mathcal{Z}_{\mathbf{C}'}(\theta) \right] } \int \delta(\zeta_1 -x) \mathcal{Z}_{\mathbf{C}'}(\theta) \mathcal{P}_N \left(\zeta_1, \dots, \zeta_N  \right)   \mathrm{d}\zeta_1 \dots \mathrm{d} \zeta_N \, .&&
    \end{align}   
Now if we \emph{choose} $\mathcal{Z}_{\mathbf{C}'}(\theta)$  such that for large $N$ this function explicitly depend on the position of the largest eigenvalue/singular value $\zeta_1 \equiv \zeta_1 ( \mathbf{C}')$ but is self-averaging  with respect to the other  $\zeta_2, \dots, \zeta_N$ and thus independent of them, then  the integral over these variables is nothing else than the probability of Eq.\ \eqref{eq:Pi_C.2}  we want to estimate. At large $N$ the density of eigenvalues (or singular values) of $\mathbf{C}'$ is the same as  $\mathbf{C}$, but the position of top eigenvalue/singular value  might different. In other words, we have:
    \begin{align}
    \label{eq:Tilt.zeta1.2}
       \mathrm{Prob.}\left[\zeta_1(\mathbf{C}') \simeq x \right] &=  \mathrm{e}^{ N \left[  J_C(x, \theta) - F_C(\theta) - \Pi_C(x) \right] +o(N) } \, . &&
\end{align}   
where we have introduced  the \emph{quenched free energy} as the large $N$ limit
\begin{align}
\label{eq:defQuenchedFE}
    J_C(x, \theta) &:\approx \frac{1}{N} \log \mathcal{Z}_{\mathbf{C}| \{\zeta_1=x\}}(\theta) \, , &&
\end{align}
where $\{\zeta_1=x\}$ indicates that the limit is taken with the constraint that the top eigenvalue (or singular value) is fixed at $x$; and similarly the \emph{annealed free energy} as: 
\begin{align}
\label{eq:defAnnealedFE}
     F_C(\theta) &:\approx \frac{1}{N} \log \mathbb{E} \mathcal{Z}_{\mathbf{C}}(\theta) \, . &&
\end{align}
Given a value of the free parameter $\theta$, what is the typical value $x^*(\theta)$ of the largest eigenvalue (or singular value) of the matrix $\mathbf{C}'$ as $N$ goes to infinity? By Jensen inequality, the function in the bracket of the RHS of Eq.\ \eqref{eq:Tilt.zeta1} is always non-positive (otherwise the probability would be higher than one) and so the typical value $x^*(\theta)$ corresponds to the case where this function is exactly zero, namely $x^*(\theta)$ is given as the solution of:
\begin{align}
\label{eq:RateFunction_tilt.0}
    \Pi_C(x^*(\theta)) - J_C(x^*(\theta),\theta) - F_C(\theta)&= 0  \, . &&
\end{align}
Now the idea of the tilting method is to look at Eq.\ \eqref{eq:RateFunction_tilt.0}  the other way: fix any $x >\mathrm{c}_+$, can we vary the free  parameter $\theta$ until it reaches a value  $\theta^* \equiv \theta^*(x)$ such that now the event $\{ \zeta_1(\mathbf{C}') \simeq x\}$ becomes typical? If so, we can relate the unknown rate function $\Pi_C$ to the annealed and quenched free energies: 
\begin{align}
\label{eq:RateFunction_tilt}
    \Pi_C(x) &= J_C(x,\theta^*(x)) - F_C(\theta^*(x)) \, . &&
\end{align}
Since the LHS of \eqref{eq:Tilt.zeta1} is a probability measure, it is always bounded by one, so this optimal $\theta^*$  - if it exists - is given by:
\begin{align}
\label{eq:Theta_opt.1}
   \theta^*(x) &= \underset{\theta >0}{ \mathrm{argsup} } \{ I_x(\theta) := J_{C}(x,\theta) - F_{C}(\theta) \} \, . &&
\end{align}
In particular, if the supremum is a maximum, one has that $\theta^*$  is solution of: 
\begin{align}
\label{eq:Theta_opt.2}
   I'_{x}(\theta^*) := \partial_\theta  J_{C}(x,\theta)|_{\theta^*} - F'_{C}(\theta^*) &=0 \, . &&
\end{align}
Next since we have Eq.\ \eqref{eq:conv_TopToEdge.2},  by construction $\Pi_C(\mathrm{c}_+) = 0$.  One can therefore take the derivative of  Eq.\ \eqref{eq:RateFunction_tilt} with respect to $x$ to get the following integral representation for the rate function: 
\begin{align}
\label{eq:Pi_Integral.0}
    \Pi_C(x) &= \int_{\mathrm{c}_+}^x \partial_t J_C(t,\theta^*(t)) + \theta^{* \prime}(t) (\partial_\theta J_C(t,\theta)|_{\theta^*} - F'(\theta^*)) \mathrm{d}t \, , &&
\end{align}
but by Eq.\ \eqref{eq:Theta_opt.2} the second term is null, so that we have the following simple formula: 
\begin{align}
\label{eq:Pi_Integral.1}
    \Pi_C(x) &= \int_{\mathrm{c}_+}^x \partial_t J_C(t,\theta^*(t)) \mathrm{d}t &\mbox{with $ \theta^*$ solution of Eq.\ \eqref{eq:Theta_opt.2} } \, . &&
\end{align}

\subsection{Spherical Integrals as tilting functions}
\label{sec:SphericalInt_tilt}
To summarize, we need to find a tilting function $\mathcal{Z}_{\mathbf{C}}$ which in the large $N$ limit only depends on the limiting distribution  of the matrix $\mathbf{C}$ and the position of its top eigenvalue/singular value and such that:
\begin{enumerate}
    \item we can compute the partial derivatives of the quenched free energy Eq.\ \eqref{eq:defQuenchedFE}, 
    \item we can compute the derivative of the annealed free energy of Eq.\ \eqref{eq:defAnnealedFE},
    \item we can show that for each $x >\mathrm{c}_+$, there is one optimal temperature $\theta^*(x)$ solution of Eq.\ \eqref{eq:Theta_opt.2} and we compute it, 
\end{enumerate}
then we get the rate function thanks to Eq.\ \eqref{eq:Pi_Integral.1}. Based on this necessary properties,  we argue that a natural candidate for the tilting function  is given by the  \emph{spherical integral} of the operation we are considering. The reason is twofold: 
\begin{itemize}
    \item On the one hand, the quenched free energy associated to each  spherical integral has been computed before in the literature, see Refs. \cite{GuionnetMaida05,Collins07,MergnyPotters20,husson2021asymptotic,benaych2011rectangular}, and is known to satisfy a  transition depending on the parameter $\theta$, between a phase where it does not depend explicitly on the position of the top eigenvalue/singular value and a phase where it does. The asymptotics of the quenched free energy is summarized in Sec.\ \ref{sec:As_QFE_AFE.sum}, in Sec.\ \ref{sec:As_QFE_AFE.prod} and in Sec.\ \ref{sec:As_QFE_AFE.rect} for respectively the additive spherical integral, the multiplicative spherical integral, and the rectangular spherical integral.
    \item On the other hand, the spherical functions satisfy (by construction, as we will see later on) the following \emph{decomposition property:}
    \begin{align}
\label{eq:prop:AFE}
    \mathbb{E} \left[ \mathcal{Z}_{\mathbf{C}}(\theta) \right] &=  \mathbb{E}_{\mathbf{A}} \left[ \mathcal{Z}_{\mathbf{A}}(\theta) \right] \,  \mathbb{E}_{\mathbf{B}} \left[ \mathcal{Z}_{\mathbf{B}}(\theta) \right] \, . &&
\end{align}
Applying the logarithm function and dividing by $N$ Eq.\ \eqref{eq:prop:AFE},  this means that the annealed free energy is given by:
\begin{align}
\label{eq:prop:AFE.2}
    F_C(\theta) &= F_A(w_A,\theta) + F_B(w_B,\theta) \, , &&
\end{align}
where for $\mathbf{A} \sim \mathbb{P}_{V_A,w_A}$ (and similarly for  $\mathbf{B} \sim \mathbb{P}_{V_B,w_B}$), we have defined 
\begin{align}
\label{eq:prop:AFE_A}
    F_A(w_A,\theta) &:\approx  \frac{1}{N} \log \mathbb{E}_{\mathbf{A}} \mathcal{Z}_{\mathbf{A}}(\theta) \, , &&
\end{align}
and put explicitly the dependence in the position of the wall $w_A$ (resp. $w_B$). Since we have a precise description for the law of each matrix $\mathbf{A}$ and $\mathbf{B}$ separately but not for $\mathbf{C}$, this allows us to compute the annealed free energy and the results are also given  in Sec.\ \ref{sec:As_QFE_AFE.sum}, in Sec.\ \ref{sec:As_QFE_AFE.prod} and in Sec.\ \ref{sec:As_QFE_AFE.rect}.
\end{itemize}
As a consequence, once we have an expression for the (derivatives of) the annealed and quenched free energies, we only need  to prove that the optimal $\theta^*(x)$ is well defined for each (attainable) $x$ and compute it.  This is done for each case in Sec.\ \ref{sec:OptInvTemp.sum}; Sec \ref{sec:OptInvTemp.prod} and Sec.\ \ref{sec:OptInvTemp.rect} . Injecting this expression in Eq.\ \eqref{eq:Pi_Integral.1}, we can then get an expression for the rate function and the results are given in  in Sec.\ \ref{sec:RateFunction.sum}, in Sec.\ \ref{sec:RateFunction.prod} and in Sec.\ \ref{sec:RateFunction.rect}.

\vskip 0.3cm 
\noindent \textit{Remark (Spherical function as partition function of spherical spin glass model):} We have used the notations of statistical physics to denote the quantities of Eq.\ \eqref{eq:defQuenchedFE} and \eqref{eq:defAnnealedFE} as respectively the quenched and annealed free energies. The reason is that the spherical integral  $\mathcal{Z}_{\mathbf{C}}(\theta)$ can be seen as the partition function over configuration $\vect{\sigma}$ living on the sphere, with disorder $\mathbf{C}$ and inverse temperature $\theta$: 
\begin{align}
\label{eq:PartitionF}
    \mathcal{Z}_{\mathbf{C}}(\theta) &:= \langle \mathrm{e}^{N \theta \mathcal{H}(\vect{\sigma}) } \rangle \, .&&
\end{align}
The precise description of the spin glass model will depend on the nature of the operation we are considering and be given explicitly in Sec.\ \ref{sec:SSK} for the case of the sum, in \ref{sec:LSSK} for the case of the product, and in Sec.\ \ref{sec:BSSK}  for the rectangular case. Let us mention that the rate function is given as the highest difference between the quenched and the annealed free energy for a value of the top eigenvalue/singular value being fixed. Since for such systems it is known that both the quenched and annealed free energies are equal (this is the so-called 'paramagnetic phase') below a certain threshold of the inverse temperature, the optimal inverse temperature $\theta^*$ in Eq.\ \eqref{eq:Theta_opt.1} is necessarily attained after this threshold  (in the so-called 'spin-glass phase').

\vskip 0.3cm 
\noindent \textit{Remark (the case of diagonal matrices and wall at the edge):} If one of the matrix, say $\mathbf{A}$,  is replaced by a diagonal (resp. rectangular diagonal) matrix, $\tilde{\mathbf{A}}$ as described in Sec.\ \ref{sec:InvEns_w_wall}  (resp. as described in Sec.\ \ref{sec:BiInvEns_wall}), then the decomposition property \eqref{eq:prop:AFE} is now given by:
    \begin{align}
\label{eq:AFE_frozen.1}
    \mathbb{E} \left[ \mathcal{Z}_{\mathbf{C}}(\theta) \right] &=    \mathcal{Z}_{\tilde{\mathbf{A}}}(\theta)  \,  \mathbb{E}_{\mathbf{B}} \left[ \mathcal{Z}_{\mathbf{B}}(\theta) \right] \, . &&
\end{align}
Since there is no constraint on the top eigenvalue/singular value, and the limiting density of $\tilde{\mathbf{A}}$ is the same as the one of $\mathbf{A}$, we have for the free energy
\begin{align}
\label{eq:AFE_frozen.2}
    F_C(\theta) &= J_A(\mathrm{a}_+,\theta) + F_B(w_B,\theta) \, . &&
\end{align}
Now we argue that in each case, we have the following property:
\begin{align}
    \label{eq:AFE_frozen.3}
  \partial_{\theta}  F_A(w_A = \mathrm{a}_+,\theta) &=  \partial_{\theta} J_A(\mathrm{a}_+,\theta) \, , &&
\end{align}
such that the derivative of the annealed free energy of Eq.\ \eqref{eq:prop:AFE.2} and Eq.\ \eqref{eq:AFE_frozen.2} are equal.   Since no other quantity is modified, from the point of view of large deviation, it is equivalent to replacing the diagonal matrix $\tilde{\mathbf{A}}$ by a random matrix $\mathbf{A} \sim \mathbb{P}_{V,w_A = \mathrm{a}_+}$ from an  invariant ensemble with the same limiting density $\mu_A(.)$, but with a wall at the edge. The same argument holds if one considers the free sum of two diagonal matrices.

\section{Large deviation for the sum of symmetric matrices}
\label{sec:LDP_sum}
In this section, we consider the case where the matrix $\mathbf{C}$ is given as 
\begin{align}
    \label{eq:CSum}
    \mathbf{C} &= \mathbf{A} + \mathbf{B} \, , &&
\end{align}
where $\mathbf{A} \sim \mathbb{P}_{V_A,w_A} $ and $\mathbf{B} \sim \mathbb{P}_{V_B,w_B}$ are two \emph{symmetric} matrices, each taken from an invariant ensemble with a wall as defined in Sec.\ \ref{sec:InvEns_w_wall}. We aim at computing the rate function\footnote{We recall that since we are studying the symmetric case, one needs to replace the notation $ \zeta_i \left( \mathbf{C} \right)$  and $\Pi_C(x)$    in Sec.\ \ref{Sec:Tilt}  by  $\lambda_i \left( \mathbf{C} \right)$ and $\Psi_C(x)$  respectively. } $\Psi_C(x)$: 
\begin{align}
\label{eq:Psi_C.sum}
\mathbb{P}\left[\lambda_1(\mathbf{C}) \simeq x \right] &= \mathrm{exp} \left[-N \, \Psi_C(x) +o(N) \right] \quad \quad  \text{for} \, x> \mathrm{c}_+ \, ,&&
\end{align}
where $\mathrm{c}_+$ is the edge of the limiting spectrum $\mu_C$ of $\mathbf{C}$, and $\mu_C$ is described by \emph{free probability}, see the next paragraph.

\subsection{The sum of invariant matrices: free convolution and the R-transform}
\label{sec:add_free_conv}

The limiting spectral distribution $\mu_C(.)$ is a smooth density with right edge $\mathrm{c}_+$ and is given as the unique probability distribution solution of: 
\begin{align}
\label{eq:prop_Rt}
    \mathcal{R}_C(y) &= \mathcal{R}_A(y)  + \mathcal{R}_B(y)\, , &&
\end{align}
for all $y$ in the complex plane close enough to the origin, where $\mathcal{R}_A$ (resp. $\mathcal{R}_B, \mathcal{R}_C$) is the R-transform of the limiting spectral distribution $\mu_A$ (resp. $\mu_B, \mu_C$):
\begin{align}
\label{eq:def_Rt}
    \mathcal{R}_A(y)&:= g_A^{\langle -1 \rangle}(y) - \frac{1}{y} \, , &&
\end{align}
where $g_A^{\langle -1 \rangle}(.)$ denotes  the (functional) inverse of the Stieltjes transform of Eq.\ \eqref{eq:def:Stieltjes}. This inverse, and hence also the R-transform, is only defined between $(0,g_A(\mathrm{a}_+))$. However, for invariant ensemble, one can extend analytically this function for values beyond $g_A(\mathrm{a}_+)$. In this case,  this corresponds to inverting the second branch of the Stieltjes transform given by Eq.\ \eqref{eq:def:2ndBranchStieltjes} and not the Stieltjes branch itself, see App.\ \ref{sec:Ap:Stieltjes}. In the following, the R-transform has to be understood with this analytical continuation procedure. Eq.\ \eqref{eq:prop_Rt} shows that the R-transform linearizes the sum, and hence it is the random matrix analogous of the cumulant generating function of classical probability.  The limiting spectral distribution is therefore called the free (additive) convolution of $\mu_A$ and $\mu_B$ and is usually denoted by $\mu_C := \mu_A \boxplus \mu_B$ in the RMT literature. We refer the reader to Ref. \cite{Voiculescu1986,Voiculescu1997free,Mingo2017free} for more detail on the free convolution. We will be using  the two following properties of the additive free convolution:
\begin{itemize}
    \item If we denote by
    \begin{align}
\label{eq:radius_conv_sum}
    \mathrm{r}_A &:= \lim_{w_A \to \infty} \bar{g}_A(w_A) \, , &&
\end{align}
    then for $\theta \in (0, \mathrm{r}_A)$, $\mathcal{R}_A(.)$ is an increasing function of $\theta$, see App.\ \ref{sec:Ap:monot_Rt}.
    \item The Stieltjes transform at the edge of the spectrum of the matrix $\mathbf{C}$ satisfies the following inequality:
    \begin{align}
    \label{eq:prop_freeconv}
    g_C(\mathrm{c}_+) \leq &\min \left(g_A(\mathrm{a}_+), g_B(\mathrm{b}_+) \right) \,, &&
\end{align}
\end{itemize}
see App.\ \ref{sec:Ap:ineq}.

\vskip 0.3cm
\noindent \textit{Example (R-transform of GOE matrices):} For a GOE matrix inverting Eq.\ \eqref{eq:Stieltjes_SC}, one has that the inverse of the Stieljes is given by:
\begin{align}
    \label{eq:Inv_res_SC}
    g_{\mathrm{sc}}^{\langle -1 \rangle}(y) &=  \sigma^2 y + \frac{1}{y} \, , && 
\end{align}
and so the R-transform is given by:
\begin{align}
    \label{eq:R_SC}
   R_{\mathrm{sc}}(y)=  \sigma^2 y \, . && 
\end{align}

\vskip 0.3cm
\noindent \textit{Example  (R-transform of Wishart matrices):} For a Wishart matrix inverting Eq.\ \eqref{eq:Stieltjes_MP}, one has its R-transform is given by:
\begin{align}
    \label{eq:R_MP}
   R_{\mathrm{MP}_q}(y)= \frac{1}{1-qy} \, . && 
\end{align}

\subsection{The additive spherical integral and the SSK model}
\label{sec:SSK}
 Based on what we have discussed in Sec.\ \ref{sec:SphericalInt_tilt}, we argue that a natural candidate for the tilt function of Eq.\ \eqref{eq:Tilt} is given by the \emph{additive spherical function} $\mathcal{Z}_{\mathbf{C}}(\theta)$, where for \emph{any} symmetric matrix $\mathbf{M}$ and $\theta$ positive, it is defined by: 
 \begin{align}
\label{eq:def:add_SI}
     \mathcal{Z}_{\mathbf{M}}(\theta) &:=  \int_{\mathbb{S}^{N-1}} \mathrm{exp} \left[ \frac{N \theta}{2} \vect{\sigma}^{\mathsf{T}} \mathbf{M} \vect{\sigma}  \right]  \, \mathrm{d}\omega(\vect{\sigma}) \, , &&
\end{align}
where $\mathbb{S}^{N-1} := \{ \vect{\sigma} \in \mathbb{R}^N \text{ s.t } \| \vect{\sigma} \| =1 \}$ is the hypersphere of radius one and  $\mathrm{d}\omega(\vect{\sigma})$ is the compact notation for the uniform measure  over $\mathbb{S}^{N-1}$ normalized to one. It is clear that this function only depends on the eigenvalues of $\mathbf{M}$, because we can always absorb the matrix of eigenvectors in the variable of integration $\vect{\sigma}$ by a change of variable. Equivalently, the additive spherical integral can be written as
\begin{align}
\label{eq:HCIZrk1}
     \mathcal{Z}_{\mathbf{M}}(\theta)  &= \int_{\mathsf{O}(N)} \, \mathrm{e}^{  \frac{N}{2} \Tr \mathbf{O} \mathbf{M} \mathbf{O}^{\mathsf{T}} (\theta \mathbf{e} \mathbf{e}^{\mathsf{T}})  } \,  \mathrm{d} \mathbf{O}   \, , &&
\end{align}
where $\mathrm{d}\mathbf{O}$ is the normalized uniform Haar measure over $\mathsf{O}(N)$. In RMT, $\mathcal{Z}_{\mathbf{M}}(\theta)$ is also known as the rank-one Harish-Chandra-Itzykson-Zuber (HCIZ) integral, see Ref. \cite{HarishChandra1957,Itzykson1980}. The asymptotic behavior of this spherical integral is related to the  $R$-transform, as we will see later. Note that if one replaces the rank one matrix $\theta \mathbf{e} \mathbf{e}^{\mathsf{T}}$ in Eq.\ \eqref{eq:HCIZrk1} by a full rank matrix, then the asymptotic is governed by a different complex variational principle that was first derived by Matytsin, see Refs. \cite{Matytsin1994,Guionnet2002,Bun2014} and it is an open problem to understand the crossover between the two different regimes. By Haar invariance, for \emph{any} symmetric matrices $\mathbf{M}_1$ and $\mathbf{M}_2$, we have the following property: 
\begin{align}
    \label{eq:HaarInv}
    \int_{\mathsf{O}(N)}  \mathcal{Z}_{\mathbf{M}_1+\mathbf{O}\mathbf{M}_2\mathbf{O}^{\mathsf{T}}}(\theta)  \mathrm{d} \mathbf{O} &= \mathcal{Z}_{\mathbf{M}_1}(\theta) \, \mathcal{Z}_{\mathbf{M}_2}(\theta) & \, . 
\end{align}
If we now apply this relation for $\mathbf{M}_1 = \mathbf{A}$ and $\mathbf{M}_2 = \mathbf{B}$, since by Eq.\ \eqref{eq:propInvEns} we have $\mathbf{B} \overset{\text{in law}}{=} \mathbf{O}\mathbf{B} \mathbf{O}^{\mathsf{T}}$, and $\mathbf{A}$ and $\mathbf{B}$ are independent, we have after taking the average over both $\mathbf{A}$ and $\mathbf{B}$, the desired property \eqref{eq:prop:AFE}.

\vskip 0.3cm
\noindent As we have argued in Sec.\ \ref{sec:SphericalInt_tilt}, we can interpret the additive spherical as the partition function of spherical model. Indeed, we can always write Eq.\ \eqref{eq:def:add_SI} as
\begin{align}
\label{eq:PartitionF_SSK}
       \mathcal{Z}_{\mathbf{C}}(\theta) &:= \langle \mathrm{e}^{\frac{N}{2}\theta   \mathcal{H}^{\mathrm{SSK}}(\vect{\sigma}) } \rangle \, , &&
\end{align}
where to follow the standard notation in statistical physics, we  have denoted by  $\langle . \rangle \equiv \int_{\mathbb{S}^{N-1}} . \, \mathrm{d}\omega(\vect{\sigma}) $ the  uniform average over the spins living in the sphere of radius one and the  \emph{Hamiltonian} $\mathcal{H}^{\mathrm{SSK}}(.)$ is given by the quadratic form: 
\begin{align}
\label{eq:def:H_SSK}
      \mathcal{H}^{\mathrm{SSK}}(\vect{\sigma}) &:= \sum_i^N \lambda_i \left( \mathbf{M} \right) \sigma_i^2  \, . &&
\end{align}
This  model is known (see Refs. \cite{Kosterlitz,Gross1984,Cugliandolo_1995,Baik2016})  as  the ($p=2$) \emph{Spherical Sherrington-Kirkpatrick} (SSK in short) model\footnote{The SSK is usually introduced with a different convention by absorbing the $N$ in the spin variable: $\tilde{\vect{\sigma}} := \sqrt{N} \vect{\sigma}$  so that the spins lives on a sphere with radius $\sqrt{N}$, which does not change the spherical integral.}. The matrix $\mathbf{M}$ is the \emph{disordered pairwise interaction} and the parameter $\theta$ is the \emph{inverse temperature} of the model.  The SSK model  has been studied in detail in the literature, and in particular the case where $\mathbf{M}$ is a GOE matrix has received a lot of attention. In this paper, we are interested in the case where $\mathbf{M}=\mathbf{C}$ given by Eq.\ \eqref{eq:CSum}.  Qualitatively, for high temperature it is known that the system is in a paramagnetic phase and  all eigenvalue of the matrix $\mathbf{C}$ contributes roughly equally to the partition function while for low temperature  the situation is drastically different and the system is in a spin glass phase where the partition function is dominated by rare configurations which  put more weight on  the top eigenvalue $\lambda_{1}(\mathbf{C})$.

\subsection{Asymptotic behavior of the annealed and quenched free energies of the SSK model}
\label{sec:As_QFE_AFE.sum}
As explained in Sec.\ \ref{Sec:Tilt}, in order to get the rate function of the largest eigenvalue of $\mathbf{C}$, we need to compute the derivatives of the quenched and annealed free energies of Eq.\ \eqref{eq:defQuenchedFE} and Eq.\ \eqref{eq:defAnnealedFE} with the partition function given by Eq.\ \eqref{eq:def:add_SI}. 
\vskip 0.3 cm
\noindent For a matrix $\mathbf{C}$ with limiting density $\mu_C$ conditioned to have its largest eigenvalue $\lambda_1(\mathbf{C})$ fixed at the position $x$,
the partial derivatives of the quenched free energy are known \cite{GuionnetMaida05}  to be given by:
\begin{align}
\label{eq:As_QFE_sum.1}
  \partial_{\theta} J_C(x, \theta) = \frac{1}{2} \left\{
    \begin{array}{ll}
    \mathcal{R}_C(\theta) & \mbox{for }\theta \leq g_C(x) \, ,\\
\\
    x - \frac{1}{\theta} & \mbox{for }  \theta \geq g_C(x) \, ,
    \end{array}
\right. &&
\end{align}
and by:
\begin{align}
\label{eq:As_QFE_sum.2}
  \partial_{x} J_C(x, \theta) = \frac{1}{2} \left\{
    \begin{array}{ll}
  0 & \mbox{for }\theta \leq g_C(x) \, ,\\
\\
   \theta - g_C(x) & \mbox{for }  \theta \geq g_C(x) \, ,
    \end{array}
\right. &&
\end{align}
\vskip 0.3 cm
\noindent  Thanks to Eq.\ \eqref{eq:prop:AFE.2}, the computation of the (derivative of the) annealed free energy $F_C(.)$ reduced to the computation of the annealed free energies  $F_A(.)$ and $F_B(.)$ given by Eq.\ \eqref{eq:prop:AFE_A}. Those annealed free energies can be computed based on (an extension of) recent ideas developed in Ref. \cite{foini2021annealed} in the case of GOE. The proof is left in the App.\ \ref{sec:Ap:As_AFE_QFE}.  For $\mathbf{A} \sim \mathbb{P}_{V_A,w_A}$ (and similarly for $\mathbf{B} \sim \mathbb{P}_{V_B,w_B}$), the annealed free energy is given by 
\begin{align}
\label{eq:As_AFE.sumA}
 \partial_{\theta} F_A(w_A,\theta) = \frac{1}{2} \left\{
    \begin{array}{ll}
  \mathcal{R}_A(\theta) & \mbox{for }\theta \leq \bar{g}_A(w_A) \, ,\\
\\
   w_A - \frac{1}{\theta} & \mbox{for }   \theta \geq  \bar{g}_A(w_A) \, ,
    \end{array}
\right. &&
\end{align}
where  $\bar{g}_A(x)$ is the second branch of the Stieltjes transform from Eq.\eqref{eq:def:2ndBranchStieltjes}. Let us mention two important remarks:

\vskip 0.3cm
\noindent \textit{Remark (wall at the edge and diagonal matrices):} If we choose the position of the wall to be exactly at the edge: $w_A = \mathrm{a}_+$, since we have the relation
\begin{align}
    \label{eq:Stielt_at_edge}
    \bar{g}_A(\mathrm{a}_+) &= g_A(\mathrm{a}_+) \, , &&
\end{align}
Eq.\ \eqref{eq:As_AFE.sumA} reads in this case:
\begin{align}
\label{eq:As_AFE.sumA.edge}
 \partial_{\theta} F_A(w_A = \mathrm{a}_+,\theta) = \frac{1}{2} \left\{
    \begin{array}{ll}
  \mathcal{R}_A(\theta) & \mbox{for }\theta \leq  g_A(\mathrm{a}_+) \, ,\\
\\
    \mathrm{a}_+ - \frac{1}{\theta} & \mbox{for }   \theta \geq   g_A(\mathrm{a}_+)\, .
    \end{array}
\right. &&
\end{align}
Comparing Eq.\ \eqref{eq:As_AFE.sumA.edge} and Eq.\ \eqref{eq:As_QFE_sum.1} (with the index $C$ replaced by the index $A$ and for $x=\mathrm{a}_+$), we see that we have indeed the relation \eqref{eq:AFE_frozen.3} such that from the point of large deviation we can consider a fixed diagonal matrix as an invariant matrix with a wall at the edge of its distribution. 
 
\vskip 0.3cm
\noindent \textit{Remark (wall at infinity and classical invariant ensemble):} Since classical ensembles are obtained by taking the limit $w_A \to \infty$, we have for the corresponding annealed free energy:
\begin{align}
\label{eq:As_AFE.sumA.infinity}
 \partial_{\theta} F_A(w_A \to \infty,\theta) = \frac{1}{2} \left\{
    \begin{array}{ll}
  \mathcal{R}_A(\theta) & \mbox{for }\theta \leq   \mathrm{r}_A \, ,\\
\\
   \infty & \mbox{for }   \theta \geq    \mathrm{r}_A\, ,
    \end{array}
\right. &&
\end{align}
where we recall that $ \mathrm{r}_A$ is defined as the limit of the second branch of Stieltjes transform, see Eq.\ \eqref{eq:radius_conv_sum}.  One may note that the second line of Eq.\ \eqref{eq:As_AFE.sumA.infinity} is removed if $\mathrm{r}_A = \infty$ (which is for example is the case for a GOE matrix, see Eq.\ \eqref{eq:2BStieltjes_SC} for $z\to \infty$) but is present otherwise (which is for example is the case for a Wishart matrix for which $\mathrm{r}_A= \frac{1}{q}$, see Eq.\ \eqref{eq:2BStieltjes_MP} for $z\to \infty$).

\subsection{Retrieving the right large deviation for one random matrix}
\label{sec:warm_up_tilt_1RM}
The next step  in order to get the rate function is to show that there exist an optimal temperature $\theta^*(x)$  and compute it. Let's consider the case of \emph{one} random matrix $\mathbf{A}\sim \mathbb{P}_V(.)$ in a classical invariant ensemble, and let's retrieve the expression of Eq.\ \eqref{eq:Psi_1RM.2} with the tilting method as a warm-up exercise. For any $x>\mathrm{a}_+$, the optimal inverse temperature $\theta^*$ is given as the supremum of Eq.\ \eqref{eq:Theta_opt.1} with the notation $J_C(.)$ and $F_C(.)$ replaced by $J_A(.)$ and $F_A(w_A \to \infty,.)$. 

\begin{figure}
     \centering
         \includegraphics[width= 0.55\textwidth]{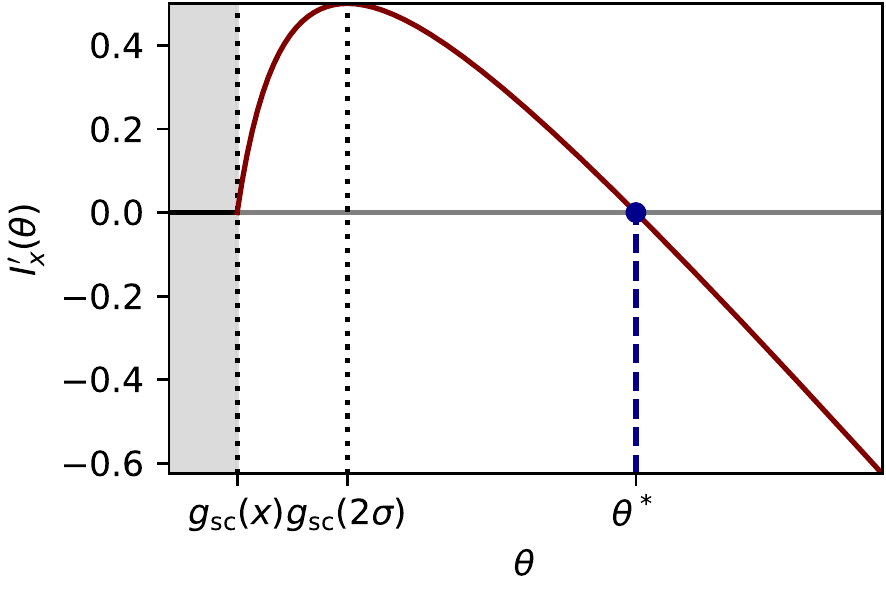}
    \caption{Derivative of the difference of the free energy in the case of one GOE random matrix with $\sigma=1$ and $x=3$, given as the argument of Eq.\ \eqref{eq:ThetaOpt_oneRM.2}. For $\theta \leq g_{\mathrm{sc}}(x)$, this function (in black) is null since the two free energies are equals, see Eq.\  \eqref{eq:paraphase.1RM}, and this corresponds to the paramagnetic phase. For $\theta \geq g_{\mathrm{sc}}(x)$, this function (in brown) is increasing and then decreasing with a maximum at $g_{\mathrm{sc}}( \mathrm{a}_+ = 2 \sigma)=1$, and this corresponds to the spin glass phase. The optimal inverse temperature (in blue) corresponds to the value where this function crosses the real axis in the spin glass phase.   } 
\label{fig:Diff_FE_GOE}
\end{figure}

\vskip 0.3cm
\noindent One may notice by integrating Eqs. \eqref{eq:As_QFE_sum.1} and \eqref{eq:As_AFE.sumA.infinity} with respect to $\theta$, that for $\theta$ between $0$  and $g_A(x)$, we have:
 \begin{align}
 \label{eq:paraphase.1RM}
          J_A(x,\theta) &=  F_A(w_A \to \infty, \theta) = \frac{1}{2} \int_{0}^{\frac{\theta}{2}} \mathcal{R}_A(\theta') \mathrm{d} \theta' & (\mbox{for } \theta \in (0, g_A(x))  \, . &&
    \end{align}
 This corresponds to the high temperature regime (or \emph{paramagnetic phase}) of the system where both the annealed and quenched free energy are equal. Necessarily, the optimal temperature $\theta^*(x)$, if there is one, cannot be in this region since from Eq.\ \eqref{eq:Theta_opt.1}, we want precisely  the difference between the two free energies  to be as high as possible. We can therefore restrict the  range of possible optimal temperature to be in the \emph{spin glass phase}, $\theta > g_A(x)$:
 \begin{align}
 \label{eq:ThetaOpt_oneRM.2}
         \theta^*(x) &= \underset{\theta>g_A(x)}{\mathrm{argsup}}  \left\{ I_{x}(\theta) := J_A(x,\theta) - F_A(w_A \to \infty, \theta)  \right\} \, .&&
 \end{align}
Let's compute the derivative with respect to $\theta$ of this function $I_{x}(\theta)$. According to Eq.\ \eqref{eq:As_QFE_sum.1} and Eq.\ \eqref{eq:As_AFE.sumA.infinity}, and the definition of the $R$-transform given by Eq.\ \eqref{eq:def_Rt}, it is simply given by: 
\begin{align}
\label{eq:dFreeEnergies_oneRM}
   I_{x}'(\theta) &= \frac{1}{2} \left( x - g_A^{\langle - 1 \rangle}(\theta) \right) & (\mbox{for } \theta  > g_A(x)) \, . &&
\end{align}
Here the function $g_A^{\langle - 1 \rangle}(\theta)=\mathcal{R}_A(\theta)+1/\theta$ contains the inverse of both branch of the Stieltjes transform. It is decreasing until it reaches the value $\theta = g_A(\mathrm{a}+)$ and then it is increasing until it reaches the (possible infinite) value, $\theta= \mathrm{r}_A =  \bar{g}_A(w_A \to  \infty)$ where it goes to infinity. Conversely,  the function $I_{x}'(\theta)$ of Eq.\ \eqref{eq:dFreeEnergies_oneRM}, seen as function of $\theta$ for $x$ fixed, starts at zero at $\theta = g_A(x)$ and then is increasing until it reaches the point $\theta =g_A(\mathrm{a}+)$ and then decreasing again, and goes to $- \infty$ as $\theta \to \mathrm{r}_A$. Thus, as one varies $\theta$ starting at  $g_A(x)$, this function is positive and then negative and only crosses the real axis once. As a consequence, the supremum in Eq.\ \eqref{eq:ThetaOpt_oneRM.2} is a maximum and this maximum is unique. This maximum $\theta^*$ is given at the unique point where the function $ I_{x}'(\theta)$ of Eq.\ \eqref{eq:dFreeEnergies_oneRM} crosses the real axis  in the region $\theta > g_A(\mathrm{a}_{+})$. In other words, finding $\theta^*$ amounts to solve the equation: 
\begin{align}
\label{eq:SolThetaOpt_oneRM}
    x  &=g_A^{\langle - 1 \rangle}(\theta^*(x))  &\mbox{for } \theta^*(x) > g_A(\mathrm{a}_+) \, ,&&
\end{align}
which is nothing else than the definition of the second branch of the Stieltjes transform, that is we have:
\begin{align}
\label{eq:SolThetaOpt_oneRM.2}
    \theta^*(x) &= \bar{g}_A(x) \, . &&
\end{align}
\vskip 0.3cm
\noindent If we now use the integral representation Eq.\ \eqref{eq:Pi_Integral.1} of the rate function, with the expression of Eq.\ \eqref{eq:As_QFE_sum.2} for the partial derivative of the quenched free energy, together with Eq.\ \eqref{eq:SolThetaOpt_oneRM.2} for the expression of $\theta^*$, we recover Eq.\ \eqref{eq:Psi_1RM.2} as expected. A plot of the function $  I_{x}'$, for $\mathbf{A}$ a GOE matrix, is given in Fig.\ \ref{fig:Diff_FE_GOE}.

\subsection{Optimal inverse temperature for the sum} 
\label{sec:OptInvTemp.sum}
\begin{figure}[h]
\centering
    \subfloat{\includegraphics[width=0.4\linewidth]{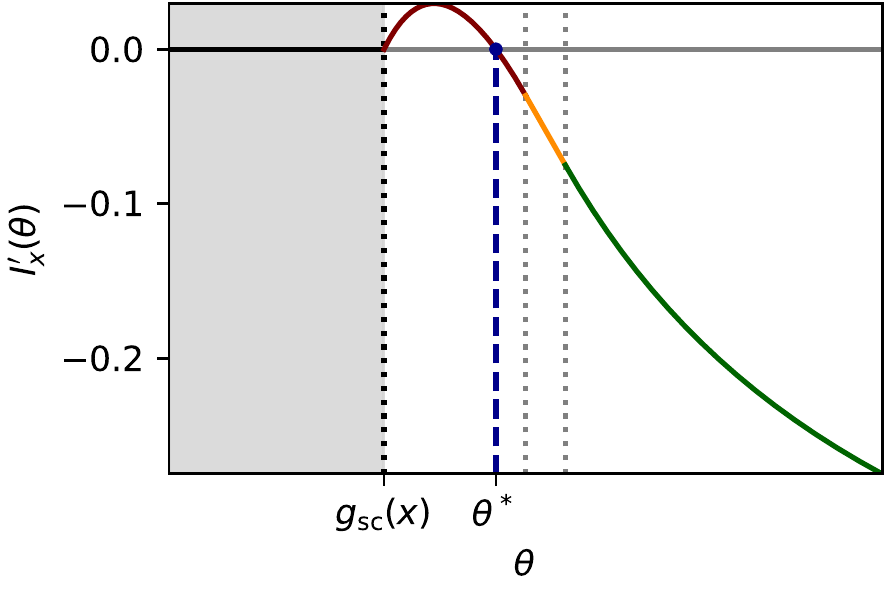}}
    \qquad
    \subfloat{\includegraphics[width=0.4\linewidth]{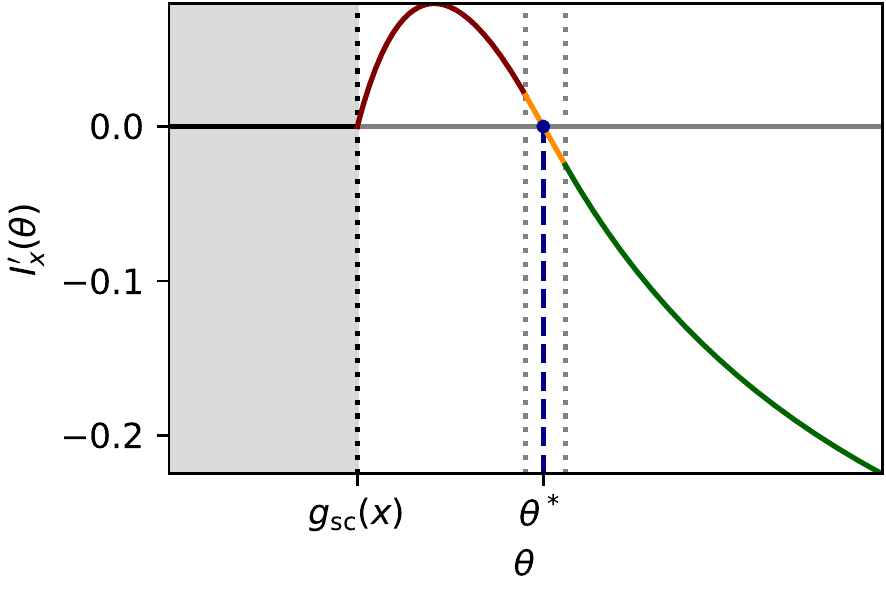}}
    \qquad
    \subfloat{\includegraphics[width=0.4\linewidth]{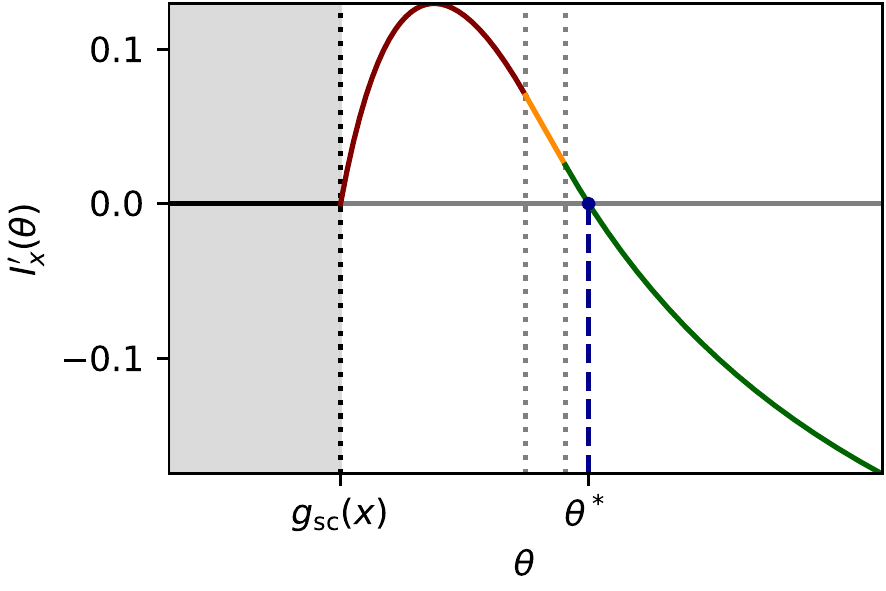}}
\caption{Representation of the function of Eq.\ \eqref{eq:DiffFE_Sum} for $\mathbf{A}$ and $\mathbf{B}$ two matrices from a GOE ensemble with a wall at their edge, with $\sigma_A =1$ and $\sigma_B = 9/10$ and for different values of $x$. Each color represents the correct expression of this piecewise continuous function in a given interval.  In the upper left, $x=2.75$ and the optimal inverse temperature is attained in the first region where the brown curve crosses the real axis. In the upper right, $x=2.85$ and the optimal inverse temperature is attained in the second region where the yellow curve crosses the real axis. In the center, $x=2.95$ and the optimal inverse temperature is attained in the third region where the green curve crosses the real axis.}
    \label{fig:Diff_FE}
\end{figure}

Let's now consider the general case given by Eq.\ \eqref{eq:CSum}.  Without loss of generality, we can consider\footnote{For general $\mathbf{A}$ and $\mathbf{B}$, having $w_A \leq w_B$ does not imply $\bar{g}_A(w_A) \leq \bar{g}_B(w_B)$ nor its converse. One can even come up with examples where $\mathbf{A}$ has a no wall ($w_A\to\infty$) while  $\mathbf{B}$ has a finite wall $w_B$ but still $\bar{g}_A(w_A) \leq \bar{g}_B(w_B)$.}
\begin{align}
   \label{eq:ga<gb}
   \bar{g}_A(w_A) &\leq \bar{g}_B(w_B) \, . &&
\end{align}
We further assume the non-trivial condition: 
\begin{align}
   \label{eq:gc<inf}
   g_C(\mathrm{c}_+) &< \infty \, , &&
\end{align}
as for $g_C(\mathrm{c}_+) = \infty$ (which necessarily implies $g_A(\mathrm{a}_+) = g_B(\mathrm{b}_+) = \infty$ by Eq.\ \eqref{eq:prop_freeconv}), the right large deviation is infinite for any $x > \mathrm{c}_+$. As in the previous section, we first want to show that the supremum in Eq.\ \eqref{eq:Theta_opt.1} is attained at a unique point, where $F_C$ is given by the sum of Eq.\ \eqref{eq:prop:AFE.2} and the annealed free energies $F_A(.)$ and $F_B(.)$ are given by Eq.\ \eqref{eq:As_AFE.sumA}. Since the function $I_{x}(\theta)$ is given as the sum of three piece-wise functions,   let's first note that we have the following set of inequalities:
\begin{align}
   \label{eq:setineq}
 g_C(x) \leq g_C(\mathrm{c}_+) \leq g_A(\mathrm{a}_+) &\leq \bar{g}_A(w_A) \leq \bar{g}_B(w_B) \, . &&
\end{align}
The first inequality is due to the fact that the Stieltjes is decreasing for $x >\mathrm{c}_+$. The second inequality is the property \eqref{eq:prop_freeconv} of the free convolution. The third is due to the second branch of the Stieltjes transform being monotonically increasing. The fourth is the previously mentioned convention of Eq.\ \eqref{eq:ga<gb}. Using the asymptotics of Eqs. \eqref{eq:As_QFE_sum.1} \eqref{eq:As_AFE.sumA} for the quenched and annealed free energies, together with the linearizing property \eqref{eq:prop_Rt} of the R-transform, one has the following behavior for the difference between the derivative of the annealed and quenched free energy:
\begin{align}
\label{eq:DiffFE_Sum}
I_{x}'(\theta) = \frac{1}{2} \left\{
    \begin{array}{llll}
  0 & \mbox{for }\theta \leq g_C(x) \, ,\\
\\
  x - g_C^{\langle -1 \rangle}(\theta) & \mbox{for  } g_C(x) \leq \theta \leq \bar{g}_A(w_A) \, ,\\
\\
  x - w_A - \mathcal{R}_B(\theta) & \mbox{for  } \bar{g}_A(w_A) \leq \theta \leq \bar{g}_B(w_B) \, ,\\
\\
  x - w_A - w_B + \frac{1}{\theta} & \mbox{for  } \theta \geq \bar{g}_B(w_B) \, .
    \end{array}
\right. &&
\end{align}
\noindent where if $\bar{g}_B(w_B) = \infty$ one has to remove the last line and similarly if $\bar{g}_A(w_A) = \infty$, one has to remove the last two lines. This function is represented in Fig.\ \ref{fig:Diff_FE} for different values of $x$.  Let's look at each interval separately.
\begin{enumerate}
    \item For $\theta < g_C(x)$, we are in the paramagnetic phase where both the annealed and the free energy are equal. Since for each $x > \mathrm{c}_+$, we want again the difference between the two to be as high as possible, the optimal inverse temperature is not in this region of the phase space. 
    \item  For $g_C(x) \leq \theta \leq \bar{g}_A(w_A)$, as we have seen in the simple case of one invariant matrix, the function $\theta \mapsto x - g_C^{\langle -1 \rangle}(\theta) $ is increasing until it reaches $g_C(\mathrm{c}_+)$ and then decreasing. 
    \item  For $\bar{g}_A(w_A) \leq \theta \leq \bar{g}_B(w_B)$, since the R-transform is increasing, the function $\theta \mapsto x - w_A - \mathcal{R}_B(\theta)$ is decreasing. 
    \item For $\bar{g}_A(w_A) \leq \theta \leq \bar{g}_B(w_B)$, the function $\theta \mapsto   x - w_A - w_B + \frac{1}{\theta}$ is decreasing.
\end{enumerate}
One can easily check that the function of Eq.\ \eqref{eq:DiffFE_Sum} is continuous at each point where its behavior changes. At $\bar{g}_A(w_A)$ it is equal to:
\begin{align}
\label{eq:theta_c1}
I_{x}'(\bar{g}_A(w_A))  &= \frac{1}{2} \left( x -g_C^{\langle -1 \rangle}( \bar{g}_A(w_A) ) \right) \, , && 
\end{align}
and at $\bar{g}_B(w_B)$, it is equal to:
\begin{align}
\label{eq:theta_c2}
I_{x}'(\bar{g}_B(w_B))  &= x - w_A - w_B + \frac{1}{\bar{g}_B(w_B)} \, .&& 
\end{align}
 To summarize, in the spin glass phase $\theta \geq g_C(x)$ the function of Eq.\ \eqref{eq:DiffFE_Sum} is continuously increasing until $\theta = g_C(\mathrm{c}_+)$ and then it is continuously decreasing.
 \vskip 0.3cm
 \noindent For $x > w_A + w_B$, it is easy to check that this function never crosses the real axis for values of $\theta > g_C(x)$, as a consequence,
\begin{align}
     \label{eq:Theta.x>a+b}
     \theta^*(x) &= \infty & \mbox{for } x > w_A +w_B \, . &&
\end{align}
This is expected because for the sum of two matrices we have the classical inequality:
 \begin{align}
     \label{eq:c<a+b}
   \lambda_1(\mathbf{C}) &\leq \lambda_1(\mathbf{A}) 
 + \lambda_1(\mathbf{B}) \, , &&
\end{align} 
   and since by definition of the walls, $\lambda_1(\mathbf{A}) \leq w_A $ and $\lambda_1(\mathbf{B}) \leq w_B$, the top eigenvalue of $\mathbf{C}$ cannot exceed $w_A + w_B$. So we find that the rate function is infinite for $x>w_A+w_B$.
\vskip 0.3cm
\noindent Otherwise, for values of $x < w_A + w_B$, this function always crosses the real axis once in this region. The correct equation for $\theta^*(x)$ - the point where the function $I'_{x}(.)$ touches the real axis, see Eq.\ \eqref{eq:Theta_opt.2} - depends  on if the value of this function at $\bar{g}_A(w_A)$ or $\bar{g}_B(w_B)$ is above or below zero, and hence on the value of $x$. There exist three possible cases, separated by two critical points, $x_{c_1}$ and $x_{c_2}$  defined respectively as the solution of the RHS of Eq.\ \eqref{eq:theta_c1} and  the RHS of Eq.\ \eqref{eq:theta_c2}) being equal to zero, that is:
 \begin{align}
 \label{eq:xc1}
     x_{c_1} &:= g_C^{\langle -1 \rangle}( \bar{g}_A(w_A) ) = w_A + \mathcal{R}_B(\bar{g}_A(w_A)) \, ,&&
 \end{align}
\ and 
  \begin{align}
  \label{eq:xc2}
x_{c_2}&:= w_A + w_B - \frac{1}{\bar{g}_B(w_B)} \, .&&
 \end{align}
 Note that $x_{c_1}\leq x_{c_2}$ as we have postulated Eq.\ \eqref{eq:ga<gb}. 
 We have:
 \begin{enumerate}
     \item for $\mathrm{c}_+ < x< x_{c_1}$, the optimal inverse temperature is attained in the region $g_C(x) \leq \theta \leq \bar{g}_A(w_A)$ and  so replacing in Eq.\ \eqref{eq:Theta_opt.2}  the expression of the difference of the free energies by the  top line of  Eq.\ \eqref{eq:DiffFE_Sum}, it is  solution of the same equation \eqref{eq:SolThetaOpt_oneRM} as the one in the simple one invariant random matrix case (with $g_A^{\langle -1 \rangle}$ replaced by $g_C^{\langle -1 \rangle}$) and thus we have:
     \begin{align}
     \label{eq:Thetax.1}
         \theta^*(x) &= \bar{g}_C(x) \, ,&&  
     \end{align}
     where $\bar{g}_C(.)$ is defined as the inverse of $g_C^{\langle -1 \rangle}(.)$ for values beyond $g_C(\mathrm{c}_+)$.
     \item For $x_{c_1}<x<x_{c_2}$, the optimal inverse temperature is attained in the region $\bar{g}_A(w_A) \leq \theta \leq \bar{g}_B(w_B)$ and so from the expression of the second line of the RHS of Eq.\ \eqref{eq:DiffFE_Sum}, it is solution of:
    \begin{align}
    \label{eq:Thetax.2}
         \mathcal{R}_B\left(\theta^*(x) \right) &=  x - w_A \, ,&&  
     \end{align}
     since the R-transform is continuously increasing, it has an inverse which we denote by $\mathcal{R}_B^{\langle - 1 \rangle}$ so that the optimal temperature is given by:
     \begin{align}
         \theta^*(x) &= \mathcal{R}_B^{\langle - 1 \rangle} \left( x - w_A \right) \, . &&
     \end{align}
     \item For $x_{c_2}<x<w_A +w_B$, $\theta^*(x)$ is attained in the region  $ \theta \geq \bar{g}_B(w_B)$ and so solving the third line of Eq.\ \eqref{eq:DiffFE_Sum} being equal to zero, we have:
     \begin{align}
         \theta^*(x)& = \frac{1}{w_A +w_B -x} \, .&&
     \end{align}
 \end{enumerate}
 One can check that the  piecewise function $\theta^*(x)$ is actually continuously increasing.

\vskip 0.3 cm
\subsection{Expression for the rate function} 
\label{sec:RateFunction.sum}
\begin{figure}
    \centering
    \subfloat{\includegraphics[width= 0.45\textwidth]{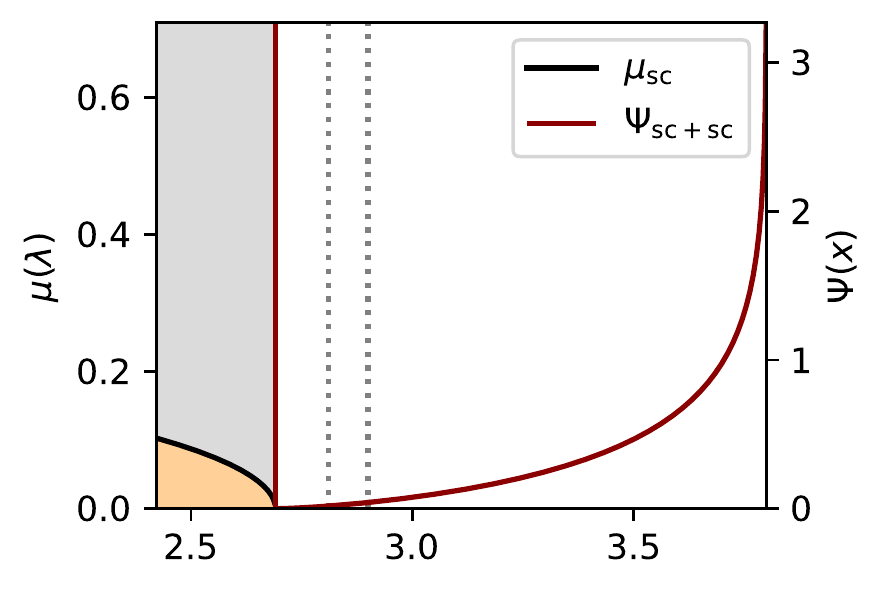}}
    \qquad
    \subfloat{\includegraphics[width=0.45\linewidth]{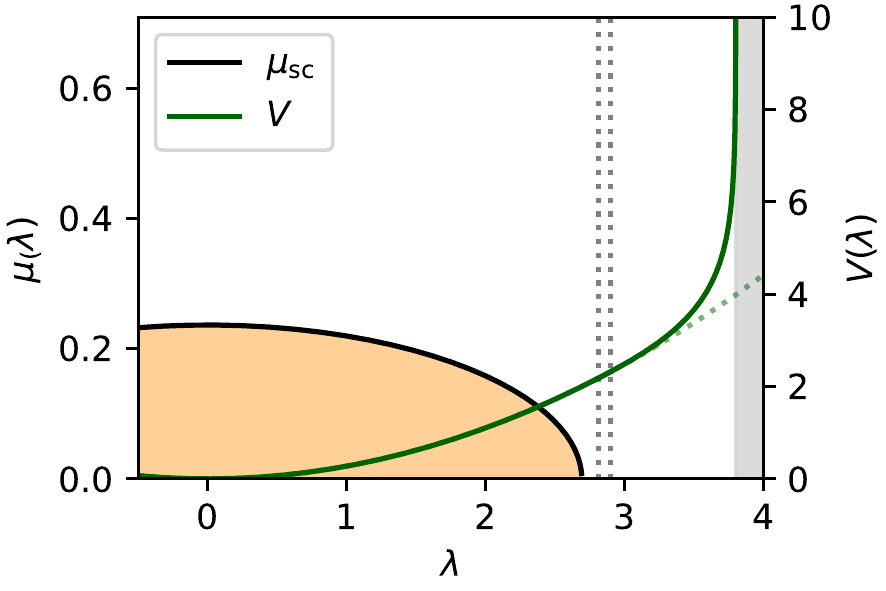}}
    \caption{One the left, the rate function of the largest eigenvalue of the sum of   $\mathbf{A}$ and $\mathbf{B}$ from a GOE ensemble with a wall at their edge, with $\sigma_A =1$ and $\sigma_B = \sigma = 9/10$, see Eq.\ \eqref{eq:RateFunctionSC+SC}. On the right, the effective potential, Eq.\ \eqref{eq:effective_potential},  for the same problem. Note that for $x<x_{c_1}$, the effective potential is given by $V(x)=x^2/(2(1+\sigma^2))$ and that nothing special happens at $x=c_+=2\sqrt{1+\sigma^2}$. The potential is non-analytic at the two points $x_{c_1}=2+\sigma^2$ and $x_{c_2}=2+\sigma$ (the two vertical dashed lines) and diverges at $w_A+w_b=2+2\sigma$ after which it is infinite (shaded area). Although it is impossible to see from this graph, it deviates from the quadratic potential (dotted green line) for $x>x_{c_1}$.}
\label{fig:RateFunction_scsc}
\end{figure}

Now that we have the expression for the optimal temperature, we can get the expression for the (right) rate function $\Psi_C(.)$ thanks to Eq.\ \eqref{eq:Pi_Integral.1}. This gives:
\begin{align}
\label{eq:RateFunction_Sum}
\Psi_C(x) =\left\{
    \begin{array}{lll}
  \displaystyle \frac{1}{2} \int_{\mathrm{c}_+}^x \left( \bar{g}_C(t) - g_C(t)\right) \mathrm{d}t   &   \mbox{for } \mathrm{c}_+ \leq x \leq x_{c_1} \, ,\\
\\
  \displaystyle K_1 + \frac{1}{2} \int_{x_{c_1}}^x  \left( \mathcal{R}_B^{\langle - 1 \rangle} \left( t - w_A \right) - g_C(t) \right)  \mathrm{d}t  & \mbox{for } x_{c_1} \leq x \leq x_{c_2} \, ,\\
\\
   \displaystyle   K_2  + \frac{1}{2} \log \left( \frac{1}{w_A + w_B - x} \right) -  \frac{1}{2} \int_{x_{c_2}}^x g_C(t)  \mathrm{d}t  & \mbox{for  } x_{c_2} \leq x \leq w_A+w_B \, .
    \end{array}
\right. &&
\end{align}
and is infinite for values of $x$ outside $(\mathrm{c}_+,w_A + w_B)$. The constant $K_1$ (resp. $K_2$) is given when $x_{c_1} < \infty$ (resp. when $x_{c_2} < \infty$)  by:
\begin{align}
    \label{eq:K1_add}
    K_1 &:= \frac{1}{2} \int_{\mathrm{c}_+}^{x_{c_1}} \left( \bar{g}_C(t) - g_C(t)\right)  \mathrm{d}t \, , &&
\end{align}
and by:
\begin{align}
    \label{eq:K2_add}
    K_2 &:= K_1 + \frac{1}{2} \log \left( \frac{1}{\bar{g}_B(w_B)} \right) +  \frac{1}{2} \int_{x_{c_1}}^{x_{c_2}} \left( \mathcal{R}_B^{\langle - 1 \rangle} \left( t - w_A \right) - g_C(t) \right)  \mathrm{d}t \, , &&
\end{align}
such that the rate function is continuous (and actually at least $C^2$) at each critical point. 
\vskip 0.3cm
\noindent \noindent  \textit{Remark (Effective potential):} The effective potential Eq.\ \eqref{eq:effective_potential} can be computed from the rate function and gives that $V(x)/2$ is equal to the same expression as Eq.\ \eqref{eq:RateFunction_Sum} with a plus sign instead of a minus sign in front of $g_C(t)$ (including in the constants Eqs.\ \eqref{eq:K1_add} and \eqref{eq:K1_add}). The potential is only defined up to an arbitrary constant chosen here such that $V(\mathrm{c}_+)=0$. An example of effective potential in plotted in Fig.\ref{fig:RateFunction_scsc}.

\vskip 0.3cm 
\noindent  \textit{Remark (Tracy-Widom '3/2'-scaling near the edge):} Since the first regime matches the ones of the classical case of one random matrix given by Eq.\ \eqref{eq:Psi_1RM.2}, we retrieve in particular the Tracy-Widom '3/2'-scaling of Eq.\ \eqref{eq:3o2scaling_Proba} near the edge for the free convolution of non-degenerate densities, as recently investigated in Ref. \cite{ji2021}.

\vskip 0.3cm
\noindent \textit{Remark (Number of critical points):} In general, if one is considering the sum of two random matrices taken from an invariant ensemble with a wall, then the rate function has two possible critical points. However, if $\bar{g}_A(w_A) = \bar{g}_B(w_B)$, then from Eqs. \eqref{eq:xc1} \eqref{eq:xc2}, one can see that two critical points merge, and we have at most one critical point. In particular, this happens when one is considering the free sum $\tilde{ \mathbf{A}} + \mathbf{O} \tilde{ \mathbf{A}} \mathbf{O}^{\mathsf{T}}$, where $ \tilde{ \mathbf{A}}$ is a fixed diagonal matrix.
Since $x_{c_1}$ diverges if $w_A\to\infty$ while $x_{c_2}$ diverges if either $w_A\to\infty$ or $w_B\to\infty$, the free sum of a two fixed diagonal and a fluctuating matrix (no wall) has at most one critical point while the sum of two fluctuating matrices never has a critical point, in this case the rate function is the same as that of an invariant ensemble Eq.\ \eqref{eq:Psi_1RM.2}.

\vskip 0.3cm
\noindent \textit{Remark (Interpretation of the three regimes):} For $c_+\leq x \leq x_{c_1}$ the rate function is the same as the one from an invariant ensemble. In this regime, the eigenvalues (including rare large ones) of the matrix $\mathbf{C}$ behave exactly as an invariant ensemble with the potential (and its analytical continuation outside the segment $[c_-,c_+]$) $V_C(.)$ compatible with the limiting density $\mu_C(.)$, that is, $V_C(.)$ is given by Eq.\ \eqref{eq:Tricomi}. In particular, walls (if any) do not modify the rate function in this regime.  For $x_{c_1}\leq x \leq x_{c_2}$, the wall $w_A$ starts to matter, the derivative of the rate function is now the same as if the matrix $\mathbf{A}$ were replaced by a rank-1 matrix with eigenvalue $w_A$ but with the still correct $g_C(x)$ (see App.\ \ref{sec:Ap:rk1_deformation}). Finally, for 
$x_{c_1}\leq x \leq w_A+w_B$, both walls matter and the derivative of the rate function is now the same as for the sum of two rank-one matrices with eigenvalues $w_A$ and $w_B$, again up to the correct $g_C(x)$ (see App.\ \ref{sec:Ap:RateFunction.rk1rk1}). In particular, very close to the maximal value $w_A+w_B$, we have:
\begin{align}
\mathbb{P} \left[ \lambda_1 \left( \mathbf{C} \right) \simeq  (w_A+w_B)(1-\epsilon) \right] &\approx \epsilon^{N/2} & \mbox{ for } \epsilon \ll 1 \, ,  &&
\end{align}
which is the asymptotic probability of two random vectors of having a squared overlap of order $1-\epsilon$ in dimension $N$.

\vskip 0.3 cm 
\noindent \textit{Example (Free sum of two diagonal semi-circle matrices):} In this paragraph, we will compute explicitly the rate function $\Psi_C(x)$ for a matrix $\mathbf{C} = \tilde{\mathbf{A}} + \mathbf{O} \tilde{\mathbf{B}}\mathbf{O}^{\mathsf{T}}$, where $\tilde{\mathbf{A}}$ and $\tilde{\mathbf{B}}$ are two fixed diagonal matrices with the semi-circle distribution of Eq.\ \eqref{eq:SC_dist} as their limiting spectrum. Without loss of generality, let's consider that their variance are given by $\sigma_A =1$ and by $\sigma_B^2 \equiv \sigma^2 \leq 1$ respectively. The computation is equivalent to the sum of two invariant random matrices in quadratic potentials $V_{i}(x)=x^2/(2\sigma_{i}^2)$ and a wall at $w_{i}=2\sigma_{i}$ for $i=A,B$ respectively. The free convolution of two semi-circle distributions is again  a semi-circle distribution, with variance the sum of the variance. In other words, the limiting law and Stieltjes transform of the matrix $\mathbf{C}$ is given respectively by Eq.\ \eqref{eq:SC_dist} and Eq.\ \eqref{eq:Stieltjes_SC} with $\sigma_C = \sqrt{1+\sigma^2}$.  Since we have $g_A(2) \leq g_B(2 \sigma)$ and the R-transform of the semi-circle distribution is given by Eq.\ \eqref{eq:R_SC}, the optimal inverse temperature is given by:
\begin{align}
\label{eq:ThetaOpt_sum_SC}
    \theta^*(x)=\left\{
    \begin{array}{lll}
   \frac{x+\sqrt{x^2-4(1+\sigma^2)}}{2(1+\sigma^2)} & \mbox{for } 2\sqrt{1+\sigma^2} \leq x \leq 2+\sigma^2 \, ,\\
\\
    \frac{x-2}{\sigma^2} & \mbox{for } 2+\sigma^2\leq x \leq 2+\sigma \, ,\\
\\
    \frac{1}{2+2\sigma-x } & \mbox{for } 2+\sigma\leq x \leq 2+2\sigma \, . 
    \end{array}
\right. &&
\end{align}
and so the rate function is given for $x \in [2 \sqrt{1+ \sigma^2},2+2 \sigma]$ by
\begin{align}
\label{eq:RateFunctionSC+SC}
    \Psi_{\mathrm{sc}+\mathrm{sc}}(x)=\left\{
    \begin{array}{lllll}
 \displaystyle \frac{x \sqrt{x^2 - 4 (1+\sigma^2)}}{4(1+\sigma^2)} +  \log \left( \frac{2 \sqrt{1+\sigma^2}}{\sqrt{x^2 - 4(1+\sigma^2)}+x} \right) & \mbox{for } 2\sqrt{1+\sigma^2} \leq x \leq 2+\sigma^2 \, ,\\
\\
\displaystyle \frac{(x-2)^2}{4 \sigma^2 (1+\sigma)^2} + \frac{4\sigma^2 + x^2 -8x +12 +x \sqrt{x^2-4(1+\sigma^2)} }{8 (1+\sigma^2)} \\
\displaystyle + \frac{1}{4} \log \left( \frac{x^2  -2(1+\sigma^2) -x \sqrt{x^2-4(1+\sigma^2)} }{2(1+\sigma^2)^2} \right) & \mbox{for } 2+\sigma^2 \leq x \leq 2+\sigma \, ,\\
\\
\displaystyle     \frac{1}{4} \log \left( \frac{(2+\sigma)(2+\sigma - \sqrt{\sigma(4 - 3\sigma)})-2(1+\sigma^2)}{2} \right) \\
\displaystyle + \frac{6(1+\sigma^2) - x^2 +x\sqrt{x^2-4(1+\sigma^2)}}{8 (1+\sigma^2)} \\
\displaystyle +  \frac{1}{2} \log \left( \frac{\sigma \left(2+\sigma + \sqrt{\sigma(4-3\sigma)}\right)}{(1+\sigma^2)(2(1+\sigma) -x)\left( x + \sqrt{x^2 -4(1+\sigma^2)} \right)} \right) & \mbox{for } 2+\sigma \leq x \leq 2+2\sigma   \, .
    \end{array}
\right. &&
\end{align}
and is infinite otherwise. This function has been plotted in Fig.\ \ref{fig:RateFunction_scsc} for $\sigma=9/10$. 

\section{Large deviation for the product of symmetric matrices}
\label{sec:LDP_prod}
In this section we consider the case where the matrix $\mathbf{C}$ is given as the symmetric product:\footnote{One may note that since we are interested in the eigenvalues, we can equivalently consider the product $\mathbf{A} \mathbf{B}$ since this matrix is similar to the matrix $\mathbf{C}$ and hence has the same eigenvalues. However, unlike $\mathbf{C}$, the matrix $\mathbf{A} \mathbf{B}$ is a priori not symmetric.}
\begin{align}
\label{eq:prod_model}
    \mathbf{C} &= \sqrt{\mathbf{A}} \mathbf{B}  \sqrt{\mathbf{A}} \, ,&&
\end{align}
where $\mathbf{A} \sim \mathbb{P}_{V_A, w_A}$ and $\mathbf{B} \sim \mathbb{P}_{V_B, w_B}$ are two  positive semi-definite random matrices. In the large $N$ limit, the limiting density $\mu_C$ of $\mathbf{C}$ is described by the multiplicative free convolution of the next section and our goal is to compute the rate function $\Psi_C(x)$:\footnote{We recall once again that since we are studying the  case of symmetric matrices, one needs to replace the notation $ \zeta_i \left( \mathbf{C} \right)$  and $\Pi_C(x)$ by respectively  $ \zeta_i \left( \mathbf{C} \right)$ and  $\Psi_C(x)$ in Sec.\ \ref{Sec:Tilt}.}
\begin{align}
\label{eq:Psi_C.prod}
\mathbb{P}\left[\lambda_1(\mathbf{C}) \simeq x \right] &\approx \mathrm{exp} \left[-N \, \Psi_C(x) +o(N) \right] \quad \quad  \text{for} \, x> \mathrm{c}_+ \, ,&&
\end{align}
describing the right large deviation of the top eigenvalue of $\mathbf{C}$ far from its typical value given by the edge $\mathrm{c}_+$ of $\mu_C$. Let us make the following important remark concerning the case of the product of rectangular matrices. 
\vskip 0.3cm
\noindent \textit{Remark (product of rectangular random matrices): } We argue that the case of the product of two rectangular matrices reduced to the symmetric cases. Indeed, if we have two rectangular matrices $\mathbf{A}$ and $\mathbf{B}$ of size $(N \times M)$ and $(M \times K)$ respectively, then the non-zero singular values of the product are given by: 
\begin{align}
\label{eq:svprod}
    s_i \left( \mathbf{A} \mathbf{B}  \right) &= \sqrt{\lambda_i \left(  \mathbf{A} \mathbf{B}\mathbf{B}^{\mathsf{T}} \mathbf{A}^{\mathsf{T}} \right) } = \sqrt{\lambda_i \left(  \sqrt{\mathbf{A}^{\mathsf{T}}\mathbf{A}} \mathbf{B}\mathbf{B}^{\mathsf{T}} \sqrt{\mathbf{A}^{\mathsf{T}}\mathbf{A}} \right) } \, . &&
\end{align}
Since $ \mathbf{A}^{\mathsf{T}}\mathbf{A}$ and $\mathbf{B}\mathbf{B}^{\mathsf{T}}$ are two $(M \times M)$ symmetric matrices we recover the case of Eq.\ \eqref{eq:prod_model}.

\subsection{Product of invariant matrices: the multiplicative free convolution and the S-transform}
\label{sec:mult_free_conv}
The limiting density $\mu_C(.)$ is given as the unique probability distribution solution of: 
\begin{align}
\label{eq:prop_St}
    \tilde{\mathcal{S}}_C(y) &= \tilde{\mathcal{S}}_A(y) \, \tilde{\mathcal{S}}_B(y) \, ,&&
\end{align}
for all $z$ in the complex plane close enough to the origin, where $\tilde{\mathcal{S}}_A$ (resp. $\tilde{\mathcal{S}}_B, \tilde{\mathcal{S}}_C$) is the (modified)  \emph{S-transform}\footnote{The standard S-transform $\mathcal{S}$ is usually defined in the literature as the reciprocal of our S-transform: $\tilde{\mathcal{S}}(z)  = \frac{1}{\mathcal{S}(z)}$. As we will see, the convention $\tilde{\mathcal{S}}(.)$ will appear more naturally. } of the limiting spectral distribution $\mu_A$ (resp. $\mu_B, \mu_C$):
\begin{align}
\label{eq:def_St}
   \tilde{\mathcal{S}}_A(y) &:= \frac{y}{y+1} t_A^{\langle -1 \rangle}(y)  \, , &&
\end{align}
where  $t_A$ (resp. $t_B, t_C$), is the \emph{T-transform} of the limiting distribution $\mu_A$ (resp. $\mu_B, \mu_C$), which is defined for all $z \in \mathbb{C} \setminus [\mathrm{c}_-, \mathrm{c}_+]$ by: 
\begin{align}
 \label{eq:def_tt}
    t_A(z) &:= z g_A(z) - 1 = \int_{\mathrm{a}_{-}}^{\mathrm{a}_+} \frac{\lambda \, \mu_A(\lambda)}{z-\lambda} \mathrm{d}\lambda   \, . &&
\end{align}
Similarly, we define the second branch of the T-transform as: 
\begin{align}
 \label{eq:def_2btt}
    \bar{t}_A(z) &:= z \bar{g}_A(z) - 1   \, . &&
\end{align}
The function $\bar{t}_A(.)$ is the inverse of  $t_A^{\langle -1 \rangle}(.)$ continued to values higher than $t_A(\mathrm{a}_+)$.  As a consequence of Eq.\ \eqref{eq:prop_St}, the logarithm of the S-transform linearizes the multiplication and hence it is the random matrix analogous of the (log of the) Mellin transform of classical probability.  The limiting spectral distribution is therefore called the \emph{free multiplicative convolution} of $\mu_A$ and $\mu_B$ and is usually denoted by $\mu_C := \mu_A \boxtimes \mu_B$ in the RMT literature. We will be using the following properties of the multiplicative free convolution: 
\begin{itemize}
    \item If we denote by 
    \begin{align}
        \label{eq:ROC_S}
        \mathrm{s}_{A} &:= \lim_{w_A \to \infty} \bar{t}_A (w_A) \, , &&
    \end{align}
    then for $\theta \in (0, \mathrm{s}_A)$, $\tilde{\mathcal{S}}_A(.)$ is an increasing function of $\theta$, see App.\  \ref{sec:Ap:monot_St}.
    \item The T-transform  of the matrix $\mathbf{C}$ satisfies the following inequality at the edge:
        \begin{align}
        \label{eq:tc<ta^tb}
        t_C(\mathrm{c}_+) &\leq \min \left( t_A(\mathrm{a}_+),t_B(\mathrm{b}_+) \right)  \, , &&
    \end{align}
    see App.\ \ref{sec:Ap:ineq}.
\end{itemize}

\vskip 0.3cm
\noindent \textit{Example (S-transform of a Wishart matrix):} For a  Wishart matrix,  from its Stieltjes transform given by Eq.\ \eqref{eq:Stieltjes_MP}, one can easily get the T-transform defined by Eq.\ \eqref{eq:def_tt} and invert it to get the S-transform  of Eq.\ \eqref{eq:def_St}:
\begin{align}
    \label{eq:St_MP}
     \tilde{\mathcal{S}}_{\mathrm{MP}_q}(y) &= 1 + qy  \, . &&
\end{align}

\subsection{The multiplicative spherical integral and the LSSK model}
\label{sec:LSSK}
For the multiplicative case,  we  argue that a good candidate for the tilt is the \emph{multiplicative spherical integral} $\mathcal{Z}_{\mathbf{C}}(\theta)$ , where for \emph{any} positive semi-definite matrix $\mathbf{M}$ and $\theta  \geq 0$, this spherical integral is given by:
\begin{align}
\label{eq:PartitionF_LSSK}
    \mathcal{Z}_{\mathbf{M}}(\theta) &:= \int_{\mathbb{S}^{N-1}} \mathrm{d}\omega(\vect{\sigma}) \,  (\vect{\sigma}^{\mathsf{T}} \mathbf{M} \vect{\sigma} )^{ \frac{N}{2} \theta} \, . &&
\end{align}
where we recall that $\mathbb{S}^{N-1}$ is the hyper-sphere of radius one  and $ \mathrm{d}\omega(.)$ is the unit uniform measure over $\mathbb{S}^{N-1}$. By Haar property, for any two positive semi-definite matrices $\mathbf{M}_1$ and $\mathbf{M}_2$, this function satisfies:
\begin{align}
     \int_{\mathsf{O}(N)}  \mathcal{Z}_{\sqrt{\mathbf{M}_1} \mathbf{O} \mathbf{M}_2 \mathbf{O}^{\mathsf{T}}  \sqrt{\mathbf{M}_1} }(\theta)   \mathrm{d}\mathbf{O} &=    \mathcal{Z}_{\mathbf{M}_1}(\theta) \,    \mathcal{Z}_{\mathbf{M}_2}(\theta) \, ,&&
\end{align}
which gives for $\mathbf{M}_1=\mathbf{A}$ and $\mathbf{M}_2 = \mathbf{B}$ taken from an invariant ensemble and after averaging, the desired decomposition property of Eq.\ \eqref{eq:prop:AFE}.

\vskip 0.3cm
\noindent We can again interpret  $\mathcal{Z}_{\mathbf{M}}(\theta)$ as the partition function of a spherical model:
\begin{align}
\label{eq:PartitionF_LSSK.2}
    \mathcal{Z}_{\mathbf{M}}(\theta) &:= \langle \mathrm{e}^{\frac{N}{2} \theta \mathcal{H}^{\mathrm{LSSK}}(\vect{\sigma}) } \rangle \, ,&&
\end{align}
with the Hamiltonian:
\begin{align}
\label{eq:H_LSSK}
     \mathcal{H}^{\mathrm{LSSK}}(\vect{\sigma} )&:= \log \left( \sum_{i=1}^N \lambda_i(\mathbf{M}) \sigma_i^2 \right)  \, .&&
\end{align} 
Since $\mathbf{M}$ is the symmetric product of definite positive matrices, its eigenvalues are positive, so this Hamiltonian is well defined. In this paper, we are interested in the case $\mathbf{M}=\mathbf{C}$ given by Eq.\ \eqref{eq:prod_model}. Due to the logarithmic term, we denote this model as the \emph{Logarithmic Spherical Sherrington-Kirkpatrick} (LSSK for short) model. The behavior  of  $\mathcal{Z}_{\mathbf{C}}(\theta)$ for $N$ large has been recently investigated in Refs. \cite{MergnyPotters20,husson2021asymptotic}  and are given in the following section. Due to its similarity with the original SSK model, one should expect to have a similar behavior, with a paramagnetic phase at high temperature and a spin glass phase at low temperature. 

\subsection{Asymptotic behavior of the quenched and annealed free energies of the LSSK model}
\label{sec:As_QFE_AFE.prod}
We summarize here the asymptotic behavior of (the derivatives of) the quenched and annealed free energies.

\vskip 0.3cm
\noindent For  $\mathbf{C}$ conditioned to have its largest eigenvalue fixed at the position $x$, the partial derivatives of the quenched free energy of Eq.\ \eqref{eq:defQuenchedFE} with $\mathcal{Z}_{\mathbf{C}}(\theta)$ given by Eq.\ \eqref{eq:PartitionF_LSSK} satisfy a phase transition. In the high temperature regime, it has been shown in Refs. \cite{MergnyPotters20,husson2021asymptotic}  that the derivative with respect to the parameter $\theta$ of the quenched free energy is for small enough $\theta$  the logarithm of the S-transform. One can also get the behavior for low temperature where there is a saturation, and we have:
\begin{align}
\label{eq:As_QFE_prod.1}
\partial_{\theta} J_C(x, \theta) = \frac{1}{2} \left\{
    \begin{array}{ll}
    \log \tilde{\mathcal{S}}_C(\theta) & \mbox{for }\theta \leq t_C(x) \, ,\\
\\
       \log \left(   \frac{x\theta}{\theta +1} \right)  & \mbox{for }  \theta \geq t_C(x) \, .
    \end{array}
\right. &&
\end{align}
Similarly, the partial derivative with respect to $x$ is given by:
\begin{align}
\label{eq:As_QFE_prod.2}
  \partial_{x} J_C(x, \theta) = \frac{1}{2} \left\{
    \begin{array}{ll}
  0 & \mbox{for }\theta \leq t_C(x) \, ,\\
\\
    \frac{\theta +1}{x}  - g_C(x) & \mbox{for }  \theta \geq t_C(x) \, .
    \end{array}
\right. &&
\end{align}
For the derivative of the annealed free energy $F_C(.)$, one only needs the  of $F_A(w_A,.)$ and $F_B(w_B,.)$ separately since we have the decomposition of Eq.\ \eqref{eq:prop:AFE.2}, which are given by (see App.\ \ref{sec:Ap:As_AFE_LSSK} for a derivation):
\begin{align}
\label{eq:As_AFE_prod}
 \partial_{\theta} F_A(w_A,\theta) = \frac{1}{2} \left\{
    \begin{array}{ll}
  \log \tilde{\mathcal{S} }_A(\theta) & \mbox{for }\theta \leq \bar{t}_A(w_A) \, ,\\
\\
     \log \left(   \frac{w_A\theta}{\theta +1} \right)  & \mbox{for }  \theta \geq \bar{t}_A(w_A) \, ,
    \end{array}
\right. &&
\end{align}
and similarly for $F_B(w_B,.)$. Note that if the wall is at the edge $w_A=\mathrm{a}_+$, since $t_A(\mathrm{a}_+) = \bar{t}_A (\mathrm{a}_+)$, we have Eq.\ \eqref{eq:AFE_frozen.3} and hence form the large deviation of the top eigenvalue, we can indeed consider fixed diagonal matrices as invariant matrices with a wall at the edge. Conversely,  classical invariant ensembles correspond to the limit $w_A \to \infty$ from which we see that the annealed free energy is equal to the logarithm of the S-transform for $\theta<\mathrm{s}_A$, with $\mathrm{s}_A$ given by the limit of \eqref{eq:ROC_S}, and is otherwise infinite.

\subsection{Optimal temperature for the product}
\label{sec:OptInvTemp.prod}
Without any loss of generality, let's assume
\begin{align}
\label{eq:ta<tb}
  \bar{t}_A(w_A) &\leq \bar{t}_B(w_B)  \, , &&
\end{align}
and 
\begin{align}
\label{eq:tc<inf}
  \bar{t}_C(\mathrm{c}_+) &\leq \infty  \, . &&
\end{align}
Our goal is to show that the supremum in Eq.\ \eqref{eq:Theta_opt.1} is attained at a unique point by looking at the derivative $I_{x}'(.)$ with respect to $\theta$. Paying attention to the bounds in Eqs. \eqref{eq:As_QFE_prod.1} and \eqref{eq:As_AFE_prod}, one has the following behavior:
\begin{align}
\label{eq:DiffFE_Prod}
I_{x}'(\theta) = \frac{1}{2} \left\{
    \begin{array}{llll}
  0 & \mbox{for }\theta \leq t_C(x) \, ,\\
\\
  \log  \dfrac{x}{t_C^{\langle -1 \rangle}(\theta)}   & \mbox{for  } t_C(x) \leq \theta \leq \bar{t}_A(w_A) \, ,\\
\\
 \log \dfrac{x}{w_A \tilde{\mathcal{S}}_B(\theta)} & \mbox{for  } \bar{t}_A(w_A) \leq \theta \leq \bar{t}_B(w_B) \, ,\\
\\
 \log  \dfrac{x(\theta+1)}{w_A w_B \theta} & \mbox{for  } \theta \geq \bar{t}_B(w_B) \, .
    \end{array}
\right. &&
\end{align}
Based on similar monotonous argument as in the additive case of Sec.\ \ref{sec:OptInvTemp.sum} , one can show that for $\theta \geq t_C(x)$ this function is continuously increasing until it reaches the point $t_C(\mathrm{c}_+)$ and then it is continuously decreasing. For values of $x > w_A \, w_B$, it never crosses the real axis, and we have $\theta^*(x)=\infty$. Otherwise, it crosses the real axis exactly one time and the equation determining $\theta^*(x)$ depends on the position of $x$ with respect to the two critical points $x_{c_1}$ and $x_{c_2}$ defined by
\begin{align}
\label{eq:xc1_prod}
x_{c_1}& := t_C^{\langle -1 \rangle}(\bar{t}_A(w_A)) = w_A  \, \tilde{\mathcal{S}}_B\left( \bar{t}_A(w_A) \right) \, ,&&
\end{align}
and by:
 \begin{align}
 \label{eq:xc2_prod}
x_{c_2}& := w_A w_B \, \frac{\bar{t}_B(w_B)}{\bar{t}_B(w_B) +1} \, .&&
\end{align}

\begin{enumerate}
     \item for $\mathrm{c}_+ < x< x_{c_1}$, $\theta^*$ is attained in $(t_C(\mathrm{c}_+), \bar{t}_A(w_A))$ and so setting the second line of the RHS of Eq.\ \eqref{eq:DiffFE_Prod} being equals to zero, gives:
     \begin{align}
     \label{thetax_prod.1}
         \theta^*(x) &= \bar{t}_C(x) \, ;&&  
     \end{align}
     \item for $x_{c_1}<x<x_{c_2}$, the optimal inverse temperature is attained in the region $\bar{t}_A(w_A) \leq \theta \leq \bar{t}_B(w_B)$ and $\theta^*(x)$ is solution of the third line of the RHS of Eq.\eqref{eq:DiffFE_Prod}  being equal to zero, that is:
     \begin{align}
      \label{thetax_prod.2}
         \theta^*(x) &= \tilde{\mathcal{S}}_B^{\langle - 1 \rangle} \left( \frac{x}{w_A} \right) \, ; &&
     \end{align}
     \item for $x_{c_2}<x<w_A w_B$, $\theta^*(x)$ is attained in the region  $ \theta \geq \bar{t}_B(w_B)$ and so from Eq.\ \eqref{eq:DiffFE_Prod} it is given by:
     \begin{align}
      \label{thetax_prod.3}
         \theta^*(x)& = \frac{x}{w_A w_B -x} \, .&&
     \end{align}
 \end{enumerate}
 
 \subsection{Expression for the rate function}
 \label{sec:RateFunction.prod}
Using  the expressions of the previous section for the optimal temperature together with the expression of the partial derivative of the quenched free energy of Eq.\ \eqref{eq:As_QFE_prod.2} in Eq.\ \eqref{eq:Pi_Integral.1}, we have that the rate function is given by:

\begin{align}
\label{eq:RateFunction_Prod}
\Psi_C(x) =\left\{
    \begin{array}{lll}
  \displaystyle \frac{1}{2} \int_{\mathrm{c}_+}^x  \left( \bar{g}_C(t) - g_C(t)  \mathrm{d}t \right)   &   \mbox{for } \mathrm{c}_+ \leq x \leq x_{c_1} \, ,\\
\\
  K_1 +  \displaystyle \frac{1}{2} \int_{x_{c_1}}^x \left( \frac{\tilde{\mathcal{S}}_B^{\langle - 1 \rangle} \left(\frac{t}{w_A} \right)+1}{t} - g_C(t) \right)  \mathrm{d}t  & \mbox{for } x_{c_1} \leq x \leq x_{c_2} \, ,\\
\\
 K_2 + \displaystyle \frac{1}{2} \log \left( \dfrac{x}{w_A w_B -x} \right) - \frac{1}{2} \int_{x_{c_2}}^x g_C(t)  \mathrm{d}t  & \mbox{for  } x_{c_2} \leq x \leq w_A w_B \,.
    \end{array}
\right. &&
\end{align}
and is infinite for other values of $x$. The constant $K_1$ is again given by Eq.\ \eqref{eq:K1_add}  with now $x_{c_1}$ given Eq.\ \eqref{eq:xc1_prod} (if $x_{c_1} < \infty)$ and $K_2$ is defined  by:
\begin{align}
\label{eq:K2_prod}
K_2 &:= K_1 + \frac{1}{2}  \log \left( \frac{1}{\bar{t}_B(w_B)} \right) +  \frac{1}{2} \int_{x_{c_1}}^{x_{c_2}} \left( \frac{\tilde{\mathcal{S}}_B^{\langle - 1 \rangle} \left(\frac{t}{w_A} \right)+1}{t} -  g_C(t) \right)  \mathrm{t}s  &&
\end{align}
if $x_{c_2}<\infty$.

\vskip 0.3cm
\noindent \textit{Example (Rate Function for Generalized Wishart):} Let's consider the case  in Eq.\ \eqref{eq:prod_model} where $\mathbf{B}$ is a (White) Wishart with shape ratio $q$ and $\mathbf{A}$ is a fixed diagonal  (positive semi-definite) matrix. In this case, we have $w_A = \mathrm{a}_+$ and $w_B = \bar{t}_B(w_B) = \infty$. As a consequence, there is just one critical point and using the expression \eqref{eq:def_St} for the  S-transform of the Wishart matrix, it is given by:
\begin{align}
\label{eq:xc1_GenWishart}
    x_{c_1} &= \mathrm{a}_+( 1 + q \, t_A(\mathrm{a}_+) )\, .&&
\end{align}
and using Eq.\ \eqref{eq:St_MP} for the S-transform, we have for the rate function:
\begin{align}
\label{eq:RateFunction_GenWishart}
\Psi_C(x) =\left\{
    \begin{array}{lll}
  \displaystyle \frac{1}{2} \int_{\mathrm{c}_+}^x \left( \bar{g}_C(t) - g_C(t) \right)  \mathrm{d}t   &   \mbox{for } \mathrm{c}_+ \leq x \leq x_{c_1} \, ,\\
\\
  \displaystyle K_1 + \frac{1}{2} \left( - \frac{1}{q} - t_A(\mathrm{a}_+) + \frac{x}{q \mathrm{a}_+} + \left( 1- \frac{1}{q} \right) \log \left( \frac{x}{x_{c_1}}\right) -  \int_{x_{c_1}}^x g_C(t)  \mathrm{d}t \right) & \mbox{for } x \geq x_{c_1} \, ,\\
    \end{array}
\right. &&
\end{align}
which is up to a change in the notation, the results obtained in Ref. \cite{Maillard21}.

\section{Large deviation for the top singular value of sum of rectangular matrices}
\label{sec:LDP_Rect}
In this section, we consider the case where the matrix $\mathbf{C}$ is given as 
\begin{align}
    \label{eq:CSum_rect}
    \mathbf{C} &= \mathbf{A} + \mathbf{B} \, , &&
\end{align}
where $\mathbf{A} \sim \mathbb{P}_{V_A,w_A} $ and $\mathbf{B} \sim \mathbb{P}_{V_B,w_B}$ are two \emph{rectangular} matrices, each taken from a bi-invariant ensemble with a wall as defined in Sec.\ \ref{sec:BiInvEns_wall}. We aim at computing the rate function $\Phi_C(x)$:\footnote{We recall that since we are studying the rectangular case, one needs to replace the notation $ \zeta_i \left( \mathbf{C} \right)$  and $\Pi_C(x)$    in Sec.\ \ref{Sec:Tilt}  by  $s_i \left( \mathbf{C} \right)$ and $\Phi_C(x)$  respectively. } 
\begin{align}
\label{eq:Phi_C.sum}
\mathbb{P}\left[s_1(\mathbf{C}) = x \right] &\approx \mathrm{exp} \left[-N \, \Phi_C(x) +o(N) \right] \quad \quad  \text{for} \, x> \mathrm{c}_+ \, ,&&
\end{align}
where $\mathrm{c}_+$ is the edge of the limiting density of singular values $\rho_C$ of $\mathbf{C}$, described by the \emph{rectangular free convolution}, see the next paragraph. Note that unlike the case of the product of rectangular matrices, this case does not boil down to consider the sum of symmetric matrices, since: 
\begin{align}
\label{eq:sv_sum_neq_eigvals}
    s_i \left( \mathbf{A} + \mathbf{B} \right) &= \sqrt{\lambda_i \left( \mathbf{A} \mathbf{A}^{\mathsf{T}} + \mathbf{B} \mathbf{B}^{\mathsf{T}} + \mathbf{A} \mathbf{B}^{\mathsf{T}}  + \mathbf{B} \mathbf{A}^{\mathsf{T}} \right)} \neq \sqrt{\lambda_i \left( \mathbf{A} \mathbf{A}^{\mathsf{T}} + \mathbf{B} \mathbf{B}^{\mathsf{T}}  \right)} \, . &&
\end{align}
 For specific values of $q$ (namely $q=0$, corresponding to \emph{long matrices} and $q=1$, corresponding to square matrices), it is known - as we will see - that this rectangular free convolution is related to the additive free convolution of Sec.\ \ref{sec:add_free_conv}.

\subsection{Rectangular free convolution}
\label{sec:Rect_conv}
The LSVD $\rho_C$ of $\mathbf{C}$ of Eq.\ \eqref{eq:CSum_rect}  is given by the \emph{rectangular free convolution (with shape ratio $q$)} \cite{BenaychGeorges2008,BenaychGeorges2009}, denoted by $\rho_A \boxplus_{q} \rho_B$ in the RMT literature, which is -- similarly to the additive and multiplicative free convolution -- defined via a linearizing transform called the (modified)\footnote{The standard convention for the \emph{rectangular R-transform}, $\mathcal{R}^{(q)}_A$, is related to our rectangular C-transform $\tilde{\mathcal{C}}^{(q)}_A$, via $ \tilde{\mathcal{C}}^{(q)}_A(t) =: \mathcal{R}^{(q)}_A(t^2)/t $. As we will see, the convention $\tilde{\mathcal{C}}^{(q)}_A$ will appear more naturally.} \emph{rectangular C-transform (with shape ratio $q$)}:
\begin{align}
\label{Ap:prop:RectRt}
    \tilde{\mathcal{C}}^{(q)}_C(y) &=   \tilde{\mathcal{C}}^{(q)}_A(y)+   \tilde{\mathcal{C}}^{(q)}_B(y) \, . &&
\end{align}
This transform is given by the formula:
\begin{align}
\label{Ap:def:RectRt}
    \tilde{\mathcal{C}}^{(q)}_A(y)& :=    \frac{U\left( y \, d_A^{\langle -1 \rangle}(y) \right)}{y} \, , &&
\end{align}
where $U(.)$ is defined by:
\begin{align}
\label{Ap:def:U}
U(y) &:= \frac{- 1 - q + \sqrt{(1-q)^2 +4 q y^2 }}{2 q} \, .&&
\end{align}
Note that  $U(.)$ is an increasing function of $y$ whose inverse transform is given by the simple formula:
\begin{align}
\label{Ap:def:InvU}
    U^{\langle -1 \rangle}(z) &= \sqrt{(1+z)(1+ qz)} \, . &&
\end{align}
The function $d_A^{\langle -1 \rangle}(.)$ in Eq.\ \eqref{Ap:def:RectRt} is the inverse functional of the \emph{D-transform}, which is the rectangular counterpart of the Stieltjes transform (resp. T-transform) in the additive (resp. multiplicative) case, defined by:
\begin{align}
\label{Ap:def:Dt}
    d_A(z) &:=  \sqrt{\left( \int  \frac{z}{z^2-s^2} \rho_A(s)\mathrm{d}s \right) \, \left( q  \int  \frac{z}{z^2-s^2} \rho_A(s)\mathrm{d}s  + \frac{1 - q}{z} \right)} \, .&&
\end{align}
 Note that if we denote by $g_{AA^{\mathsf{T}}}(.)$ the Stieltjes transform of the measure $\mu_{AA^{\mathsf{T}}}(.) =\frac{\rho_A(\sqrt{.})}{2 \sqrt{.}}$, the limiting Stieltjes of the matrix $\mathbf{A} \mathbf{A}^{\mathsf{T}}$, then we have:
\begin{align}
\label{Ap:prop:Dt}
     d_A(z) &= \sqrt{q z^2 \left( g_{AA^{\mathsf{T}}}(z^2) \right)^2 + (1-q)  g_{AA^{\mathsf{T}}}(z^2)} \, .&&
\end{align}
\vskip 0.3cm
\noindent \textit{Remark (long matrices ($q \to 0$) and additive free convolution):} In the limit $q \to 0$, corresponding to the case of ($N \times M)$ rectangular \emph{long} matrices with $1 \ll N \ll M$, we have for the function $U$ and the D-transform,
\begin{align}
\label{eq:U.q0}
    U(y) &\underset{q \to 0}{\to} y^2 -1 \, , &&
\end{align}
and 
\begin{align}
\label{eq:Dt.q0}
    d_A(z) &\underset{q \to 0}{\to} \sqrt{g_{AA^{\mathsf{T}}}(z^2)} \, , &&
\end{align}
and so the inverse of the D-transform is given by:
\begin{align}
\label{eq:InvDt.q0}
    d_A^{\langle -1 \rangle}(y) &\underset{q \to 0}{\to} \sqrt{g_{AA^{\mathsf{T}}}^{\langle -1 \rangle}(y^2)} \, . &&
\end{align}
As a consequence, the rectangular C-transform is related to the R-transform by
\begin{align}
\label{Ap:RectRt.q0}
    \tilde{\mathcal{C}}^{(0)}_A(y)& = y \, \mathcal{R}_{AA^{\mathsf{T}}} \left( y^2\right)  \, , &&
\end{align}
and so by the linearizing property of the C-transform and the R-transform we have:
\begin{align}
\label{Ap:RectRt.q0.1}
    \tilde{\mathcal{C}}^{(0)}_C(y) &= y \, \mathcal{R}_{AA^{\mathsf{T}}+BB^{\mathsf{T}}} \left( y^2\right)  \, . &&
\end{align}
In other words for long matrices, if one is looking at the LSVD of the sum, one can replace the symbol '$\neq$' in Eq.\ \eqref{eq:sv_sum_neq_eigvals} by an equality!
\vskip 0.3cm
\noindent \textit{Remark (square matrices ($q=1$) and symmetrized density):} For $q=1$, corresponding to (asymptotic) square matrices, the function $U$ is simply given by:
\begin{align}
\label{eq:U.q1}
    U(y) &\underset{q \to 1}{\to} y -1 \, , &&
\end{align}
and the D-transform of  Eq.\ \eqref{Ap:def:Dt} considerably simplifies (for $z>\mathrm{a}_+$) into:
\begin{align}
    \label{Ap:Dt.q1}
     d_A(z) &=   \int  \frac{z}{z^2-s^2} \rho_A(s)\mathrm{d}s =  \int \frac{ \mu_{\hat{A}}(\lambda)}{z-\lambda} \mathrm{d}\lambda =: g_{\hat{A}}(z)  && (\mbox{for } q=1) \, .
\end{align}
where we recall that $\mu_{\hat{A}}$ is the symmetrized density of $\rho_A$, given by Eq.\ \eqref{eq:symm_density}. The C-transform  of Eq.\ \eqref{Ap:def:RectRt} reads in this case:
\begin{align}
\label{Ap:prop:RectRt.q1}
 \tilde{\mathcal{C}}^{(1)}_A(y) &= \mathcal{R}_{\hat{A}}(y) \, ,&&
\end{align}
and by linearizing property, we have that the LSVD of the matrix $\mathbf{C}$ is given as the unique probability measure on $\mathbb{R}_+$ such that:
\begin{align}
\label{Ap:prop:RectRt.q1.2}
 \tilde{\mathcal{C}}^{(1)}_C(y) &= \mathcal{R}_{\hat{A}+\hat{B}}(y) \, .&&
\end{align}
In other words, the singular values of the sum of two (bi-free) square matrices is given asymptotically by the additive free convolution  of Sec.\ \ref{sec:add_free_conv} of their respective symmetrized singular value densities.
\vskip 0.3cm 
\noindent \textit{Example (Gaussian rectangular random matrices):} Let's consider the case of Gaussian rectangular matrices with LSVD given Eq.\ \eqref{eq:LSVD_gauss}. Using Eq.\ \eqref{Ap:prop:Dt} with the expression of Eq.\ \eqref{eq:St_MP} for the Stieltjes transform of the Mar\v{c}enko-Pastur distribution, one gets the following expression for the D-transform (for $z>0$):
\begin{align}
    \label{eq:dt_Gauss}
    d_A(z) &= \frac{1}{\sigma^2} \sqrt{\frac{z^2 -(1+q) \sigma^2 - \sqrt{z^4 - 2(1+q) \sigma^2 z^2 + (1-q)^2 \sigma^4}}{2q}} \, ,&&
\end{align}
whose inverse is given by
\begin{align}
    \label{eq:inv_dt_Gauss}
    d_A^{\langle -1 \rangle}(y) &=  \frac{\sqrt{(1+ \sigma^2 y^2)(1+q \sigma^2 y^2)}}{y} \, . &&
\end{align}
The argument inside the square-root function is nothing else than the inverse $U^{\langle -1 \rangle}$ evaluated at $ \sigma^2 y^2$, see Eq.\ \eqref{Ap:def:InvU}. Using Eq.\ \eqref{Ap:def:RectRt}, the rectangular C-transform of the Gaussian rectangular matrix is given simply by:
\begin{align}
\label{eq:Ct_Gauss}
     \tilde{\mathcal{C}}^{(q)}_A(y) &= \sigma^2 y \, . && 
\end{align}
\vskip 0.3cm
\noindent In the following, we will be using the two following properties of the rectangular free convolution:
\begin{itemize}
    \item we argue that the function $ \tilde{\mathcal{C}}^{(q)}_A(y)$ is a continuous increasing function see App.\ \ref{sec:Ap:monot} 
    \item The D-transform at the edge of $\rho_C$ satisfies the following inequality:
    \begin{align}
        \label{eq:dc<da^db}
        d_C(\mathrm{c}_+) &\leq \min \left(  d_A(\mathrm{a}_+),  d_B(\mathrm{a}_+) \right) \, &&
    \end{align}
    see App.\ \ref{sec:Ap:ineq}.
\end{itemize}
\subsection{Rectangular spherical integral and the BSSK model}
\label{sec:BSSK}
Similar to the cases of the sum of symmetric matrices, we choose the tilt function for the problem of the sum of rectangular matrices to be given by the \emph{rectangular spherical integral} $\mathcal{Z}_{\mathbf{C}}(\theta)$ defined for  \emph{any} rectangular matrix $\mathbf{M}$ and $\theta \geq 0$ by:
\begin{align}
\label{eq:PartitionF_BSSK.int}
    \mathcal{Z}_{\mathbf{M}}(\theta) &:= \int_{\mathbb{S}^{N -1}} \mathrm{d}w(\vect{\sigma}_1)   \int_{\mathbb{S}^{M -1}} \mathrm{d}w(\vect{\sigma}_2) \, \mathrm{exp}\left[ \sqrt{N M} \theta\vect{\sigma}^{\mathsf{T}}_1  \mathbf{M} \vect{\sigma}_2  \right] \, . &&
 \end{align}
For any rectangular matrices $\mathbf{M}_1$ and $\mathbf{M}_2$,  this function satisfies the following property: 
 \begin{align}
 \label{eq:prop_rect_SI}
     \int_{\mathsf{O}(N)}  \int_{\mathsf{O}(M)}  \mathcal{Z}_{\mathbf{M}_1  +\mathbf{U} \mathbf{M}_2 \mathbf{V}^{\mathsf{T}}}(\theta)   \mathrm{d}\mathbf{U} \mathrm{d}\mathbf{V} &=    \mathcal{Z}_{\mathbf{M}_1}(\theta) \,    \mathcal{Z}_{\mathbf{M}_2}(\theta) \, ,&&
\end{align}
which evaluated for $\mathbf{M}_1 = \mathbf{A}$ and $\mathbf{M}_2 = \mathbf{B}$ given as rectangular bi-invariant random matrices,  gives after integration over the laws of $\mathbf{A}$ and $\mathbf{B}$, the decomposition property of Eq.\ \eqref{eq:prop:AFE}.
\vskip 0.3cm
\noindent This rectangular spherical can again be understood as the partition function of a spherical model with inverse temperature $\theta$. Indeed, if we denote by
\begin{align}
    \label{eq:def:sig_rect}
    \vect{\sigma} &:= \frac{1}{\sqrt{2}} \left[ \vect{\sigma}_1, \vect{\sigma}_2 \right] \in \mathbb{S}^{N+M-1} \, , &&
\end{align}
$\mathcal{Z}_{\mathbf{C}}(\theta)$ can be written as: 
\begin{align}
\label{eq:PartitionF_BSSK}
    \mathcal{Z}_{\mathbf{M}}(\theta) &:= \langle \mathrm{e}^{\sqrt{NM} \theta \mathcal{H}^{\mathrm{BSSK}}(\vect{\sigma}) } \rangle \, ,&&
\end{align}
with the Hamiltonian 
\begin{align}
\label{eq:H_BSSK}
    \mathcal{H}^{\mathrm{BSSK}}(\theta) &:= \vect{\sigma}^{\mathsf{T}} \begin{pmatrix}
   \mathbf{0} & \mathbf{M}\\
    \mathbf{M}^{\mathsf{T}} & \mathbf{0} 
\end{pmatrix} \vect{\sigma}  \, ,&&
\end{align}
and by bi-invariance, this can be also written as:
\begin{align}
\label{eq:H_BSSK.1}
    \mathcal{H}^{\mathrm{BSSK}}(\theta) &= \sum_{i=1}^N s_i(\mathbf{M}) \sigma_i \sigma_{N+i}.&&
\end{align}
This model is known \cite{Auffinger2014,Barra2014,Baik2020} as the \emph{Bipartite Spherical Sherrington-Kirkpatrick} (BSSK in short) spin model, due to the graph structure of the interaction matrix: each coordinate of one family vector interacts only with members of the other family.

\subsection{Asymptotic behavior of the annealed and quenched free energy of the BSSK model}
\label{sec:As_QFE_AFE.rect}
To have the rate function, we first need to compute the derivatives of the quenched and annealed free energy of the BSSK model of Eq.\ \eqref{eq:defQuenchedFE} and Eq.\ \eqref{eq:defAnnealedFE} respectively with the tilt function $\mathcal{Z}_{\mathbf{C}}$ given by Eq.\ \eqref{eq:PartitionF_BSSK.int}.
\vskip 0.3cm
\noindent For $\mathbf{C}$ conditioned to have its largest singular value at the position $x$, the quenched annealed free energy is known to be related to the rectangular C-transform for small enough value of the inverse temperature $\theta$. For higher value, it depends also on the position $x$, see the derivation in the Appendix, and we have in full generality:
\begin{align}
\label{eq:As_QFE_rect.1}
\partial_{\theta} J_C(x, \theta) =\left\{
    \begin{array}{ll}
    \tilde{\mathcal{C}}^{(q)}_C(\theta) & \mbox{for }\theta \leq d_C(x) \, ,\\
\\
       \frac{U(\theta x)}{\theta}  & \mbox{for }  \theta \geq d_C(x) \, .
    \end{array}
\right. &&
\end{align}
Similarly, the partial derivative with respect to $x$ is given by:
\begin{align}
\label{eq:As_QFE_rect.2}
  \partial_{x} J_C(x, \theta) =\left\{
    \begin{array}{ll}
  0 & \mbox{for }\theta \leq d_C(x) \, ,\\
\\
    \frac{ \sqrt{(1-q)^2 +4q \theta^2 x^2 } - \sqrt{(1-q)^2 +4q d_A(x)^2 x^2} }{2qx} & \mbox{for }  \theta \geq t_C(x) \, .
    \end{array}
\right. &&
\end{align}
\vskip 0.3cm
\noindent The derivative of the annealed free energy $F_C$ is the sum of the derivative of the  annealed free energy $F_A$ and $F_B$ which are given by:
\begin{align}
\label{eq:As_AFE_rect}
\partial_{\theta} F_A(w_A,\theta) =\left\{
    \begin{array}{ll}
  \tilde{\mathcal{C}}_A^{(q)}(\theta) & \mbox{for }\theta \leq \bar{d}_A(w_A) \, ,\\
\\
    \frac{ U(\theta w_A)}{\theta}  & \mbox{for } \theta \geq \bar{d}_A(w_A) \, ,
    \end{array}
\right. &&
\end{align}
and similarly for $F_B$. If the wall is at the edge $w_A=\mathrm{a}_+$, since $d_A(\mathrm{a}_+) = \bar{d}_A (\mathrm{a}_+)$, we have Eq.\ \eqref{eq:AFE_frozen.3} and hence form the large deviation of the top singular value, we can  consider fixed rectangular diagonal matrices as bi-invariant matrices with a wall at the edge.

\subsection{Optimal temperature for the rectangular case}
\label{sec:OptInvTemp.rect}
Without loss of generality, we assume: 
\begin{align}
    \label{eq:da<db}
    \bar{d}_A(w_A)&\leq \bar{d}_B(w_B) \, , &&
\end{align}
and the non-trivial condition: 
\begin{align}
    \label{eq:dc<inf}
    d_C(\mathrm{c}_+)&< \infty \, . &&
\end{align}
In this case, the derivative with respect to $\theta$ of the function $I_x(.)$ in the supremum of Eq.\ \eqref{eq:Theta_opt.1} is given by:
\begin{align}
\label{eq:DiffFE_Rect}
I_x'(\theta) =\left\{
    \begin{array}{llll}
  0 & \mbox{for }\theta \leq d_C(x) \, ,\\
\\
 \frac{U(\theta x)}{\theta} - \tilde{\mathcal{C}}^{(q)}_C(\theta) & \mbox{for } d_C(x) \leq \theta \leq \bar{d}_A(w_A) \, ,\\
\\
\frac{U(\theta x)}{\theta} - \frac{U(\theta w_A)}{\theta} - \tilde{\mathcal{C}}^{(q)}_B(\theta) & \mbox{for  } \bar{d}_A(w_A) \leq \theta \leq \bar{d}_B(w_B) \, ,\\
\\
\frac{U(\theta x)}{\theta} - \frac{U(\theta w_A)}{\theta} - \frac{U(\theta w_B)}{\theta} & \mbox{for  } \theta \geq \bar{d}_B(w_B) \, .
    \end{array}
\right. &&
\end{align}
By property of the rectangular free convolution, for $\theta>d_C(x)$, this function is first increasing with $\theta$ until it reaches the value $\theta =d_C(\mathrm{c}_+)$  and then it is decreasing with $\theta$. One can check that it crosses the real axis if $x$ is in the interval $[\mathrm{c}_+,w_A+w_B]$, and in this case, the position of the optimal inverse temperature $\theta^*$ depends on two critical points $x_{c_1}$ and $x_{c_2}$. The first one is given by
\begin{align}
    \label{eq:xc1_rect}
    x_{c_1} &:= d_C^{\langle -1 \rangle}(\bar{d}_A(w_A)) \, .&&
\end{align}
Unfortunately, unlike the sum and the product of symmetric matrices, the expression for $x_{c_1}$ in terms of the rectangular C-transform is quite involved. if we introduce the function:
\begin{align}
\label{eq:def:fq}
    f_q(z)&:= \frac{1}{2} \sqrt{(1-q)^2 +4 q z^2} \, , &&
\end{align}
to ease the notation, then we have: 
\begin{align}
    \label{eq:xc1_rect.2}
    x_{c_1} &= \displaystyle \sqrt{w_A^2 + \tilde{\mathcal{C}}_B^{(q)}(\bar{d}_A(w_A)) \left( q \tilde{\mathcal{C}}_B^{(q)}(\bar{d}_A(w_A)) + \dfrac{f_q(w_A \bar{d}_A(w_A))}{\bar{d}_A(w_A)} \right) } \, .&&
\end{align}
The other critical point $x_{c_2}$ is given by:
\begin{align}
\label{eq:xc2_rect}
 x_{c_2} &:= \frac{1}{\bar{d}_B(w_B)} U^{\langle -1 \rangle} \left( U( w_A \bar{d}_B(w_B)) +  U( w_B \bar{d}_B(w_B)) \right) \,  , &&
\end{align}
where we recall that $U^{\langle -1 \rangle}$ is given by Eq.\ \eqref{Ap:def:InvU}. Eq.\ \eqref{eq:xc2_rect} can be written in semi-explicit form with the function $f_q$ of Eq.\ \eqref{eq:def:fq}:
\begin{align}
\label{eq:xc2_rect.2}
 x_{c_2} &= \frac{\sqrt{ \left(f_q\left( w_A \bar{d}_B(w_B) \right) +  f_q\left( w_B \bar{d}_B(w_B) \right) -1 \right) \left( f_q\left( w_A \bar{d}_B(w_B) \right) +  f_q\left( w_B \bar{d}_B(w_B) \right) -q \right) }}{\sqrt{q} \bar{d}_B(w_B) } \, .
\end{align}
\vskip 0.3cm
\noindent 
\begin{enumerate}
     \item for $\mathrm{c}_+ \leq x \leq x_{c_1}$, $\theta^*$ is attained in $(d_C(\mathrm{c}_+), \bar{d}_A(w_A))$ and hence $\theta^*$ is solution of  the second line of the RHS of Eq.\ \eqref{eq:DiffFE_Rect} being equals to zero. Using Eq.\ \eqref{Ap:def:RectRt} to express the C-transform in terms of the function $U$, one gets:
     \begin{align}
     \label{eq:thetax_rect.1}
        U( \theta^* x) &= U(d^{\langle -1 \rangle }_C(\theta^*) \theta^*) \, ;&&  
     \end{align}
     and so by applying the (monotonous) function $U^{\langle -1 \rangle}(.)$ to this equation and dividing by $\theta^*$, one gets to solve the equation:
      \begin{align}
     \label{eq:thetax_rect.1.2}
        d^{\langle -1 \rangle }_C(\theta^*) &=x & \left( \mbox{for } \theta^* \in (d_C(\mathrm{c}_+) , \bar{d}_A(w_A)) \right) \, .&&  
     \end{align}
 The solution is given by the second branch of the D-transform:
      \begin{align}
     \label{eq:thetax_rect.1.3}
         \theta^*(x) &= \bar{d}_C(x) \, .&&  
     \end{align}
     
     \item for $x_{c_1} \leq x \leq x_{c_2}$, the optimal inverse temperature is attained in the region $\bar{d}_A(w_A) \leq \theta \leq \bar{d}_B(w_B)$. $\theta^* \equiv \theta^*(x)$ is solution of the third line of the RHS of Eq.\eqref{eq:DiffFE_Rect}  being equal to zero so that it satisfies:
     \begin{align}
      \label{eq:thetax_rect.2}
      U(x \theta^*) &:=  U(w_A \theta^*) + \tilde{\mathcal{C}}^{(q)}_B(\theta^*) \theta^*   \, ; &&
     \end{align}
     If one applies the function $U^{\langle -1 \rangle}$ on each side, one gets after simplification an analytical expression for the function $\theta^* \mapsto x(\theta^*)$:
     \begin{align}
   \label{eq:thetax_rect.2.1}
              x(\theta^*)&= \sqrt{w_A^2 + \tilde{\mathcal{C}}_B^{(q)}(\theta^*) \left( q \tilde{\mathcal{C}}_B^{(q)}(\theta^*) + \frac{ f_q \left( w_A \theta^* \right) }{\theta^*} \right) } \, , &&
     \end{align}
     which is by definition the inverse of the function $\theta^*(x)$. Now unfortunately for general values of the parameter $q$,  we do not have a simple analytical formula for the optimal inverse temperature function of  the position $x$ and so  we simply denote by $F_1$ the solution of Eq.\ \eqref{eq:thetax_rect.2.1}  with unknown $\theta^*$ for $x$ between $x_{c_1}$ and $x_{c_2}$.

     \item for $x_{c_2} \leq x \leq w_A + w_B$, $\theta^*(x)$ is attained in the region  $ \theta \geq \bar{d}_B(w_B)$ and so from Eq.\ \eqref{eq:DiffFE_Rect} and after simplification, one gets the following analytical expression for the function  $\theta^* \mapsto x(\theta^*)$:
     \begin{align}
      \label{eq:thetax_rect.3}
      x(\theta^*) &= \frac{\sqrt{(f_q(w_A \theta^*) +f_q(w_B \theta^*) -1)(f_q(w_A \theta^*) + f_q(w_B \theta^*)-q)}}{\sqrt{q} \theta^*}  \, .&&
      \end{align}
      In full generality, one can isolate one of the radical functions and take the square of the newly obtain equation and repeat the process until it becomes a polynomial equation. In our setting and for general $q,w_A,w_B$, one would obtain that $(\theta^*)^2$ is one of the zeros of a polynomial of degree $8$, and hence there is no hope of finding an analytical expression for $\theta^*$. As a consequence, we simply denote by $F_2$ the solution (in $\theta)$ of Eq.\ \eqref{eq:thetax_rect.3} for $x$ higher than $x_{c_2}$.   Now for  specific values of $w_A$ and $w_B$, for example $w_A =w_B$,  or $q$, for example $q=0$ or $q=1$, Eq.\ \eqref{eq:thetax_rect.3} becomes, after some work, a quadratic or even linear equation for $(\theta^*)^2$ (or $\theta$), as we will see. 
 \end{enumerate}
 \vskip 0.3cm
 \noindent \textit{Remark (Simplification for the case of long ($q=0$) matrices):} \begin{itemize}
     \item For $x_{c_1} \leq x \leq x_{c_2}$, Eq.\ \eqref{eq:thetax_rect.2.1} for the optimal inverse temperature simplifies into:
     \begin{align}
         \label{eq:OptTheta.Rect.q0}
          x &= \sqrt{w_A + \frac{ \tilde{\mathcal{C}}_B^{(0)} \left( \theta^* \right)}{\theta^*} } \, , &&
     \end{align}
     and since the rectangular C-transform of a long matrix is related to R-transform by Eq.\ \eqref{Ap:RectRt.q0}, we have:
          \begin{align}
         \label{eq:OptTheta.Rect.q0.1}
          \theta^*(x) &= \sqrt{\mathcal{R}^{\langle -1 \rangle}_{BB^{\mathsf{T}}}(x^2 -w_A^2)}  \, . &&
     \end{align}
      \item For $x_{c_2} \leq x \leq w_A +w_B$, Eq.\ \eqref{eq:thetax_rect.3} for the optimal inverse temperature simplifies into:
           \begin{align}
         \label{eq:OptTheta.Rect.q0.2}
          x &= \frac{\sqrt{ {\theta^*}^2 \left( w_A^2 +w_B^2 \right) -1}}{\theta^*}  \, , &&
     \end{align}
     such that the optimal temperature is given by:
     \begin{align}
         \label{eq:OptTheta.Rect.q0.3}
          \theta^*(x) &= \frac{1}{\sqrt{w_A^2 +w_B^2 -x^2}}  \, . &&
     \end{align}
 \end{itemize}

 \vskip 0.3cm
 \noindent \textit{Remark (Simplification for the case of square ($q=1$) matrices):} 
 \begin{itemize}
     \item For $x_{c_1} \leq x \leq x_{c_2}$, Eq.\ \eqref{eq:thetax_rect.2.1} for the optimal inverse temperature simplifies into:
     \begin{align}
            \label{eq:thetax_rect.2.3}
            x-w_A &=  \mathcal{C}_B^{(1)}(\theta^*)   \, , &&
     \end{align}
    and since for $q=1$, the rectangular C-transform is the R-transform of the symmetrized density (see Eq.\ \eqref{Ap:prop:RectRt.q1}), we have:
        \begin{align}
        \label{eq:thetax_rect.2.4}
           \theta^*(x) &=  \mathcal{R}_{\hat{B}}^{\langle -1 \rangle}(x-w_A)  \,  . &&
     \end{align}
      \item For $x_{c_2} \leq x \leq w_A +w_B$, Eq.\ \eqref{eq:thetax_rect.3} for the optimal inverse temperature simplifies into:
             \begin{align}
            \label{eq:thetax_rect.3.1}
            \left( x \theta^*(x) \right)^2 &= \left( \theta^*(x) \left(w_A + w_B \right) -1 \right)^2 \, , &&
     \end{align}
     and hence we have:
        \begin{align}
        \label{eq:thetax_rect.3.2}
           \theta^*(x) &= \frac{1}{w_A + w_B -x} \,  . &&
     \end{align}
 \end{itemize}

\subsection{Expression for the rate function}
\label{sec:RateFunction.rect}
Using the general expression of Eq.\ \eqref{eq:Pi_Integral.1} for the rate function with the expression of Eq.\ \eqref{eq:As_QFE_rect.2} and the expression of the optimal inverse temperature of the previous section, we have: 
\begin{align}
\label{eq:RateFunction_Rect}
\Psi_C(x) =\left\{
    \begin{array}{lll}
  \displaystyle \frac{1}{2} \int_{\mathrm{c}_+^2}^{x^2}  \left( \bar{g}_{CC^{\mathsf{T}}}(t)  -g_{CC^{\mathsf{T}}}(t) \right)  \mathrm{d}t   &   \mbox{for } \mathrm{c}_+ \leq x \leq x_{c_1} \, ,\\
\\
   K_1 +  \displaystyle \int_{x_{c_1}}^x \dfrac{ f_q \left(  F_1(t)t \right)  - f_q \left(  d_C(t) t \right) }{qt}  \mathrm{d}t   & \mbox{for } x_{c_1} \leq x \leq x_{c_2} \, ,\\
\\
 K_2 +  \displaystyle \int_{x_{c_2}}^x \dfrac{ f_q \left(  F_2(t)t \right)  - f_q \left(  d_C(t) t \right) }{qt}  \mathrm{d}t   & \mbox{for  } x_{c_2} \leq x \leq w_A + w_B \,.
    \end{array}
\right. &&
\end{align}
where we recall that $x_{c_1}$, $x_{c_2}$ are defined by Eq.\ \eqref{eq:xc1_rect} and Eq.\ \eqref{eq:xc2_rect} and $F_1$ (resp. $F_2$) is defined as the (correct) solution with unknown $\theta^*$ of Eq.\ \eqref{eq:thetax_rect.2.1} (resp. Eq.\  \eqref{eq:thetax_rect.3}) and $K_1$ and $K_2$ are the constants such that this rate function is continuous. To get the top line of Eq.\ \eqref{eq:RateFunction_Rect} we have the property of Eq.\ \eqref{Ap:prop:Dt} relating the D-transform and the Stieltjes transform. 
\vskip 0.3cm
\noindent \textit{Remark (Rate function for the sum of long ($q \to 0$) matrices):} In this case, we have:
\begin{align}
\label{eq:RateFunction_Rect.q0}
\Phi_C(x) = \Psi_{AA^{\mathsf{T}}+BB^{\mathsf{T}}}(x^2) = \left\{
    \begin{array}{lll}
  \displaystyle \frac{1}{2} \int_{\mathrm{c}_+^2}^{x^2}   \left( \bar{g}_{AA^{\mathsf{T}}+BB^{\mathsf{T}}}(t)  -  g_{AA^{\mathsf{T}}+BB^{\mathsf{T}}}(t) \right) \mathrm{d}t   &   \mbox{for } \mathrm{c}_+ \leq x \leq x_{c_1} \, ,\\
\\
 \displaystyle   K_1 +  \frac{1}{2} \int_{x_{c_1}^2}^{x^2} \left( \mathcal{R}_{BB^{\mathsf{T}}}^{\langle -1 \rangle}(t-w_A^2) - g_{AA^{\mathsf{T}}+BB^{\mathsf{T}}}(t) \right) \mathrm{d}t   & \mbox{for } x_{c_1} \leq x \leq x_{c_2} \, ,\\
\\
\displaystyle  K_2 +  \frac{1}{2} \log \left(\dfrac{1}{w_A^2+w_B^2 -x^2} \right) -  \frac{1}{2} \int_{x_{c_2}^2}^{x^2}  g_{AA^{\mathsf{T}}+BB^{\mathsf{T}}}(t) \mathrm{d}t   & \mbox{for  } x_{c_2} \leq x \leq w_A + w_B \,.
    \end{array}
\right. &&
\end{align}

\vskip 0.3cm
\noindent \textit{Remark (Rate function for the sum of square ($q=1$) matrices):} In this case, we have:
\begin{align}
\label{eq:RateFunction_Rect.q1}
\Phi_C(x) =2\Psi_{\hat{A}+\hat{B}}(x) =  \left\{
    \begin{array}{lll}
  \displaystyle \int_{\mathrm{c}_+}^x    \bar{g}_{\hat{A}+\hat{B}}(t)  -  g_{\hat{A}+\hat{B}}(t) \mathrm{d}t   &   \mbox{for } \mathrm{c}_+ \leq x \leq x_{c_1} \, ,\\
\\
  \displaystyle  K_1 +   \int_{x_{c_1}}^x \left( \mathcal{R}_{\hat{B}}^{\langle -1 \rangle}(t-w_A) - g_{\hat{A}+\hat{B}}(t) \right) \mathrm{d}t   & \mbox{for } x_{c_1} \leq x \leq x_{c_2} \, ,\\
\\
\displaystyle   K_2 +  \log \left(\frac{1}{w_A+w_B -x} \right) -\int_{x_{c_2}}^x  g_{\hat{A}+\hat{B}}(t) \mathrm{d}t   & \mbox{for  } x_{c_2} \leq x \leq w_A + w_B \,.
    \end{array}
\right. &&
\end{align}

\vskip 0.3cm
\noindent \textit{Example (Free sum of fixed diagonal quarter-circle distribution):} Let's consider two diagonal square matrices $\tilde{\mathbf{A}}$ and $\tilde{\mathbf{B}}$ with LSVD given by Eq.\ \eqref{eq:LSVD_Ginibre} where without loss of generality we take the variances to be respectively equal to $\sigma_A =1$ and $\sigma_B \leq 1$ then we have that the rate function $\Phi_{\mathrm{qc}+\mathrm{qc}}$ associated to the large deviation of the top singular value of the sum $\tilde{\mathbf{A}}+ \mathbf{U}\tilde{\mathbf{B}} \mathbf{V}^{\mathsf{T}}$ is given by:
\begin{align}
    \Phi_{\mathrm{qc}+\mathrm{qc}}(x) &= 2 \Psi_{\mathrm{sc}+\mathrm{sc}}(x) \, , 
\end{align}
with $\Psi_{\mathrm{sc}+\mathrm{sc}}$ given by Eq.\ \eqref{eq:RateFunctionSC+SC}.

\section{Conclusion}
\label{sec:Conclusion}
In this paper, we have derived the right large deviation function for:
\begin{enumerate}
    \item the top eigenvalue of the sum of two \emph{arbitrary} symmetric matrices;
    \item the top eigenvalue of the product of two \emph{arbitrary} symmetric matrices;
    \item the top singular of the sum of two \emph{arbitrary} rectangular matrices;
\end{enumerate}
where by arbitrary we mean that we can take the matrices to be either taken from a rotationally invariant ensemble (resp. a bi-invariant for the case of rectangular random matrices) or to be a randomly rotated fixed diagonal matrix (resp. a fixed diagonal rectangular matrices). The results rely on a direct link with spherical spin models and are summarized in Sec.\ \ref{sec:RateFunction.sum} for the case of the sum of symmetric matrices, in Sec.\ \ref{sec:RateFunction.prod} for the case of the product of symmetric matrices and in Sec.\ \ref{sec:RateFunction.rect} for the case of the sum of rectangular matrices. In each case, we find that the rate function has up to three different regimes, and we give an interpretation of the behavior in each regime. Let us mention right away that this construction directly extends to complex matrices: for unitary invariant random matrices (resp. complex rectangular matrices), it is known that the derivatives of the quenched and annealed free energies are twice the ones of the real case and so does the rate functions. A natural question is to extend the above construction to tackle the 'left' large deviation, for which the speed of convergence of the large deviation is $N^2$. We leave this problem for future research.

\vskip 0.3cm
\noindent \textbf{Acknowledgements: } P.M.  would like to warmly thank Tristan Gautié for fruitful discussions at the early stage of the project. P.M.  would also like to thank Florent Benaych-Georges for discussions regarding the properties of the rectangular free convolution and Satya Majumdar for several discussions regarding Coulomb gas and more generally large deviation problems in Random Matrix Theory. Both authors would like to thank Laura Foini and Jorge Kurchan for sharing their work on annealed averages. 

\bibliographystyle{ieeetr}
\bibliography{biblio}

 \newpage

\appendix

\section{Properties of transforms of RMT and Free Probability}
\label{sec:Ap:prop}
\subsection{The Stieltjes transform and its inverse}
\label{sec:Ap:Stieltjes}
For a random matrix $\mathbf{A} \sim \mathbb{P}_{V}$ taken from an invariant ensemble with empirical spectral distribution $\mu_{\mathbf{A}}(\lambda)= \frac{1}{N}\sum_{i=1}^N \delta(\lambda - \lambda_i(\mathbf{A))} \to \mu_A$, we recall that its Stieltjes transform is given by:
\begin{align}
    \label{Ap:Stieltjes_FiniteN}
    g_{\mathbf{A}}(z)&:= \Tr (z - \mathbf{A})^{-1} = \int \frac{\mu_{\mathbf{A}}(\lambda)}{z-\lambda} \mathrm{d}\lambda \to g_A(z) := \int_{\mathrm{a}_-}^{\mathrm{a}_+} \frac{\mu_A(\lambda)}{z-\lambda} \mathrm{d}\lambda \, .&&
\end{align}
Conversely, one obtains the density $\mu_A(.)$ from the knowledge of the Stieltjes transform thanks to the so-called Sokhotski-Plemelj inversion formula:
\begin{align}
    \label{Ap:Plemelj}
    \mu_A(\lambda) &=  \frac{1}{\pi} \mathfrak{Im } \, g_A(\lambda - \mathrm{i} 0^+) \, .&&
\end{align}
It is clear from the expression \eqref{Ap:Stieltjes_FiniteN} of the Stieltjes transform that for real $z>\mathrm{a}_+$, $g_A(.)$ is a continuous decreasing function, with asymptotic behavior
\begin{align}
\label{Ap:As_Stieltjes}
    g_A(z) &\underset{z \to \infty}{\sim} \frac{1}{z} \, . &&
\end{align}
Since this function is decreasing its admits an inverse $g_A^{\langle -1 \rangle}$ defined on $\left(0, g_A(\mathrm{a}_+)\right)$ which is also decreasing. Now, it turns out that one can generally extend analytically this function for values outside  this interval, which we also denote by $g_A^{\langle -1 \rangle}$.  If one can do so, necessarily the inverse function cannot continue to be decreasing, otherwise we can invert it again to obtain a real function, which would be in contradiction with the inversion formula \eqref{Ap:Plemelj}. Note that this in turn gives us a way to find the top edge of the distribution:
\begin{itemize}
    \item if $g_A(\mathrm{a}_+) < \infty$, then $\mathrm{a}_+ = g_A^{\langle -1 \rangle}(y^*)$ with $y^*$ such that $(g_A^{\langle -1 \rangle})'(y^*)=0$ ; 
\item otherwise, $\mathrm{a}_+ = \lim_{x \to \infty} g_A^{\langle -1 \rangle}(x)$.
\end{itemize}

\begin{figure}
     \centering
         \includegraphics[width= 0.55\textwidth]{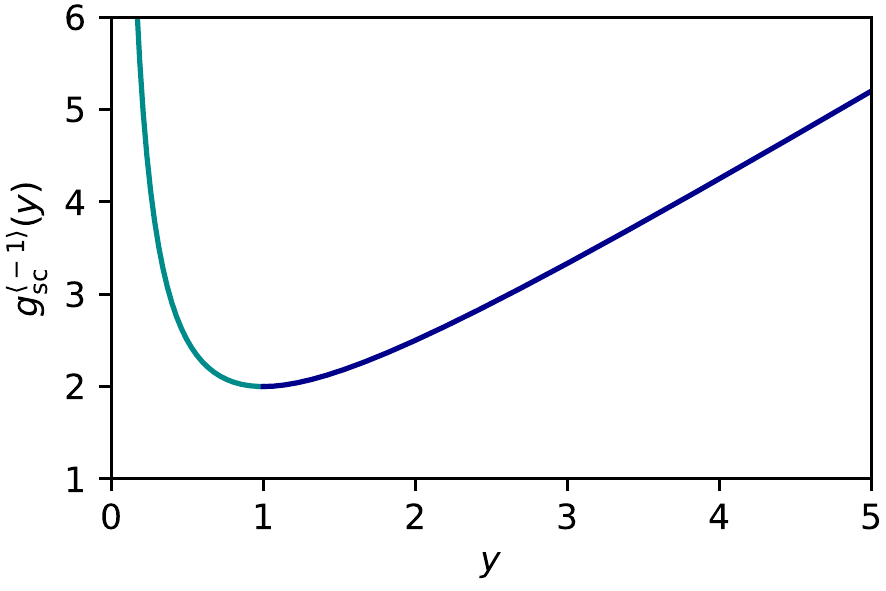}
    \caption{Inverse of the Stieltjes transform of the semi-circle distribution with $\sigma=1$. For $y \leq g_{\mathrm{sc}}(2\sigma) = 1$, this function (in cyan) in the inverse of the Stieltjes transform while for $y \geq 1$, this function (in blue) is the inverse of the second branch of the Stieltjes.  } 
\label{fig:Invg_sc}
\end{figure}

\noindent Now, in the case of invariant ensemble, the analytical continuation of the inverse of the Stieltjes admits a natural interpretation. Without loss of generality, let's consider the case $g_A(\mathrm{a}_+)<\infty$, since one can immediately see that there is no such analytical continuation of the inverse in the case $g_A(\mathrm{a}_+)= \infty$. For $V(.)$ an analytic potential on the real line, convex after the top edge $\mathrm{a}_+$ ,  and $z$ a complex variable outside the support of the distribution $\mu_A$, the corresponding Stieltjes transform $g_A \equiv g_A(z)$ is known to satisfy the following algebraic equation: 
\begin{align}
\label{Ap:eq:algebraic_eq_stieltjes}
    g_A^2 - V'(z) g_A + P(z) &= 0\,  ,&&
\end{align}
where the function $P$ is defined as:
\begin{align}
\label{Ap:eq:def:P_of_z}
    P(z) &:= \lim_{N \to \infty} \sum_{i=1}^N \frac{V'(z)-V'(\lambda_i)}{z-\lambda_i}  = \int_{\mathrm{a}_-}^{\mathrm{a}_+} \frac{V'(z) - V'(\lambda)}{z - \lambda} \mu_A(\lambda) \mathrm{d}\lambda  \,  .&&
\end{align}
Eq.\ \eqref{Ap:eq:algebraic_eq_stieltjes} is of second order in $g_A$ and therefore admits two solutions. Using the asymptotic  behavior of Eq.\ \eqref{Ap:As_Stieltjes} for the Stieltjes transform, one gets that the correct solution is given by the so-called Br{\'e}zin-Itzykson-Parisi-Zuber (BIPZ) formula \cite{Brezin1978}:
\begin{align}
\label{eq:BIPZ}
    g_A(z) &= \frac{V'(z)}{2} - \frac{\sqrt{{V'}^2(z)-4 P(z)}}{2} \, .&&
\end{align}
Interestingly, one can naturally look at the other non-physical solution of Eq.\ \eqref{Ap:eq:algebraic_eq_stieltjes} given by:
\begin{align}
\label{Ap:eq:def:2B_Stieltjes}
    \bar{g}_A(z) &= \frac{V'(z)}{2} + \frac{\sqrt{{V'}^2(z)-4 P(z)}}{2} \, , &&
\end{align}
which is known as the \emph{second branch of the Stieltjes transform}. For $z \geq \mathrm{a}_+$, this function starts at $\bar{g}_A(\mathrm{a}_+) = g_A(\mathrm{a}_+)$ and is then continuously increasing with asymptotic behavior given by:
\begin{align}
\label{Ap:eq:As_2B_Stieltjes}
    \bar{g}_A(z) \underset{z \to \infty}{\sim} V'(z) \, . &&
\end{align}
Note that this gives Eq.\ \eqref{eq:def:2ndBranchStieltjes} in the main text. If now look at the algebraic equation \eqref{Ap:eq:algebraic_eq_stieltjes} the other way by  fixing the value of $g_A =y$ for some $y$ in $(0,g_A(\mathrm{a}_+))$, the corresponding $z(y) \equiv z$ is by definition the inverse $g_A^{\langle -1 \rangle}(y)$. If now  the parameter $y$ is higher than $g_A(\mathrm{a}_+)$ (but lower than $\mathrm{r}_A := \lim_{z \to \infty} V'(z)$), Eq.\ \eqref{Ap:eq:algebraic_eq_stieltjes} is the implicit equation for  the analytical continuation of $g^{\langle -1 \rangle}$. Since this regime corresponds to the second branch of Stieltjes, we have a natural interpretation for the analytical continuation of $g_A^{\langle -1 \rangle}$ beyond the point $g_A(\mathrm{a}_+)$:  it is the inverse function of the second branch of the Stieltjes transform. Now, for a general smooth density $\mu_A$ (not necessarily arising from the eigenvalue distribution of an invariant ensemble), we define the second branch of its Stieltjes as the inverse  of the $g_A^{\langle -1 \rangle}$ beyond $g_A(\mathrm{a}_+)$. We have represented the function  $g_A^{\langle -1 \rangle}$ associated to semi-circle distribution for $\sigma=1$ in Fig.\ \ref{fig:Invg_sc}.

\noindent Similarly to the Stieltjes transform, one has the same behavior for the T-transform of Eq.\ \eqref{eq:def_tt}  and D-transform of Eq.\ \eqref{Ap:def:Dt}. In particular, the analytical continuation of the inverse of the T-transform $t_A^{\langle -1 \rangle}(.)$ (resp. D-transform $d_A^{\langle -1 \rangle}(.)$ ) beyond the point $t_A(\mathrm{a}_+)$  (resp. $d_A(\mathrm{a}_+)$) corresponds respectively to invert the second branch of the T-transform $\bar{t}_A(.)$ satisfying:
\begin{align}
\label{Ap:2branch_tt}
    \bar{t}_A(z) &= z \bar{g}_A(z) -1 \, , &&
\end{align}
and to invert the second branch of the D-transform satisfying:
\begin{align}
\label{Ap:2branch_Dt}
  \bar{d}_A(z) &= \sqrt{q z^2 \left( \bar{g}_{AA^{\mathsf{T}}}(z^2) \right)^2 + (1-q)  \bar{g}_{AA^{\mathsf{T}}}(z^2)} \, .  &&
\end{align}
\subsection{monotonic behavior of the linearizing transforms}
\label{sec:Ap:monot}
In this section, we show that the R-transform, S-transform and C-transform are monotonously increasing. 

\subsubsection{for the R-transform}
\label{sec:Ap:monot_Rt}
For any $y \in \left(0,\mathrm{r}_A \right)$, we recall that the R-transform is given by:
\begin{align}
    \label{Ap:eq:Rt}
    \mathcal{R}_A(y) &:= g_A^{\langle -1 \rangle}(y) - \frac{1}{y} \, ,&&
\end{align}
where $ g_A^{\langle -1 \rangle}(y)$ is the inverse of $ g_A$ on $(0,g_A(\mathrm{a}_+))$ and is the inverse of $\bar{g}_A$ on $(g_A(\mathrm{a}_+),\mathrm{r}_A)$, as described in the previous Sec.\ \ref{sec:Ap:Stieltjes}. 
\begin{itemize}
    \item For $y \in (g_A(\mathrm{a}_+),\mathrm{r}_A)$, one may note that the R-transform is clearly (strictly) monotonously increasing since it is the sum of (strictly) increasing function.
    \item For $y \in (0,g_A(\mathrm{a}_+))$, let's take a look at the derivative  of the R-transform:
\begin{align}
    \label{Ap:eq:d_Rt}
    \mathcal{R}_A'(y) &= \frac{1}{g_A' \left( g_A^{\langle -1 \rangle}(y) \right) } + \frac{1}{y^2} \, ,&&
\end{align}
but for any $z>\mathrm{a}_+$, we have:
\begin{align}
\label{Ap:eq:-gp<g2}
    - g_A' \left( z \right) &= \int \left( \frac{1}{z -\lambda} \right)^2 \mu_A(\lambda) \mathrm{d} \lambda \geq \left(\int \frac{\mu_A(\lambda)}{z -\lambda}  \mathrm{d} \lambda \right)^2 = g_A^2(z) \, . &&
\end{align}
If now one applies this inequality for $z = g_A^{\langle -1 \rangle}(y)$ with $y \in (0,g_A(\mathrm{a}_+))$, the RHS of Eq.\ \eqref{Ap:eq:-gp<g2} is simply given by $y^2$ since $g_A(g_A^{\langle -1 \rangle}(y) ) =y$ and so:
\begin{align}
    \label{Ap:eq:d_Rt>0}
    \mathcal{R}_A'(y) &\geq 0 \, . &&
\end{align}
such that the R-transform is indeed monotonously increasing. If $\mu_A$ is not a Dirac mass, then the above inequalities are strict, such that one can indeed take the inverse of the R-transform. 
\end{itemize}

\subsubsection{for the S-transform}
\label{sec:Ap:monot_St}
For any $y \in \left(0,\mathrm{s}_A \right)$, we recall that the (modified) S-transform is given by:
\begin{align}
    \label{Ap:eq:St}
    \tilde{\mathcal{S}}_A(y) &:= t_A^{\langle -1 \rangle}(y) \frac{y}{y+1} \, .&&
\end{align}
\begin{itemize}
    \item For $y \in \left( t_A(\mathrm{a}_+),\mathrm{s}_A \right)$, The (modified) S-transform is a continuous increasing function as the product of positive and increasing functions.
    \item For $y \in \left(0, t_A(\mathrm{a}_+) \right)$, the derivative of the S-transform is given by
\begin{align}
    \label{Ap:eq:d_St}
     \left( \tilde{\mathcal{S}}_A \right)'(y) &= \frac{1}{(y+1)^2 t_A' \left( t_A^{\langle -1 \rangle}(y) \right) } \left( t_A^{\langle -1 \rangle}(y) \, t_A' \left( t_A^{\langle -1 \rangle}(y) \right) + y(y+1)  \right) \, .&&
\end{align}
Since the derivative of the T-transform is given by 
\begin{align}
 \label{Ap:eq:d_Tt}
 t_A'(z)  &= - \int_{\mathrm{a}_-}^{\mathrm{a}_+} \frac{\lambda}{ \left(z-\lambda \right)^2 } \mu_A(\lambda) \mathrm{d}\lambda \leq 0 \, , &&
\end{align}
the term in front of the parenthesis is always non-positive. Next expressing the Stieltjes  transform in terms of the T-transform in Eq.\ \eqref{Ap:eq:-gp<g2}, thanks to the definition of the T-transform of Eq.\ \eqref{eq:def_tt}, we have the following inequality:
\begin{align}
 \label{Ap:eq:d_Tt<}
 - z t_A'(z)  \geq t_A(z) \left( t_A(z) +1 \right) \, , &&
\end{align}
which for $z=t_A^{\langle -1 \rangle}(y)$ gives from Eq.\ \eqref{Ap:eq:d_St} that the derivative of the S-transform is always positive: 
\begin{align}
 \label{Ap:eq:d_St>0}
  \left( \tilde{\mathcal{S}}_A \right)'(y)  \geq 0 \, . &&
\end{align}
\end{itemize}
As a consequence, the S-transform is always monotonously increasing. For a non-degenerate (that is different from a Dirac mass) distribution $\mu_A$, one can again check that the inequalities are strict such that the S-transform is strictly continuously increasing and hence admits a well-defined inverse. 
\subsubsection{for the rectangular C-transform}
\label{sec:monot_Ct}
We recall that the rectangular C-transform is given by:
\begin{align}
\label{Ap:def:RectRt.2}
    \tilde{\mathcal{C}}^{(q)}_A(y)& :=    \frac{U\left( y \, d_A^{\langle -1 \rangle}(y) \right)}{y} =: W(y, d_A^{\langle -1 \rangle}(y))\, , &&
\end{align}
with 
\begin{align}
\label{Ap:eq:def:W}
    W(y,z') &:= \frac{U(y \, z')}{y} \, . &&
\end{align}
and the  function $U$ is given by Eq.\ \eqref{Ap:def:U}. 
\begin{itemize}
    \item For $y \in \left( d_A(\mathrm{a}_+),\mathrm{c}_A := \lim_{w_A \to \infty} \bar{d}_A(w_A) \right)$ and $z'>\mathrm{a}_+$, one can check that the function $W(y,z')$ is increasing with each variable while the other one is fixed (that is, $ z' \mapsto W_{y}(z') := W(y,z')$ and $ y \mapsto W_{z'}(y) := W(y,z')$  are continuously increasing). Furthermore, for $y \in \left( d_A(\mathrm{a}_+),\mathrm{c}_A \right)  $, $d_A^{\langle -1 \rangle}(y)$ is also (continuously) increasing, thus the rectangular C-transform is continuously increasing as the composition of continuously increasing functions.
    \item For $y \in \left(0, d_A(\mathrm{a}_+) \right)$, the derivative of the rectangular C-transform is given by:
\begin{align}
    \label{Ap:eq:d_Ct}
     \left(  \tilde{\mathcal{C}}^{(q)}_A \right)'(y) &= \frac{(1+q)}{2q y^2} + \frac{\left[ \frac{4 \, q \, y \, d_A^{\langle -1 \rangle}(y)}{d_A'\left(  d_A^{\langle -1 \rangle}(y) \right)} - (1-q)^2 - 4q \left(   d_A^{\langle -1 \rangle}(y) \right)^2 \right]}{2qy^2 \sqrt{(1-q)^2+4q  \left(d_A^{\langle -1 \rangle}(y) \right)^2 }} \, .&&
\end{align}
Next for $z> \mathrm{a}_+ > 0$, we can rewrite Eq.\ \eqref{Ap:eq:-gp<g2} as: 
\begin{align}
    \label{Ap:eq:-gp<g2.z2}
     - \frac{\mathrm{d}}{\mathrm{d}z} \left( g_{AA^{\mathsf{T}}} (z^2) \right) &\geq 2 z g_{AA^{\mathsf{T}}}(z^2) \, . &&
\end{align}
From the definition of Eq.\ \eqref{Ap:def:Dt} of the D-transform, we can express $g_{AA^{\mathsf{T}}}$ in terms of the D-transform:
\begin{align}
    \label{Ap:eq:g_of_d}
    g_{AA^{\mathsf{T}}}(z^2) &= \frac{-1+q + \sqrt{(1-q)^2 + 4 q z^2 d_A(z)}}{2 q z^2} && \, .
\end{align}
Injecting this into the inequality of  Eq.\ \eqref{Ap:eq:-gp<g2.z2} and then setting $z=d_A^{\langle -1 \rangle}(y)$ leads hopefully after several simplifications to the positivity of the term in bracket of Eq.\ \eqref{Ap:eq:d_Ct}. Since all the other terms are positive, the rectangular C-transform is a (continuously) increasing function.

\end{itemize}

\subsection{Inequalities at the edge of the free convolutions}
\label{sec:Ap:ineq} 
The goal of this section is to prove the set of inequalities given by the Eqs. \eqref{eq:prop_freeconv} \eqref{eq:tc<ta^tb} \eqref{eq:dc<da^db}. The proof of this statement in the additive case is also given in Ref. \cite{GuionnetMaida20}. The idea is based on \emph{subordination relations}: 
\begin{itemize}
    \item For the case of the sum of symmetric matrices of Sec.\ \ref{sec:LDP_sum}, let's assume without any loss of generality $g_A(\mathrm{a}_+) \leq g_B(\mathrm{b}_+)$. Using the definition of Eq.\ \eqref{eq:def_Rt} of the R-transform, we can write the linearizing property of Eq.\ \eqref{eq:prop_Rt} as an implicit equation for $g_C$. For any $z\geq \mathrm{c}_+$, we have:
    \begin{align}
    \label{Ap:eq:sub_sum}
        g_C(z) &= g_A \left( z - R_B \left( g_C(z) \right)  \right) \, . &&
    \end{align}
    Since $z \geq \mathrm{c}_+$, $g_C(z) \in \mathbb{R}$ and so for the RHS of Eq.\ \eqref{Ap:eq:sub_sum} to be real, we must have  $z - R_B \left( g_C(z) \right) \geq \mathrm{a}_+$. By decreasing property of $g_A$, this implies:
      \begin{align}
    \label{Ap:eq:sub_sum.2}
        g_C(z) &\leq  g_A \left( \mathrm{a}_+  \right) \, , &&
    \end{align}
    and  if we now set $z = \mathrm{c}_+  + \epsilon$ and take $\epsilon \to 0^+$, this give the desired property of Eq.\ \eqref{eq:prop_freeconv}.
    \item Similarly, in the case of the product of positive semi-definite matrix of Sec.\ \ref{sec:LDP_prod}, we assume without loss of generality $t_A(\mathrm{a}_+) <t_B(\mathrm{b}_+)$  and the subordination relation reads for $z \geq \mathrm{c}_+$:
    \begin{align}
    \label{Ap:eq:sub_prod}
        t_C(z) &= t_A \left( \frac{z}{\tilde{\mathcal{S}}_B \left( t_C(z) \right)} \right) \leq t_A(\mathrm{a}_+) \, , 
    \end{align}
    where the inequality in Eq.\ \eqref{Ap:eq:sub_prod} is due to the same reasoning as in the additive case and taking again  the limit $z = \mathrm{c}_+ + \epsilon$ with $\epsilon \to 0^+$ gives Eq.\ \eqref{eq:tc<ta^tb}. 
    \item for the cases of the sum of rectangular matrices, we have the following subordination relation for $z \geq \mathrm{c}_+$: 
    \begin{align}
        \label{eq:sub_rect}
        d_C(z) &= d_A \left( \frac{U^{\langle -1 \rangle} \left( U(d_C(z) z) -\tilde{\mathcal{C}}_B^{(q)}(d_C(z)) d_C(z) \right)}{d_C(z)}\right)  \, , &&
    \end{align}
    which gives Eq.\ \eqref{eq:dc<da^db} for similar reason.
    
\end{itemize}

\section{Asymptotic of the quenched and annealed free energy}
\label{sec:Ap:As_AFE_QFE}
The goal of this section is to describe the main steps to get the asymptotic behavior of the quenched and free energies of each model. For the annealed free energy, we primarily insist on the additive symmetric case (SSK model)  as the way to get the annealed free energy for the LSSK model and BSSK model is similar.
\subsection{For the SSK model (additive spherical integral)}
In this section, we focus on the SSK model of Sec.\ \ref{sec:LDP_sum}. 

\subsubsection{Complex integral representation and quenched free energy}
\label{sec:Ap:As_QFE_SSK}
The starting point of the computation of the free energies is a complex integral representation of the partition function $\mathcal{Z}_{\mathbf{M}}(\theta)$ of Eq.\ \eqref{eq:PartitionF_SSK}.  If we denote again by $\vect{\lambda} =(\lambda_1, \dots, \lambda_N)$, the eigenvalues of a symmetric matrix $\mathbf{M}$, by removing the constraint over the sphere by introducing a Lagrange multiplier $z$ and using a Gaussian integration, we can write the partition function as:
\begin{align}
\label{Ap:PartitionF_SSK_int}
  \mathcal{Z}_{\mathbf{M}}(\theta) &= \frac{1}{2\pi \mathrm{i}} \int_{\mathscr{C}} \mathrm{e}^{\frac{N}{2} G_N(\vect{\lambda},z,\theta)} \mathrm{d}z \, , &&
\end{align}
where $\mathscr{C}$ is a vertical line in the complex plane that goes to the right of all the eigenvalues and
\begin{align}
\label{Ap:Arg_Exp_SSK}
   G_N(\vect{\lambda},z,\theta) &:= z \theta  -1 - \log \theta  - \frac{1}{N} \sum_{k=1}^N
\log(z -\lambda_i) + O\left(\frac{1}{N}\right) \, . &&
\end{align}
The integral representation is true for any symmetric matrix. If we now set $\mathbf{M}$ to be equal to the matrix $\mathbf{C}$ conditioned to have its top eigenvalue at the position $x$, $\mathbf{M} =\mathbf{C}| \{ \lambda_1  = x \}$, and take the large $N$ limit, we have the following saddle-point approximation:
\begin{align}
\label{PartitionF_SSK_int.SP}
      \mathcal{Z}_{\mathbf{C} | \{ \lambda_1  = x \}}(\theta) &\approx \frac{1}{K}  \mathrm{e}^{\frac{N}{2} G_N(x,\lambda_2,\dots, \lambda_N,z^*,\theta)} \, , &&
\end{align}
with $K$ a constant and $z^*$ the solution of the saddle point equation given by setting the partial derivative with respect to $z$ in the expression of Eq.\ \eqref{Ap:Arg_Exp_SSK} to be equal to zero. A careful analysis shows that this saddle point $z^*$ may exhibit a possible saturation at $x$, depending on the value of the parameter $\theta$, as shown in Ref. \cite{GuionnetMaida05} and this gives the behavior of the partial derivatives of Eq.\ \eqref{eq:As_QFE_sum.1} and Eq.\ \eqref{eq:As_QFE_sum.2}.

\subsubsection{Computation of the annealed free energy}
\label{sec:Calcul_AFE_SSK}
We now turn to the limiting behavior of the annealed free energy of the SSK model given by Eq.\ \eqref{eq:As_AFE.sumA} for a random matrix $\mathbf{A} \sim \mathbb{P}_{V,w_A}$, based on ideas developed in Ref. \cite{foini2021annealed}. Setting $\mathbf{M}= \mathbf{A}$ in Eq.\ \eqref{Ap:PartitionF_SSK_int} and taking the average over the law of Eq.\ \eqref{eq:JointDensityEigvals}  for the eigenvalues of $\mathbf{A}$, we have the following integral representation:
\begin{align}\label{eq:ssk_integral}
   \mathbb{E}_{\mathbf{A}}\mathcal{Z}_{\mathbf{A}}(\theta) &=  \frac{1}{2\pi \mathrm{i}}  \int_{ \mathbb{R}^N \times \mathscr{C}} \mathrm{e}^{\frac{N}{2}  H_N(\vect{\lambda},z,\theta)  } \mathbb{I}_{ \{ \lambda_i \leq w_A \} }  \mathrm{d}\vect{a} \mathrm{d}z  \, , &&
\end{align}
with: 
\begin{align}
   H_{N}(\vect{\lambda},z,\theta)  &:=  - \sum_{i=1}^N V(\lambda_i)  + \frac{1}{N} \sum_{i,j | j \neq i} \log | \lambda_i - \lambda_j| - \frac{1}{N} \sum_{i=1}^N \log(z-\lambda_i) + z \theta  -1 -\log \theta +O \left( \frac{1}{N} \right) \, . &&
\end{align}
This can be understood as the Hamiltonian of a system of $N+1$ particles\footnote{The variable $z$ superficially looks like another eigenvalue repelled by all the other ones and with its own linear potential. But closer inspection reveals that the force between $z$ and the $\lambda_i$ is actually attractive. What is even stranger is that the equilibrium position of $z$ is a local minimum of the probability. The reason for this is that in the integral of Eq.\ \eqref{eq:ssk_integral}, $z$ is integrated on a vertical line in the complex plane, so the second derivative in that direction should be positive for the integral to converge.}.  In the large $N$ limit, we argue that the integral is dominated by the most probable configuration given as the solution of  the set of saddle-point equations:
\begin{align}
\label{Ap:SP_AFE.sum}
\left\{
    \begin{array}{ll}
\partial_{\lambda_i^*} H_{N}(\vect{\lambda}^*,z^*,\theta)=0& \mbox{for } i = 1, \dots, N \, ,\\
\\
\partial_{z^*}  H_{N}(\vect{\lambda}^*,z^*,\theta)=0& \, .
    \end{array}
\right. &&
\end{align}
that is:
\begin{align}
\label{Ap:SP_AFE.sum.2}
\left\{
    \begin{array}{ll}
V'(\lambda_i^*) =\frac{2}{N} \sum_{j=1|j \neq i}^N \frac{1}{\lambda_i^* -\lambda_j^*} + \frac{1}{N} \frac{1}{z^* -\lambda_i^*} & \mbox{for } i = 1, \dots, N  \, ,\\
\\
\theta = \frac{1}{N} \sum_{i=1}^N \frac{1}{z^*-\lambda_i^*} & \, .
    \end{array}
\right. &&
\end{align}
These two equations have to be understood with the additional constraints:
\begin{align}
\label{Ap:SP_AFE.sum.constraint}
\left\{
    \begin{array}{ll}
 \lambda_1^*  \leq z^* &  \, ,\\
\\
 \lambda_N^* \leq \dots \leq \lambda_1^* \leq w_A & \, .
    \end{array}
\right. &&
\end{align}
We need to distinguish three different cases:
\begin{enumerate}
    \item For $\theta \leq g_A(\mathrm{a}_{+})$,  the bottom line of Eq.\ \eqref{Ap:SP_AFE.sum.2} can be satisfied with the $\lambda_i^*$'s in their classical positions. By direct inversion we find:
    \begin{align}
    \label{Ap:sol_SP.1}
        z^*(\theta) &= g_A^{\langle -1 \rangle}(\theta) \, , &&
    \end{align}
    and by self-averaging property, this gives the same result as in the quenched case. 
    \item  For $ g_A(\mathrm{a}_{+}) \leq \theta \leq \bar{g}_A(w_A) $, for the RHS of the bottom line of Eq.\ \eqref{Ap:SP_AFE.sum.2} to be equal to $\theta$, one has to have the distance between $z^*$ and $\lambda_1^*$ be of order $O\left( \frac{1}{N} \right)$ so Eq.\  \eqref{Ap:SP_AFE.sum.2} (bottom)  has to be understood as: 
    \begin{align}
    \label{Ap:eq_for_sol_SP.2}
    \theta &\simeq g_A(z^*) + \frac{1}{N} \frac{1}{z^*-\lambda_1^*} + O\left(  \frac{1}{N} \right) \, .&&
    \end{align}
    Next, since $V(.)$ is analytic, we can approximate the potential and interaction term in Eq.\ \eqref{Ap:SP_AFE.sum.2} for $i=1$ by: 
    \begin{align}
    \label{Ap:Taylor.1}
    V'(\lambda_1^*) &= V'(z^*) + O\left(\frac{1}{N} \right) \, ,&& \\
    \frac1N \sum_{j=2}^N \frac{1}{\lambda_1^* -\lambda_j^*}&\simeq g_A(z^*) + O \left( \frac{1}{N} \right)\, . &&
    \end{align}
    injecting this in the top line of Eq.\ \eqref{Ap:SP_AFE.sum.2} for $i=1$ one gets: 
    \begin{align}
    \label{Ap:Taylor.2}
    V'(z^*) &= 2 g_A(z^*) + \frac1N \frac{1}{z^*-\lambda_1^*} + O \left(\frac{1}{N} \right) \, . &&
    \end{align}
    Now making the difference of Eq.\ \eqref{Ap:eq_for_sol_SP.2} and Eq.\ \eqref{Ap:Taylor.2} to eliminate the term $\frac{1}{z^*-\lambda_1^*}$ and neglecting term of order $O \left( \frac{1}{N} \right)$ one obtain a simple self consistent equation for the unknown $z^*$: 
    \begin{align}
    \label{eq:sol_SP.2.1}
        V'(z^*) - g_A(z^*) &= \theta \, , &&
    \end{align}
    which using Eq.\ \eqref{eq:def:2ndBranchStieltjes} reads:
    \begin{align}
        \label{eq:sol_SP.2.2}
        \bar{g}_A(z^*) &= \theta \, . &&
    \end{align}
    where $\bar{g}_A(.)$ is the second branch of the Stieltjes transform as described in App.\ \ref{sec:Ap:Stieltjes}. Following the properties of the second branch of the Stieltjes transform, inverting Eq.\ \eqref{eq:sol_SP.2.2} yields:
    \begin{align}
        \label{eq:sol_SP}
        z^*(\theta) &= g_A^{\langle -1 \rangle}(\theta) \, . &&
    \end{align}
    where $g_A^{\langle -1 \rangle}(.)$ is here the analytical continuation beyond the point $g_A(\mathrm{a}_+)$ of the inverse of the Stieltjes. This explains the result in the main text.
    \item for $\theta \geq \bar{g}_A(w_A)$, the position of the top eigenvalue becomes fixed at the wall and since the distance between $z^*$ and this top of eigenvalue is infinitely small in the large $N$ limit,  we have again a saturation, but now at the position $w_A$ instead of $x$ in the quenched case:
    \begin{align}
        \label{Ap:eq:sol_SP.3}
        z^*(\theta) &= w_A \, . &&
    \end{align}
    \end{enumerate}
Once we have the expression for the saddle points $z^*$, we get the annealed free energy for the SSK model by taking the logarithm and this gives the expression of Eq.\ \eqref{eq:As_AFE.sumA} in the main text.

\vskip 0.3cm
\noindent \textit{Remark (GOE matrices and Gaussian integration):} For $\mathbf{A}$ a GOE matrix with variance $\sigma^2$, the annealed free energy can be directly obtained by Gaussian integration. By rotationally invariance, the average of the partition function is simply the moment generating function of one of the diagonal element, say $A_{11}$.  Since this element is a Gaussian random variable with variance $2\sigma^2/N$, we have: 
\begin{align}
    \mathbb{E}_{\mathbf{A}} \left[ \mathcal{Z}_{\mathbf{A}}(\theta) \right] &= \int  \frac{\mathrm{e}^{\frac{N \theta}{2} A_{11} - \frac{N}{4 \sigma^2} A_{11}^2 }}{\sqrt{4 \pi \sigma^2 /N}}   \mathrm{d}A_{11} = \mathrm{e}^{\frac{N}{2} \frac{\sigma^2 \theta^2}{2} }  \, . &&
\end{align}
Using the expression of Eq.\ \eqref{eq:R_SC} for the R-transform, this gives indeed Eq.\ \eqref{eq:As_AFE.sumA} for the annealed free energy. 

\vskip 0.3cm
\noindent \textit{Remark (Wishart matrices and Gaussian integration):} For $\mathbf{A}$ a Wishart matrix of shape parameter $q = N/M$,  the annealed free energy can be also directly obtained by Gaussian integration. Indeed, we have $\mathbf{A} \overset{\text{in law}}{=} \frac{1}{M} \sum_{m=1}^M \vect{x}_m^{\mathsf{T}} \vect{x}_m$, where the $\{\vect{x}_m \}_{m = 1,\dots, M}$ are $M$ independent $N$-dimensional standard Gaussian vectors. Since $\mathbf{A}$ is rotationally invariant, we can remove the integral over the sphere and fix $\vect{\sigma} = \vect{e}_1 = (1,0,\dots,0)$ without loss of generality. This gives: 
\begin{align}
    \mathbb{E}_{\mathbf{A}} \left[ \mathcal{Z}_{\mathbf{A}}(\theta) \right] &= \int  \mathrm{e}^{\frac{N \theta}{2 M} \vect{e}_1^{\mathsf{T}} \left( \frac{1}{M} \sum_{m=1}^M \vect{x}_m^{\mathsf{T}} \vect{x}_m  \right)\vect{e}_1   } \prod_{i=1}^M  \frac{\mathrm{e}^{ - \frac{|\vect{x}_m|^2}{2} } }{(2 \pi)^{-\frac{M}{2}} } \mathrm{d} \vect{x}_m  = \left( \int \frac{\mathrm{e}^{ - \frac{1}{2}  \vect{x}^{\mathsf{T}} ( \mathbf{I} - q\theta  \vect{e}_1 \vect{e}_1^{\mathsf{T}} ) \vect{x}     }}{(2 \pi)^{-\frac{1}{2}}}  \mathrm{d} \vect{x} \right)^{M} \, . &&
\end{align}
Now for $\theta \geq 1/q$, the integral inside the bracket is diverging and hence we get $F_A'(\theta) = \infty$ in this case. Otherwise, we can do the Gaussian integration, and we have: 
\begin{align}
   \frac{1}{N} \log  \mathbb{E}_{\mathbf{A}} \left[ \mathcal{Z}_{\mathbf{A}}(\theta) \right] &= - \frac{1}{2q} \log( 1- q \theta)  && \left( \mbox{for } \theta < \frac{1}{q} \right) \, .
\end{align}
Using the expression of Eq.\ \eqref{eq:R_MP} for the R-transform of a Wishart matrix, this gives indeed Eq.\ \eqref{eq:As_AFE.sumA} for the annealed free energy. A similar computation can be one for generalized Wishart matrices.

\subsection{For the LSSK model (multiplicative spherical integral)}
\label{sec:Ap:As_AFE_LSSK}
The behavior in the multiplicative case for which the partition function  $\mathcal{Z}_{\mathbf{M}}(\theta)$ of the LSSK given by Eq.\ \eqref{eq:PartitionF_LSSK} is very similar to the additive case, so we only outline the main steps to compute the free energies. One can show see Ref. \cite{MergnyPotters20} that this spherical integral admit the following complex integral representation:
\begin{align}
\label{Ap:PartitionF_LSSK_int}
 \mathcal{Z}_{\mathbf{M}}(\theta) &= \frac{1}{2\pi \mathrm{i}} \int_{\mathscr{C}} \mathrm{e}^{\frac{N}{2} G_N(\vect{\lambda},z,\theta)} \mathrm{d}z \, , &&
\end{align}
with 
\begin{align}
\label{Ap:argExp_LSSK}
   G_N(\vect{\lambda},z,\theta) &=  z \theta - \frac{1}{N} \sum_{i=1}^N \log (1 - \lambda_i \mathrm{e}^{-z} )  + K +  \theta \log \theta - (1+ \theta) \log( 1 + \theta)  + O\left(\frac{1}{N}\right) \,, &&
\end{align}
with $K$ a constant and $\mathscr{C}$ an integral in the complex plane crossing the real axis at a point $\gamma$ such that $\mathrm{e}^{\gamma} $ is on the right of all eigenvalues. Setting $\mathbf{M} = \mathbf{C} | \{ \lambda_1 = x \}$ and taking the large $N$ limit, we can  use once again the  saddle-point approximation, and this gives Eq.\ \eqref{eq:As_QFE_prod.1} and  Eq.\ \eqref{eq:As_QFE_prod.2} for the quenched free energy.
\vskip 0.3cm
\noindent For the annealed free energy, one can  repeat the exact same method as in Sec.\ \ref{sec:Calcul_AFE_SSK} for the additive case by looking at the saddle points of Eq.\ \eqref{Ap:PartitionF_LSSK_int} averaged over the law of the $\lambda_i$'s. One has to consider the three different cases $\theta \leq t_A(\mathrm{a}_+)$, $t_A(\mathrm{a}_+) \leq \theta \leq \bar{t}_A(w_A)$ and $\theta \geq \bar{t}_A(w_A)$. The first case is the same solution as the quenched free energy, the second is the analytical continuation of the first case and the third is the saturated case giving Eq.\ \eqref{eq:As_AFE_prod}.

\subsection{For the BSSK model (rectangular spherical integral)}
\label{sec:Ap:Rect}
In this section, we consider the asymptotic behavior of the rectangular spherical integral $\mathcal{Z}_{\mathbf{M}}(\theta)$ of Sec.\ \ref{sec:LDP_Rect}  defined by Eq.\ \eqref{eq:PartitionF_BSSK.int}. In particular, we derive in detail the computation of the quenched free energy.

\subsubsection{complex integral representation}
\label{sec:Ap:As_CIR.rect}
Removing the constraints on the spheres by introducing Lagrange multipliers $z_1, z_2$, we have: 
\begin{align}
\label{Ap:PartitionF_BSSK_int}
 \mathcal{Z}_{\mathbf{M}}(\theta) &= \frac{\Gamma \left( \frac{N}{2} \right)}{\pi^{\frac{N}{2}}} \frac{\Gamma \left( \frac{M}{2} \right)}{\pi^{\frac{M}{2}}} \left(\frac{1}{4 \pi \mathrm{i}}  \right)^2  \int_{\mathscr{C}_1,\mathscr{C}_2} \mathrm{d}z_1 \mathrm{d}z_2  \int_{\mathbb{R}^{N+M}} \mathrm{d}\vect{y} \, \mathrm{e}^{\frac{z_1}{2} + \frac{z_2}{2} - \frac{1}{2} \vect{y}^{\mathsf{T}} \mat{Q}\vect{y}} \, , &&
\end{align}
where the matrix $\mathbf{Q} \equiv \mathbf{Q}(z_1, z_2,\theta, \mathbf{s})$  is given by:
$$
\label{Ap:matM}
\mat{Q} =  \begin{pmatrix}
z_1 \mat{I}_{N} & - \sqrt{N M} \theta \Diag(\vect{s})  & \mat{0}_{M-N,N}\\
- \sqrt{N M} \theta \Diag(\vect{s})  & z_2 \mat{I}_{N}  & \mat{0}_{M-N,N}\\
\mat{0}_{M-N,N} & \mat{0}_{M-N,N}  & z_2 \mat{I}_{M-N} 
\end{pmatrix} \, .
$$
 with $\vect{s} \equiv (s_1(\mathbf{M}), \dots, s_{N}(\mathbf{M}))$ the vector of singular values.
By Gaussian integration, we have: 
\begin{align}
\label{Ap:PartitionF_BSSK_int.2}
 \mathcal{Z}_{\mathbf{M}}(\theta) &= \Gamma \left( \frac{N}{2} \right)\Gamma \left( \frac{M}{2} \right) \left(\frac{1}{4 \pi \mathrm{i}}  \right)^2  \int_{\mathscr{C}_1,\mathscr{C}_2} \, \mathrm{d}z_1 \mathrm{d}z_2 \,   \mathrm{e}^{z_1/2 + z_2/2} (\mathrm{det} \mathbf{M})^{-\frac{1}{2}} \, , &&
\end{align}
so we need to compute the determinant of this matrix $\mathbf{Q}$. Expanding twice along the right-bottom block, we have:  
\begin{align}
\det \mat{Q} &=  \det(z_2 \mat{I}_{M-N})  \det
\begin{pmatrix}
z_1 \mat{I}_{N} & - \sqrt{N M} \theta \Diag(\vect{s})  &\\
-  \sqrt{N M} \theta \Diag(\vect{s})  & z_2 \mat{I}_{N}  & \\ \end{pmatrix} \label{Ap:detM.1} \, , &&\\
\det \mat{Q} &=z_2^{M-N} \det(z_2 \mat{I}_{N}) \det \left(z_1 \mat{I}_{N}  - \left( \sqrt{N M} \theta \right)^2 \frac{1}{z_2} \Diag(\vect{s}^2) \right) \label{Ap:detM.2}\, , &&\\
\det \mat{Q} &=z_2^{M-N} \prod_{i=1}^{N} \left( z_1 z_2 - N M \theta^2  s_i^2 \right) \label{Ap:detM.3} \, . &&
\end{align}
which gives: 
\begin{align}
\label{Ap:PartitionF_BSSK_int.3}
 \mathcal{Z}_{\mathbf{M}}(\theta) &= \Gamma \left( \frac{N}{2} \right) \Gamma \left( \frac{M}{2} \right) \left(\frac{1}{4 \pi \mathrm{i}}  \right)^2 \int_{\mathscr{C}_1,\mathscr{C}_2} \mathrm{d}z_1 \mathrm{d}z_2 \,  z_2^{\frac{N -M}{2} } \mathrm{e}^{  \frac{z_1}{2} + \frac{z_2}{2}  - \frac{1}{2} \sum_{i=1}^{N} \log \left( z_1 z_2 - N M \theta^2 s_i^2 \right)} \,.
\end{align}
Let's do the change of variable $(z_1 \to N \theta z_1, \mathrm{d}z_1 \to N \theta \mathrm{d} z_1)$ , $(z_2 \to M \theta z_2 ,\mathrm{d} z_2  \to M \theta \mathrm{d} z_2 ) $ we have: 
\begin{align}
\label{eq:PartitionF_BSSK_int.3.1}
 \mathcal{Z}_{\mathbf{M}}(\theta) &=  \frac{ \Gamma\left( \frac{N}{2} \right) \Gamma\left( \frac{M}{2} \right)}{\left(  N \theta \right)^{ \frac{N}{2} - 1} \left( M \theta \right)^{ \frac{M}{2} - 1}}\left(\frac{1}{4 \pi \mathrm{i}}  \right)^2 \int_{\mathscr{C}_1, \mathscr{C}_2} \mathrm{d}z_1 \mathrm{d}z_2  \mathrm{e}^{ \frac{N \theta}{2} z_1 + \frac{M \theta}{2}  z_2+ \frac{N-M}{2} \log( z_2 ) - \frac{1}{2} \sum_{i=1}^{N} \log \left( z_1 z_2 -  s_i^2 \right)} \, . &&
\end{align}
If we now consider the large $N$ limit with the double scaling limit of Eq.\ \eqref{Ap:double_scaling_limit} and use Stirling approximation to get the behavior of the multiplicative constant, we have the following (double) integral representation: 
\begin{align}
\label{Ap:PartitionF_BSSK_int.4}
 \mathcal{Z}_{\mathbf{M}}(\theta) &= \left(\frac{1}{4 \pi \mathrm{i}}  \right)^2 \int_{\mathscr{C}_1, \mathscr{C}_2} \mathrm{d}z_1 \mathrm{d}z_2  \mathrm{e}^{ \frac{N}{2} G(z_1,z_2,\theta) } \, , &&
\end{align}
with: 
\begin{align}
\label{Ap:ArgExp_BSSK}
G(z_1,z_2,\theta)  &:= \theta z_1 + \frac{\theta}{q} z_2  - \frac{1-q}{q}\log z_2 - \frac{1}{N} \sum_{i=1}^{N} \log \left( z_1 z_2 -  s_i^2 \right) - \frac{1+ q}{q} (1+\log \theta)  + O \left( \frac{1}{N}\right) \, . &&
\end{align}
\subsubsection{saddle-point computations}
\label{sec:Ap:SP.rect}
Let's consider the case where in the large $N$ limit the singular value density of $\mathbf{M}$ is given by the  $\rho_C$, the rectangular free convolution of $\rho_A$ and $\rho_B$, but its top singular value is fixed at the position $x$. In this large $N$ limit, the  complex integral of Eq.\ \eqref{Ap:PartitionF_BSSK_int.4} is dominated by the saddle points $ (z_1^*, z_2^*) \equiv (z_1^*(\theta), z_2^*(\theta))$ solutions of the zero-gradient equations:

\begin{align}
 \label{Ap:Rect:gradSP}
\left\{
    \begin{array}{ll}
\partial_{z_1^*} G(z_1^*,z_2^*,\theta) &= 0  \, ,\\
\\
\partial_{z_2^*} G(z_1^*,z_2^*,\theta) &= 0  \, .
    \end{array}
\right. &&
\end{align}
that is, solution of:
\begin{align}
\label{eq:rec_SP1}
\theta -\frac{z_2^*}{N} \sum_{i=1}^N \left(z_1^* z_2^* - s_i^2 \right)^{-1} &=0  \, ,&&  
\end{align}
\begin{align}
\label{eq:rec_SP2}
\theta - \left( 1 - q \right) \frac{1}{z_2^*} -  \frac{q z_1^*}{N} \sum_{i=1}^N \left(z_1^* z_2^* -s_i^2\right)^{-1}   &=0 \, . &&
\end{align}
The equations Eq.\ \eqref{eq:rec_SP1}  and Eq.\ \eqref{eq:rec_SP2}  are coupled but can be easily taken care of by noting that: 
\begin{enumerate}
    \item if ones multiply Eq.\ \eqref{eq:rec_SP1} by $q z_1^*$ and Eq.\ \eqref{eq:rec_SP2} by $z_2^*$ and then subtract the two, one gets
\begin{align}
\label{Ap:eqSP.1}
\theta(q z_1^* -z_2^*)&= \left( q -1 \right) \, , &&
\end{align}
that is 
\begin{align}
\label{Ap:eqSP.1.1}
    z_2^* &= q z_1^* + \frac{1-q}{\theta} \, , &&
\end{align}
    \item if we put the variable $\theta$ on the RHS in both Eq.\ \eqref{eq:rec_SP1} and Eq.\ \eqref{eq:rec_SP2}  and multiply the two equations, we get:
     \begin{align}
     \label{Ap:eqSP.2}
     \theta^2 &= q z_1^* z_2^* \left( \frac{1}{N} \sum_{i=1}^{N} \frac{1}{z_1^* z_2^* - s_i^2} \right)^2 + (1-q) \sum_{i=1}^N \frac{1}{z_1^* z_2^* - s_i^2} \, . &&
 \end{align}
 that is following the expression \eqref{Ap:def:Dt} of the D-transform:
\begin{align}
\label{Ap:eqSP.2.1}
      \theta &= d_{\mathbf{M}}(\sqrt{z_1^* z_2^*}) \, . &&
 \end{align}
\end{enumerate}
Eq.\ \eqref{Ap:eqSP.1.1} and Eq.\ \eqref{Ap:eqSP.2.1}  allow one to get the behavior of the spherical integral but as in the additive and multiplicative case, we need to be careful before inverting Eq.\ \eqref{Ap:eqSP.2.1} and we have to consider two cases: 

\vskip 0.3 cm 
\noindent \textit{the case $\theta \leq d_C(x)$:} In this case, we can directly invert Eq.\ \eqref{Ap:eqSP.2.1} and we get: 
\begin{align}
\label{Ap:eqSP2.2}
     z_1^* z_2^* &= \left[ d_C^{\langle -1 \rangle}(\theta) \right]^2  \mbox{ for } \theta \leq d_A(x) \, . &&
 \end{align}
Injecting Eq.\ \eqref{Ap:eqSP.1.1}   in Eq.\ \eqref{Ap:eqSP2.2}, we get the following quadratic equation for $z_1^*$:
\begin{align}
\label{Ap:quadra_eq}
    q (z_1^*)^2 + \frac{1- q}{\theta} z_1^* -  \left[  d_C^{\langle -1 \rangle}(\theta) \right]^2 &=0 \, , &&
\end{align}
whose (correct) solution is given by:
\begin{align}
\label{Ap:solSP.1}
z_1^*(\theta) &= \frac{ -(1 - q) + \sqrt{(1- q)^2 + 4q \theta^2  \left[ d_C^{\langle -1 \rangle}(\theta)  \right]^2 } }{2q \theta} \, . &&
\end{align}
We have all the tools to compute the free energy in this regime. Let's first look at the partial derivative with respect to the variable $\theta$. We have:
\begin{align}
\label{Ap:partial_QFE_BSSK.1}
    \partial_{\theta} J_C(x,\theta) &= \frac{1}{2} \frac{\mathrm{d}}{\mathrm{d}\theta} G(z_1^*(\theta),z_2^*(\theta), \theta) \, , &&
\end{align}
but since $z_1^*$, $z_1^*$ are solutions of the zero-gradient equations \eqref{Ap:Rect:gradSP}, we have: 
\begin{align}
\label{Ap:partial_QFE_BSSK.2}
    \partial_{\theta} J_C(x,\theta) &=  \frac{1}{2} \partial_{\theta} G(z_1^*(\theta),z_2^*(\theta), \theta) \, , &&
\end{align}
which using Eq.\ \eqref{Ap:ArgExp_BSSK} gives:
\begin{align}
\label{Ap:partial_QFE_BSSK.3}
    \partial_{\theta} J_C(x,\theta) &=  \frac{1}{2} \left[ z_1^*(\theta) + \frac{
    z_2^*(\theta)}{q} -  \frac{1+ q}{q} \frac{1}{\theta} \right] \, . &&
\end{align}
Next using Eq.\ \eqref{Ap:eqSP.1.1}, we can express $z_2^*$ as a function of $z_1^*$:
\begin{align}
\label{Ap:partial_QFE_BSSK.4}
    \partial_{\theta} J_C(x,\theta) &=  z_1^*(\theta)  -  \frac{1}{\theta} \, . &&
\end{align}
so that using Eq.\ \eqref{Ap:solSP.1}, we have:
\begin{align}
\label{Ap:partial_QFE_BSSK.5}
\partial_{\theta} J_C(x,\theta) &= \frac{-1 - q + \sqrt{(1-q)^2 + 4 q \theta^2 \left[ d_C^{\langle -1 \rangle}(\theta)  \right]^2 }}{2 q \theta}  \quad    &&(\mbox{for } \theta \leq d_C(x))\, ,
\end{align}
which is by definition the rectangular C-transform of Eq.\ \eqref{Ap:def:RectRt}. It is also immediate to see that the partial derivative with respect to $x$ is null in this regime. 

\vskip 0.3 cm
\noindent  \textit{the case $\theta \geq d_C(x)$:} In this case, we have (again) a saturation. To satisfy Eq.\ \eqref{Ap:eqSP.2.1}, one must have:
\begin{align}
\label{Ap:saturationBSSK}
    \sqrt{z_1^* z_2^*} &= x \, ,&&
\end{align}
since Eq.\ \eqref{Ap:eqSP.1.1} is still valid, $z_1^*$ is solution of the same quadratic equation \eqref{Ap:quadra_eq} except that  the term $ d_A^{\langle -1 \rangle}(\theta) $ is replaced by $x$, so that we have:
\begin{align}
\label{Ap:SP1_sat}
    z_1^*(x,\theta) &= \frac{-(1-q) + \sqrt{(1- q)^2 + 4q \theta^2 x^2 } }{2q \theta} \, . &&
\end{align}
and hence, we have for the free energy:
\begin{align}
\label{Ap:partial_QFE_sat}
\partial_{\theta} J_C(x,\theta) &= \frac{-1 - q + \sqrt{(1-q)^2 + 4 q \theta^2 x^2 }}{2q \theta} && (\mbox{for } \theta \geq d_C(x))\, .
\end{align}
Let us now take a look at the derivative with respect to $x$. From the integral representation of Eq.\ \eqref{eq:PartitionF_BSSK_int.3.1}, we have:
\begin{align}
    \partial_x  J_C(x,\theta) &= \frac{1}{2} \partial_x \left[ 2 \theta z_1^*(x,\theta)  - \frac{1-q}{q} \log \left( q  z_1^*(x,\theta) + \frac{1-q}{\theta} \right) - \int \log (x^2 - s^2) \rho_C(s) \mathrm{d}s \right] \, , && 
\end{align}
which using the expression of Eq.\ \eqref{Ap:SP1_sat} for $z_1^*(x,\theta)$ gives:
\begin{align}
\label{Ap:partial_QFE_BSSK.6}
    \partial_x  J_C(x,\theta)&=  \frac{-(1-q) + \sqrt{(1-q)^2 +4q \theta^2 x^2 }}{2qx} - x g_{CC^{\mathsf{T}}}(x^2) && (\mbox{for } \theta \geq d_C(x))\, .
\end{align}
and using the definition \eqref{Ap:prop:Dt} of the D-transform, this can be also written as: 
\begin{align}
\label{Ap:partial_QFE_BSSK.7}
\partial_x  J_C(x,\theta) &=  \frac{ \sqrt{(1-q)^2 +4q \theta^2 x^2 } - \sqrt{(1-q)^2 +4q d_C(x)^2 x^2} }{2qx}  &&  (\mbox{for } \theta \geq d_C(x))\, .
\end{align}
This concludes the proof of the quenched free energy of the BSSK model. 
\vskip 0.3 cm
\noindent By averaging over the law \eqref{eq:JointDensitySv.1} of a  bi-invariant matrix, a similar computation as in the additive case shows that the annealed free energy is given by:
\begin{align}
    \partial_{\theta} F_A(w_A,\theta) &= \frac{-1 - q + \sqrt{(1-q)^2 + 4 q \theta^2 \left[ d_A^{\langle -1 \rangle}(\theta)  \right]^2 }}{q \theta} \quad && \mbox{for} \quad \theta \leq \bar{d}_A(w_A) \,  ,
\end{align}
and 
\begin{align}
     \partial_{\theta} F_A(w_A,\theta) &= \frac{-1 - q + \sqrt{(1-q)^2 + 4 q \theta^2 w_A^2 }}{q \theta} \quad &&\mbox{for} \quad \theta \geq \bar{d}_A(w_A) \, .
\end{align}

\section{Large deviation for rank-one deformation}
\label{sec:Ap:rk1_deformation}
In this section, we consider the case where the square (resp. rectangular) matrix $\mathbf{C}$ is a rank-one deformation of a random matrix $\mathbf{B}$ and study the right large deviation of its largest eigenvalue (resp. singular value).  For this type of problem, depending on the norm of the vector of the rank-one deformation, it is well known that there exists a regime where the top eigenvalue (resp. singular value) sticks to the right edge of the bulk density and another one where it pops out of the bulk density and forms an outlier. As in the full-rank deformation, we would like to characterize completely the large deviation of this top eigenvalue/singular value. It turns out that the tilting method with spherical integral allows one to get an answer for this problem for as long as we are looking at values of $x$ \emph{higher} than the typical one. In particular, this tilting method does not work when one looks at the large deviation at the left of the typical outlier, if there is one. We can however complete the picture for this left large deviation by looking at the poles of this rank-one deformation thanks to the Sherman-Morrison formula. We consider different types of rank-one deformations and separate the study in the several cases.

\subsection{Additive rank-one deformation}
\label{sec:rk1.sum}
In this subsection, the matrix $\mathbf{B} \sim \mathbb{P}_{V,w_B}(.)$, with limiting spectral density $\mu_B$ and (right) edge $\mathrm{b}_+$ and we consider the rank-one additive deformation:
\begin{align}
\label{eq:rk1_add_deformation}
    \mathbf{C}&:= \mathbf{B} + \gamma \mathbf{v}\mathbf{v}^{\mathsf{T}} \, , &&
\end{align}
where $\mathbf{v}$ is an arbitrary\footnote{If one replaces the matrix $\mathbf{B}$ by a fixed diagonal matrix of Sec.\ \ref{sec:InvEns_w_wall} then one has to consider the vector $\mathbf{v}$ to be taken uniformly over the sphere $\mathbb{S}^{N-1}$. } unit vector. In this case, it is well known \cite{Edwards_1976,Baik2005,BenaychGeorges2011}  that the top eigenvalue $\lambda_1(\mathbf{C})$ admits a so-called BBP-phase transition in the large $N$ limit.
\begin{itemize}
    \item For $\gamma \leq \frac{1}{g_B(\mathrm{b}_+)}$: the top eigenvalue sticks to the edge $\mathrm{b}_+$,
    \begin{align}
    \label{eq:BBP_add_bulk}
        \lambda_1(\mathbf{C}) &\to \mathrm{b}_+ \, . &&
    \end{align}
    \item For $\gamma \geq \frac{1}{g_B(\mathrm{b}_+)}$: the top eigenvalue pops out of the bulk and is equal to
        \begin{align}
        \label{eq:BBP_add_out}
        \lambda_1(\mathbf{C}) &\to  \lambda^* := \mathcal{R}_B \left(\frac{1}{\gamma} \right) + \gamma = g_B^{\langle -1 \rangle}\left(\frac{1}{\gamma}\right) \, . &&
    \end{align}
\end{itemize}
\subsubsection{The tilting method and its failure for left large deviation.}
\label{sec:tilt.rk1.sum}
At finite but large $N$, we can once again ask what is the probability $\mathbb{P} \left[ \lambda_1(\mathbf{C}) \simeq x \right]$ of finding the top eigenvalue \emph{far} from its  typical value $\mathrm{b}_+$ (resp. $\lambda^*$) if  $\gamma < \frac{1}{g_B(\mathrm{b}_+)}$ (resp. if $\gamma > \frac{1}{g_B(\mathrm{b}_+)}$). Since we are looking at the sum of two matrices, let's consider the tilt with the SSK partition function given by Eq.\ \eqref{eq:PartitionF_SSK}. By Haar property, we still have  the decomposition of Eq.\ \eqref{eq:prop:AFE} for the spherical integral, hence the annealed free energy of Eq.\ \eqref{eq:defAnnealedFE} is given by:
\begin{align}
\label{eq:AFE_rk1_add}
    F_C(\theta) &= F_B(w_B,\theta) + F_{\mathrm{rk1}}(\theta) \, , &&
\end{align}
where
\begin{align}
\label{eq:def_AFE_rk1_add}
    F_{\mathrm{rk1}}(\theta) &\simeq \frac{1}{N} \log \mathbb{E} \mathcal{Z}_{\gamma \mathbf{v}\mathbf{v}^{\mathsf{T}}}(\theta) \, . &&
\end{align} 
Since the spectrum of the rank-one matrix $\gamma \mathbf{v}\mathbf{v}^{\mathsf{T}}$ consist of $N-1$ zero eigenvalues and one non-zero eigenvalue equal to $\gamma$, We have $ F_{\mathrm{rk1}}(\theta) = J_{0}(\gamma,\theta)$ where $J_{0}$ is defined by replacing the general measure $\mu_A(.)$ in Eq.\ \eqref{eq:As_QFE_sum.1} by the trivial Dirac measure  at zero, $\delta(.-0)$. Next, since the Stieltjes transform of this measure is simply equal to $\frac{1}{z}$ and its R-transform is equal to $0$, this reads:
\begin{align}
\label{eq:As_AFE_rk1_add}
    (F_{\mathrm{rk1}})'(\theta)= \frac{1}{2} \left\{
    \begin{array}{ll}
    0 & \mbox{for } \theta \leq \frac{1}{\gamma}  \, ,\\
\\
   \gamma - \frac{1}{\theta} & \mbox{for } \theta \geq \frac{1}{\gamma} \, .
    \end{array}
\right.  &&
\end{align}
Finally, since  $\mathbf{C}$ is a rank-one modification of $\mathbf{B}$, its bulk density is also given by $\mu_B(.)$, hence from Eq.\ \eqref{eq:As_QFE_sum.1}, the behavior of its quenched free energy, with $x$ is the position of the top eigenvalue, reads:
\begin{align}
\label{eq:As_QFE_rk1_add}
   \partial_{\theta} J_{C}(x,\theta)= \frac{1}{2} \left\{
    \begin{array}{ll}
    \mathcal{R}_B(\theta) & \mbox{for } \theta \leq g_B(x)  \, ,\\
\\
   x - \frac{1}{\theta} & \mbox{for } \theta \geq g_B(x) \, .
    \end{array}
\right. &&
\end{align}
To continue we need to compute the optimal temperature $\theta^*(x)$ (if there is one) and hence study the behavior of the function
\begin{align}
    I_x'(\theta) &:= \partial_{\theta} J_{C}(x,\theta)- \partial_{\theta} F_B(w_B,\theta) -(F_{\mathrm{rk1}})'(\theta) \, . &&
\end{align}
 Comparing Eq.\ \eqref{eq:As_AFE.sumA}, Eq.\ \eqref{eq:As_AFE_rk1_add} and Eq.\ \eqref{eq:As_QFE_rk1_add}, the behavior of this function depends on the relative position of $\frac{1}{\gamma}$ with respect to $g_B(x)$ and $g_B(\mathrm{b}_+)$ and we need to consider three different cases.

\vskip 0.3cm
\noindent \textit{The case $g_B(x) \leq g_B(\mathrm{b}_+) \leq \frac{1}{\gamma}$ (right large deviation below the threshold):} In this case, the top eigenvalue converges to the edge, see Eq.\ \eqref{eq:BBP_add_bulk}. The difference between the quenched and free energy is given by:
\begin{align}
\label{eq:diff_FE_rk1sum.1}
   I_x'(\theta) = \frac{1}{2} \left\{
    \begin{array}{llll}
    0 & \mbox{for } \theta  \leq g_B(x) \, ,\\
\\
   x - g_B^{\langle -1 \rangle}(\theta) & \mbox{for } g_B(x) \leq  \theta \leq \frac{1}{\gamma} \, ,\\
\\
    (x-\gamma) - \mathcal{R}_B(\theta) & \mbox{for }  \frac{1}{\gamma} \leq \theta \leq \bar{g}_B(w_B) \, ,\\
\\
      x - w_B -\gamma + \frac{1}{\theta} & \mbox{for } \theta \geq \bar{g}_B(w_B) .
    \end{array}
\right. &&
\end{align}
Similarly to the full-rank case, for $\theta$ starting at $g_B(x)$, this function is continuously increasing with $\theta$ until it reaches the value $
\theta = g_B(\mathrm{b}_+)$ and then it continuously decreases. For $x<w_B+\gamma$, it must cross the real axis only one time  for $\theta \in (g_B(x),\infty)$. Now since $g_B(x) \leq g_B(\mathrm{b}_+) \leq \frac{1}{\gamma}$, this value $\theta^*(x)$ solution of Eq.\ \eqref{eq:Theta_opt.2} has a different behavior depending on the position of $x$ with respect to the critical points $x_{c_1}$ and $x_{c_2}$ defined by 
\begin{align}
    \label{eq:xc1_rk1_add}
    x_{c_1} &:= g_B^{\langle -1 \rangle}\left( \frac{1}{\gamma} \right) \, , &&
\end{align}
and by
\begin{align}
    \label{eq:xc2_rk1_add}
    x_{c_2} &:= w_B + \gamma - \frac{1}{\bar{g}_B(w_B)} \, , &&
\end{align}
and thus we get:
\begin{align}
\label{eq:ThetaOpt_rk1_add.1}
  \theta^*(x) =\left\{
    \begin{array}{lll}
    \bar{g}_B(x) & \mbox{for }  \mathrm{b}_+ \leq x  \leq x_{c_1} \, ,\\
\\
   \mathcal{R}_B^{\langle - 1\rangle}(x - \gamma) & \mbox{for } x_{c_1} \leq x \leq x_{c_2} \, , \\
   \\
   \frac{1}{w_B + \gamma - x} & \mbox{for } x_{c_2} \leq x \leq  w_B + \gamma \, .
    \end{array}
\right. &&
\end{align}
\vskip 0.3cm 
\noindent \textit{The case $g_B(x) \leq \frac{1}{\gamma} \leq g_B(\mathrm{b}_+)$ (right large deviation above the threshold):}
In this case, one has a creation of an outlier at the position $\lambda^*$ given by Eq.\ \eqref{eq:BBP_add_out}. Because we are looking at a value of $\frac{1}{\gamma} \leq g_B(\mathrm{b}_+)$ and $g_B^{\langle -1 \rangle}(.)$ is decreasing in $(0,g_B(\mathrm{b}_+))$, this corresponds to $x \geq \lambda^*$ and hence looking at large deviation at the right of the outlier. It is a straightforward computation  to show that the difference of free energies is again given by Eq.\ \eqref{eq:diff_FE_rk1sum.1}. However, since there is now the constraint $g_B(x) \leq \frac{1}{\gamma} \leq g_B(\mathrm{b}_+)$,  $\theta^*(x)$ cannot be attained in the interval $(g_B(x),\frac{1}{\gamma})$. In other words, $\theta^*$ is given by Eq.\ \eqref{eq:ThetaOpt_rk1_add.1} where the first line is removed:
\begin{align}
\label{eq:ThetaOpt_rk1_add.2}
  \theta^*(x) =\left\{
    \begin{array}{ll}
   \mathcal{R}_B^{\langle - 1\rangle}(x - \gamma) & \mbox{for } \lambda^* \leq x \leq x_{c_2} \, , \\
   \\
   \frac{1}{w_B + \gamma - x} & \mbox{for } x_{c_2} \leq x \leq  w_B + \gamma \, .
    \end{array}
\right. &&
\end{align}
where $x_{c_2}$ is again defined by Eq.\ \eqref{eq:xc2_rk1_add}.
\vskip 0.3cm
\noindent \textit{The case $ \frac{1}{\gamma} \leq g_B(x)  \leq g_B(\mathrm{b}_+) $ (failure of getting the left large deviation above the threshold):} This case corresponds  to have an outlier at $\lambda^*$ given by Eq.\ \eqref{eq:BBP_add_out} but looking at values of $x$ at the right of this outlier, $\mathrm{b}_+ \leq x \leq \lambda^*$. In this case, one finds that the difference of free energies is given by
\begin{align}
\label{eq:diff_FE_rk1sum.2}
   I_x'(\theta) = \frac{1}{2} \left\{
    \begin{array}{llll}
    0 & \mbox{for } \theta  \leq \frac{1}{\gamma} \, ,\\
\\
   - \gamma + \frac{1}{\theta} & \mbox{for} \frac{1}{\gamma} \leq  \theta \leq g_B(x) \, ,\\
\\
    (x-\gamma) - \mathcal{R}_B(\theta) & \mbox{for }  g_B(x) \leq \theta \leq \bar{g}_B(w_B) \, ,\\
\\
      x - w_B -\gamma + \frac{1}{\theta} & \mbox{for } \theta \geq \bar{g}_B(w_B) .
    \end{array}
\right. &&
\end{align}
This function is always decreasing with respect to the variable $\theta$ and a consequence  the supremum of Eq.\ \eqref{eq:Theta_opt.1} is theoretically attained at zero, but this cannot be correct since this corresponds to the original model without the tilt. As a consequence, in this regime the tilting method fails. It is nevertheless tempting to argue that the rate function should have the same expression as the one obtained for  'right' large deviation. This will turn out to be true, as shown by the 'Sherman-Morrison method' of the next section. 

\subsubsection{Large deviation beyond the threshold and the Sherman-Morrison formula}
\label{sec:SM.sum}
To get the full behavior in the presence of an outlier at $\lambda^*$, the idea is to use the Sherman-Morrison  formula \cite{Sherman}. For a symmetric matrix $\mathbf{M}$ and two vectors $\mathbf{u}_0$, $\mathbf{v}_0$, this formula is given by: 
\begin{align}
    \label{eq:SM_formula}
    (\mathbf{M} + \mathbf{u}_{0} \mathbf{v}_0^{\mathsf{T}})^{-1} &= \mathbf{M}^{-1} - \frac{\mathbf{M}^{-1}\mathbf{u}_0 \mathbf{v}_0^{\mathsf{T}} \mathbf{M}^{-1}}{1 + \mathbf{v}_0^{\mathsf{T}} \mathbf{M}^{-1} \mathbf{u}_0} \, . &&
\end{align}
This formula allows one to relate the resolvent of a  rank one perturbation of a matrix to the resolvent of the matrix plus an additional term. Namely, setting $\mathbf{M} = z \mathbf{I} - \mathbf{B}$, $\mathbf{u}_0 = - \gamma \mathbf{v}$ and $\mathbf{v}_0 = \mathbf{v}$ in Eq.\ \eqref{eq:SM_formula}, one has:
\begin{align}
    \label{eq:SM_formula.2}
    \mathbf{G}_{\mathbf{C}}(z) &=   \mathbf{G}_{\mathbf{B}}(z) + \frac{\gamma}{ 1 - \gamma \mathbf{v}^{\mathsf{T}} \mathbf{G}_{\mathbf{B}}(z) \mathbf{v}} \, \mathbf{G}_{\mathbf{B}}(z) \mathbf{v} \mathbf{v}^{\mathsf{T}} \mathbf{G}_{\mathbf{B}}(z) \, . &&
\end{align}
where $\mathbf{G}_{\mathbf{B}}(z)$ is the resolvent of $\mathbf{B}$, defined by:
\begin{align}
    \mathbf{G}_{\mathbf{B}}(z) &:= ( z \mathbf{I} - \mathbf{B})^{-1} \, , &&
\end{align}
and $ \mathbf{G}_{\mathbf{C}}(z)$ is defined similarly. The resolvent contains all the information of the matrix $\mathbf{C}$ and in particular the eigenvalues of $\mathbf{C}$ corresponds to  a value for which the resolvent is ill-defined. Now for the matrix $\mathbf{C}$ to have just one outlier out of the bulk, say at a position $ \lambda_1 (\mathbf{C}) = x$, the divergence in the resolvent of Eq.\ \eqref{eq:SM_formula.2} must come from the denominator of the RHS of Eq.\ \eqref{eq:SM_formula.2}, that is:
\begin{align}
\label{eq:outlier_SM_add}
    1 - \gamma \mathbf{v}^{\mathsf{T}} \mathbf{G}_{\mathbf{B}}(x) \mathbf{v} &= 0 \, . &&
\end{align}
As a consequence, we can write the probability of finding the top eigenvalue $\lambda_1(\mathbf{C})$ at the position $x$ in terms of an average over a Dirac function:
\begin{align}
    \label{Ap:proba_SM.1}
    \mathbb{P} \left[ \lambda_1(\mathbf{C}) \simeq x \right] &= \mathbb{E} \left[ \delta( 1 - \gamma \mathbf{v}^{\mathsf{T}} \mathbf{G}_{\mathbf{B}}(x) \mathbf{v} = 0  ) \right] \, . &&
\end{align}
Using the inverse Laplace representation of the Dirac, this can be equivalently written as:
\begin{align}
    \label{Ap:proba_SM.2}
    \mathbb{P} \left[ \lambda_1(\mathbf{C}) \simeq x \right] &= \frac{1}{K} \, \mathbb{E} \left[  \int_{\mathscr{C}_1} \mathrm{e}^{ \frac{N z_1}{2} - \frac{N \gamma z_1}{2} \mathbf{v}^{\mathsf{T}} \mathbf{G}_{\mathbf{B}}(x) \mathbf{v}  }  \mathrm{d} z_1 \right] \, , &&
\end{align}
where $K$ is a (complex) constant whose asymptotic will not contribute to the large deviation in its integral representation. For simplicity, we take the notation $K$ from one line to another, even though this constant might be different.    Next, since $\mathbf{B}$ is rotationally invariant, we can either take the vector $\mathbf{v}$ to be either fixed or random. To perform the computation, it will be convenient to take $\mathbf{v}$ uniform over the sphere. Removing the constraint over the sphere by introducing a second Lagrange multiplier $z_2$, Eq.\ \eqref{Ap:proba_SM.2} now writes:
\begin{align}
    \label{Ap:proba_SM.3}
    \mathbb{P} \left[ \lambda_1(\mathbf{C}) \simeq x \right] &= \frac{1}{K} \,  \mathbb{E} \left[  \int_{\mathscr{C}_1 \times \mathscr{C}_2 } \mathrm{e}^{ \frac{N z_1}{2} + \frac{N z_1}{2}} \left( \int_{\mathbb{R}^N}  \mathrm{e}^{ - \frac{N}{2} \mathbf{v}^{\mathsf{T}} \left(  z_2 \mathbf{Id} + \gamma z_1 \mathbf{G}_{\mathbf{B}}(x) \right) \mathbf{v}  } \mathrm{d} \mathbf{v} \right) \mathrm{d} z_1 \mathrm{d} z_2 \right] \, . &&
\end{align}
By Gaussian integration over the $N$-dimensional variable $\mathbf{v}$, we have the following integral:
\begin{align}
    \label{Ap:proba_SM.4}
    \mathbb{P} \left[ \lambda_1(\mathbf{C}) \simeq x \right] &=   \mathbb{E} \left[  \int_{\mathscr{C}_1 \times \mathscr{C}_2 } \mathrm{e}^{ \frac{N }{2}H(z_1,z_2,x)} \mathrm{d} z_1 \mathrm{d} z_2 \right] \, , &&
\end{align}
with: 
\begin{align}
 \label{Ap:SM_H}
  H(z_1,z_2,x) &:=z_1 + z_2 - \frac{1}{N} \sum_{i=1}^N \log \left( \gamma z_1 + z_2(x-b_i) \right) + \frac{1}{N} \sum_{i=1}^N  \log (x-b_i) + K + O \left(\frac{1}{N} \right)  \, , &&
  \end{align}
  where $K$ is constant independent of $z_1,z_2$ and $x$. In the large $N$ limit, this integral over the variables $z_1$ and $z_2$ are dominated by the saddle-points $z_1^*$ and $z_2^*$ solutions of:
\begin{align}
 \label{Ap:SM:gradSP}
\left\{
    \begin{array}{ll}
 \frac{\gamma}{N} \sum_{i=1}^N \frac{1}{\gamma z_1^*  + z_2^* (x -b_i)}  &= 1  \, ,\\
\\
\frac{1}{N} \sum_{i=1}^N  \frac{(x -b_i)}{\gamma z_1^*  + z_2^* (x -b_i)} &= 1  \, .
    \end{array}
\right. &&
\end{align}
Combining the two equations, we have that the two saddle points are related to one each other by:
\begin{align}
  \label{Ap:SM:gradSP.2}   
  z_1^* + z_2^* &= 1 \, . &&
\end{align}
As we will see later on, only the saddle-point $z_2^*$ will contribute to the large deviation. Substituting $z_2^*$ in the top line of Eq.\ \eqref{Ap:SM:gradSP}, we have:
\begin{align}
  \label{Ap:SM:gradSP.3}   
  \frac{z_2^*}{\gamma} &= g_{\mathbf{B}} \left(  x - \gamma + \frac{\gamma}{z_2^*} \right) \, . &&
\end{align}
As in the full-rank case of Section \ref{sec:Ap:As_AFE_QFE} and since we are taking the expectation, we need to distinguish two cases: Eq.\ \eqref{Ap:SM:gradSP.3} only makes sense if the argument of the RHS does not exceed $w_B$, that is:
\begin{itemize}
    \item if $x \leq x_{c_2}$, with $x_{c_2}$ defined by Eq.\ \eqref{eq:xc2_rk1_add}, then we don't have any saturation. Inverting Eq.\ \eqref{Ap:SM:gradSP.3} gives:
\begin{align}
  \label{Ap:SM:gradSP.4}   
  \frac{z_2^*}{\gamma} &= \mathcal{R}_B^{\langle -1 \rangle}\left(  x - \gamma  \right) \, . &&
\end{align}
\item if $x \geq x_{c_2}$, then there is a saturation, that is $z_2^*$ is given by
\begin{align}
  \label{Ap:SM:gradSP.5}   
  \frac{z_2^*}{\gamma} &=\frac{1}{w_B + \gamma -x} \, . &&
\end{align}
\end{itemize}
We have now all the tools to express the rate function in this regime. We have: 
\begin{align}
  \label{Ap:SM:RateFunction_IntegralForm}   
    \Psi_C(x) & =  -\int_{\lambda^*}^x \frac{\mathrm{d}}{\mathrm{d}s}   H(z_1^*,z_2^*,s) \mathrm{d}s  = - \int_{\lambda^*}^x \frac{\partial}{\partial s}   H(z_1^*,z_2^*,s) \mathrm{d}s \, , &&
\end{align}
since the partial derivatives with respect to $z_1^*$ and $z_2^*$ are exactly zero at the saddle points. The derivative of the function $H$ with respect to $s$ is given thanks to  Eq.\ \eqref{Ap:SM_H} by:
\begin{align}
  \label{Ap:SM:RateFunction_IntegralForm.2}   
    \frac{\partial}{\partial s}   H(z_1^*,z_2^*,s)  & = - \frac{1}{N} \sum_{i=1}^N \frac{z_2^*}{ \gamma z_1^* + z_2^* (x- b_i) } + \frac{1}{N} \sum_{i=1}^N \frac{1}{x -b_i} \, . &&
\end{align}
The first term in the RHS can be simplified since if we multiply the topline of Eq.\ \eqref{Ap:SM:gradSP} by $\frac{z_2^*}{\gamma}$ we find that this term is exactly $\frac{z_2^*}{\gamma}$. Taking the large $N$ limit, one obtain the final expression for the rate function. The result is summarized in the next section.

\subsubsection{Expression for the rate function}
\label{sec:RateFunction.rk1.sum}
\begin{itemize}
    \item if  $\gamma \leq \frac{1}{g_B(\mathrm{b}_+)}$, then for $x$ not in $[\mathrm{b}_+, w_B + \gamma]$ the rate function is infinite and is otherwise given by:
    \begin{align}
\label{eq:RateFunction_rk1_add.1}
 \Psi_C(x) =
       \left\{     \begin{array}{lll}
  \frac{1}{2} \int_{\mathrm{b}_+}^x \bar{g}_B(t) - g_B(t) \mathrm{d}t & \mbox{for }  \mathrm{b}_+ \leq x  \leq x_{c_1} \, ,\\
\\
   K_1 + \frac{1}{2} \int_{x_{c_1}}^x \left( \mathcal{R}_B^{\langle - 1\rangle}(t - \gamma) - g_B(t) \right) \mathrm{d}t  & \mbox{for } x_{c_1} \leq x \leq x_{c_2} \, , \\
   \\
   K_2 + \frac{1}{2} \log \left( \frac{1}{w_B+\gamma -x} \right) -  \frac{1}{2} \int_{x_{c_2}}^x g_B(t) \mathrm{d}t  & \mbox{for } x_{c_2} \leq x \leq  w_B + \gamma \, .
    \end{array} \right. &&
\end{align}
    \item if  $\gamma \geq \frac{1}{g_B(\mathrm{b}_+)}$  then for $x$ not in $[\mathrm{b}_+, w_B + \gamma]$ the rate function is infinite and is otherwise given by:
\begin{align}
\label{eq:RateFunction_rk1_add.2}
 \Psi_C(x) =
       \left\{     \begin{array}{ll}
   \frac{1}{2} \int_{\lambda^*}^x \left( \mathcal{R}_B^{\langle - 1\rangle}(t - \gamma) - g_B(t) \right) \mathrm{d}t & \mbox{for } \mathrm{b}_+ \leq x \leq x_{c_2} \, , \\
   \\
   K_2^*  + \frac{1}{2} \log \left( \frac{1}{w_B+\gamma -s} \right) -\frac{1}{2} \int_{x_{c_2}}^x g_B(t) \mathrm{d}t  & \mbox{for } x_{c_2} \leq x \leq  w_B + \gamma \,  ,
    \end{array} \right. &&
\end{align}
with 
\begin{align}
\label{eq:RateFunction_rk1_add.K2star}
  K_2^* &= \frac{1}{2} \int_{\lambda^*}^{x_{c_2}} \left( \bar{g}_B(t) - g_B(t)  \right)\mathrm{d}t +\frac{1}{2} \log \left(\frac{1}{\bar{g}_B(w_B)} \right)  \, . &&
    \end{align}
\end{itemize}

\vskip 0.3 cm 
\noindent \textit{Remark (Behavior near the edge):} Bellow the threshold when there is no outlier, if the density is non-degenerate, one recovers the Tracy-Widom '3/2' scaling for the rate function since the expression matches the ones of the classical case of Eq.\ \eqref{eq:Psi_1RM.2}. However above the threshold, because both the Stieltjes transform and the R-transform (and hence its shifted inverse) are analytic around $\lambda^*$, the rate function as a square behavior close to the outlier. This is expected because the fluctuations of the outliers are known to be Gaussian and of variance $N^{-\frac{1}{2}}$.

\begin{figure}[h]
\centering
    \subfloat{\includegraphics[width=0.44\linewidth]{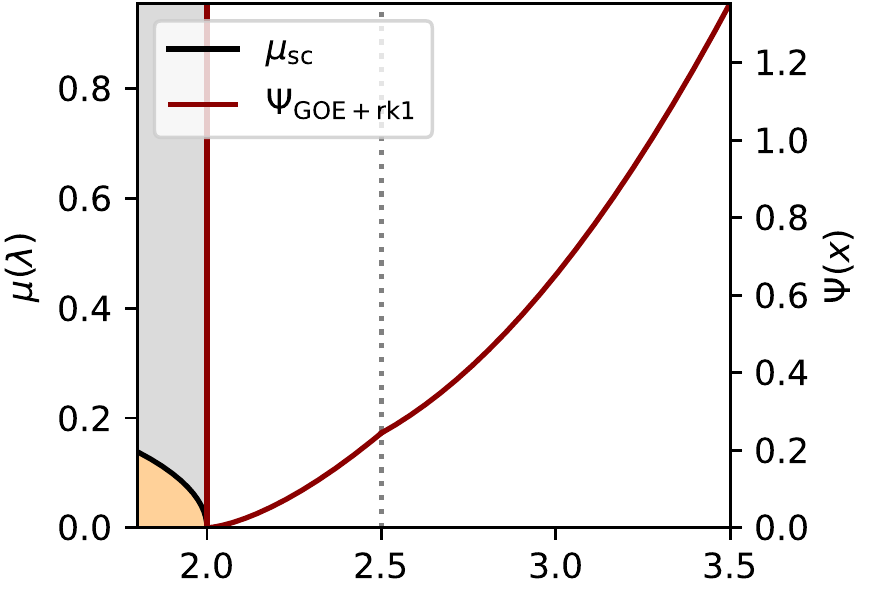}}
    \qquad
    \subfloat{\includegraphics[width=0.45\linewidth]{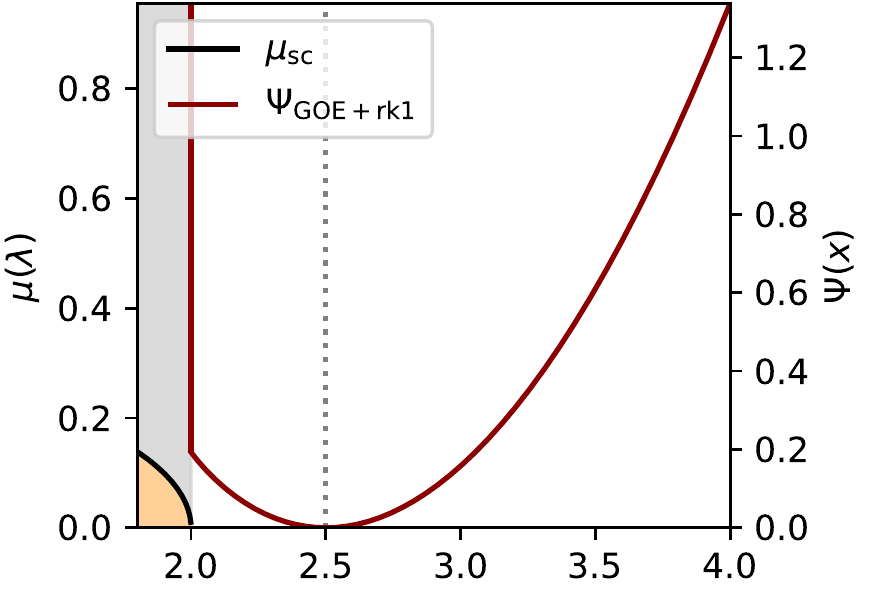}}
\caption{On the left, the Rate function (in red) of the largest eigenvalue of the sum of a GOE matrix with $\sigma =1$ and a rank-one matrix with a non-zero eigenvalue equal to $\gamma = 1/2$, as described by Eq.\ \eqref{eq:RateFunction_GOE_rk1.1}. This function admits a phase transition at $x_{c_1} =2.5$, represented by the vertical dotted line. On the right, the Rate function (in red) of the largest eigenvalue of the sum of a GOE matrix with $\sigma =1$ and a rank-one matrix with now a non-zero eigenvalue equal to $\gamma = 2$, as described by Eq.\ \eqref{eq:RateFunction_GOE_rk1.2}. In this case, one is above the BBP transition and the typical position of the largest eigenvalue of this matrix is represented in dotted line.       }
    \label{fig:RateFunction_GOErk1}
\end{figure}

\vskip 0.3 cm
\noindent \textit{Example (rank-one perturbation of GOE):} Let's consider the case where $\mathbf{A}$ is a GOE matrix. In this case, $\mathrm{b}_+ = 2\sigma$, $g_B(\mathrm{b}_+) = \frac{1}{\sigma}$, and if $\gamma  \leq \sigma$ there is no outlier but one critical point $x_{c_1} = \gamma + \frac{\sigma^2}{\gamma}$ and if $\gamma  \geq \sigma$ there is one outlier at $\lambda^* = \gamma + \frac{\sigma^2}{\gamma}$. As a consequence, the rate function is given by:

\begin{itemize}
    \item if  $\gamma  \leq \sigma$, 
\begin{align}
\label{eq:RateFunction_GOE_rk1.1}
 \Psi_C(x) =
       \left\{     \begin{array}{ll}
  \frac{x \sqrt{x^2 - 4 \sigma^2}}{4\sigma^2} + \log \left( \frac{2 \sigma}{\sqrt{x^2 - 4\sigma^2}+x} \right)  & \mbox{for }  2 \sigma \leq x  \leq  \gamma + \frac{\sigma^2}{\gamma}  \, ,\\
\\
 \frac{\left(x-(\gamma + \frac{\sigma^2}{\gamma})\right) \left(x-(3\gamma +\frac{\sigma^2}{\gamma} ) \right) +x\sqrt{x^2 - 4 \sigma^2 }}{4 \sigma^2} + \log \left( \frac{2 \sigma}{x+\sqrt{x^2 - 4\sigma^2}} \right)  & \mbox{for }  x \geq  \gamma + \frac{\sigma^2}{\gamma} \, ,
    \end{array} \right. &&
\end{align}
One can find a plot of this function for $\gamma=1/2$ and $\sigma=1$ in Fig.\ \ref{fig:RateFunction_GOErk1} (Left).
\item and if  $\gamma  \geq \sigma$,
\begin{align}
\label{eq:RateFunction_GOE_rk1.2}
 \Psi_C(x) =
    \frac{x^2 -4 \gamma x +2(\gamma^2 + \sigma^2) +x\sqrt{x^2 - 4 \sigma^2 }}{4 \sigma^2} + \log \left( \frac{2 \gamma}{x+\sqrt{x^2 - 4\sigma^2}} \right) \, . &&
\end{align}
One can find a plot of this function for $\gamma=2$ and $\sigma=1$ in Fig.\ \ref{fig:RateFunction_GOErk1} (Right).
\end{itemize}

\subsection{Multiplicative case}
\label{sec:RateFunction.rk1.prod}
Similar to the additive case, we consider in this subsection the 'multiplicative' rank-one deformation:
\begin{align}
\label{eq:rk1_prod_deformation}
    \mathbf{C} &= \sqrt{ \mathbf{I}+ \gamma \mathbf{v}\mathbf{v}^{\mathsf{T}}} \mathbf{B} \sqrt{ \mathbf{I}+ \gamma \mathbf{v}\mathbf{v}^{\mathsf{T}}} \, , &&
\end{align}
where $\mathbf{B} \sim \mathbb{P}_{V,w_B}(.)$ and is further assumed to be semi-definite positive. In the large $N$ limit, The top eigenvalue $\lambda_1(\mathbf{C})$ admits the following transition \cite{BenaychGeorges2011} : 
\begin{itemize}
    \item for $\gamma <\frac{1}{ t_B(\mathrm{b}_+)}$: the top eigenvalue sticks to the edge $\mathrm{b}_+$: 
    \begin{align}
    \label{eq:BBP_rk1_prod_bulk}
        \lambda_1(\mathbf{C}) &\to \mathrm{b}_+ \, , &&
    \end{align}
    \item for $\gamma > \frac{1}{t_B(\mathrm{b}_+)}$: the top eigenvalue pops out of the bulk and is equal to:
        \begin{align}
        \label{eq:BBP_rk1_prod_out}
        \lambda_1(\mathbf{C}) &\to  \lambda^* := t_B^{\langle -1  \rangle}\left(\frac{1}{\gamma}\right) \, . &&
    \end{align}
\end{itemize}
The proof of the large deviation in this case is similar to the additive case. One has to consider the multiplicative spherical integral and compute the free energy of the LSSK model with  disorder matrix $ \mathbf{I}+ \gamma \mathbf{v}\mathbf{v}^{\mathsf{T}}$ whose spectrum is composed of $N-1$ eigenvalues equals to $1$ and one outlier at $1+\gamma$. One has to separate the study into three cases, depending on the relative value of $\gamma$ with respect to $t_A(x)$ and $t_A(\mathrm{a}_+)$. Similarly to the additive case, the tilting method allows one to get the large deviation at the right of the typical value, whether there is an outlier or not,  but fails to get the  large deviation at the left of the outlier when there is one. To complete the picture, one needs to use the Sherman-Morrison formula of Eq.\ \eqref{eq:SM_formula} applied to our rank-one deformation of Eq.\ \eqref{eq:rk1_prod_deformation}. If if the top eigenvalue is an outlier at the position $x$, then it satisfies: 
\begin{align}
    \label{eq:outlier_SM_prod}
    1 - \gamma \mathbf{v}^{\mathsf{T}} \sqrt{\mathbf{B}} \mathbf{G}_{\mathbf{B}}(x) \sqrt{\mathbf{B}} \mathbf{v} &= 0 \, . &&
\end{align}
Using again the delta trick, one gets the  'left' large deviation. In full generality, we have the following expression for the rate function

\begin{itemize}
    \item if  $\gamma \leq \frac{1}{t_B(\mathrm{b}_+)}$, then for $x$ not in $[\mathrm{b}_+, w_B(1+\gamma)]$ the rate function is infinite and is otherwise given by:
\begin{align}
\label{eq:RateFunction_rk1_prod.1}
 \Psi_C(x) =
       \left\{     \begin{array}{lll}
   \frac{1}{2} \int_{\mathrm{b}_+}^x \bar{g}_B(t) - g_B(t) \mathrm{d}t & \mbox{for }  \mathrm{b}_+ \leq x  \leq x_{c_1} \, ,\\
\\
   K_1 +   \frac{1}{2} \int_{x_{c_1}}^x \left( \frac{\tilde{\mathcal{S}}_B^{\langle - 1\rangle}\left(\frac{t}{1+\gamma}\right)+1}{t} -   g_B(t) \right)\mathrm{d}t  & \mbox{for } x_{c_1} \leq x \leq x_{c_2} \, , \\
   \\
    K_2 +   \frac{1}{2} \log \left( \frac{x}{w_A (1+\gamma) -x} \right) -   \frac{1}{2} \int_{x_{c_2}}^x g_B(t) \mathrm{d}t  & \mbox{for } x_{c_2} \leq x \leq  w_B (1+\gamma) \, .
    \end{array} \right. &&
\end{align}
    \item if  $\gamma \geq \frac{1}{t_B(\mathrm{b}_+)}$  then for $x$ not in $[\mathrm{b}_+, w_B(1+\gamma)]$ the rate function is infinite and is otherwise given by:
\begin{align}
\label{eq:RateFunction_rk1_prod.2}
 \Psi_C(x) =
       \left\{     \begin{array}{ll}
   \frac{1}{2} \int_{\lambda^*}^x \left( \frac{\tilde{\mathcal{S}}_B^{\langle - 1\rangle}\left(\frac{t}{1+\gamma}\right)+1}{t} - g_B(t) \right) \mathrm{d}t & \mbox{for } \mathrm{b}_+ \leq x \leq x_{c_2} \, , \\
   \\
   K_2^* +  \frac{1}{2} \log \left( \frac{x}{w_B(1+\gamma) -x} \right) -  \frac{1}{2} \int_{x_{c_2}}^x g_B(t)\mathrm{d}t   & \mbox{for } x_{c_2} \leq x \leq  w_B (1+\gamma) \,  .
    \end{array} \right. &&
\end{align}
with 
\begin{align}
\label{eq:RateFunction_rk1_add.K2star.2}
  K_2^* &=   \log \left( \frac{1}{\bar{t}_B(w_B)} \right) + \displaystyle \int_{\lambda^*}^{x_{c_2}}  \left[ \frac{\tilde{\mathcal{S}}_B^{\langle - 1\rangle}\left(\frac{t}{1+\gamma}\right)+1}{t} - g_B(t) \right] \mathrm{d}t \, . &&
    \end{align}
\end{itemize}

\begin{figure}[h]
\centering
    \subfloat{\includegraphics[width=0.44\linewidth]{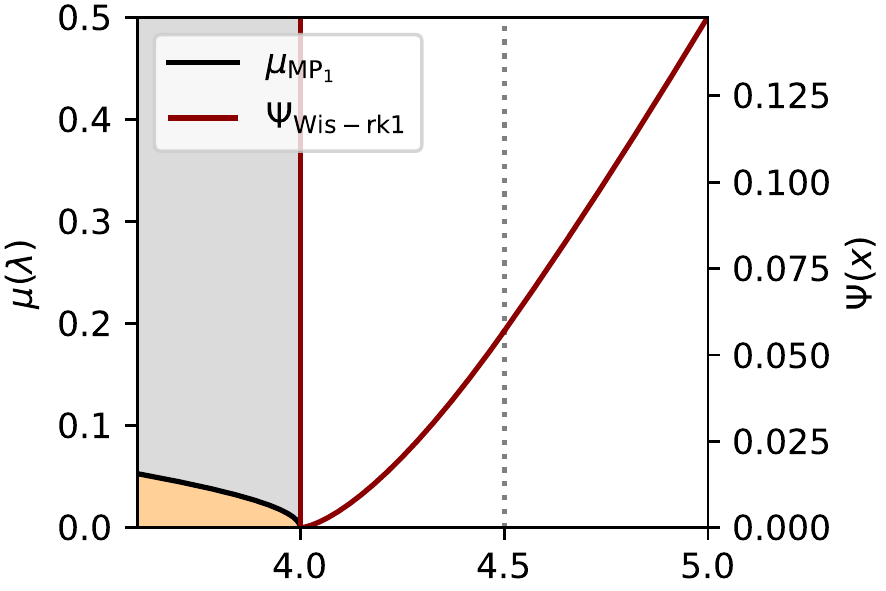}}
    \qquad
    \subfloat{\includegraphics[width=0.45\linewidth]{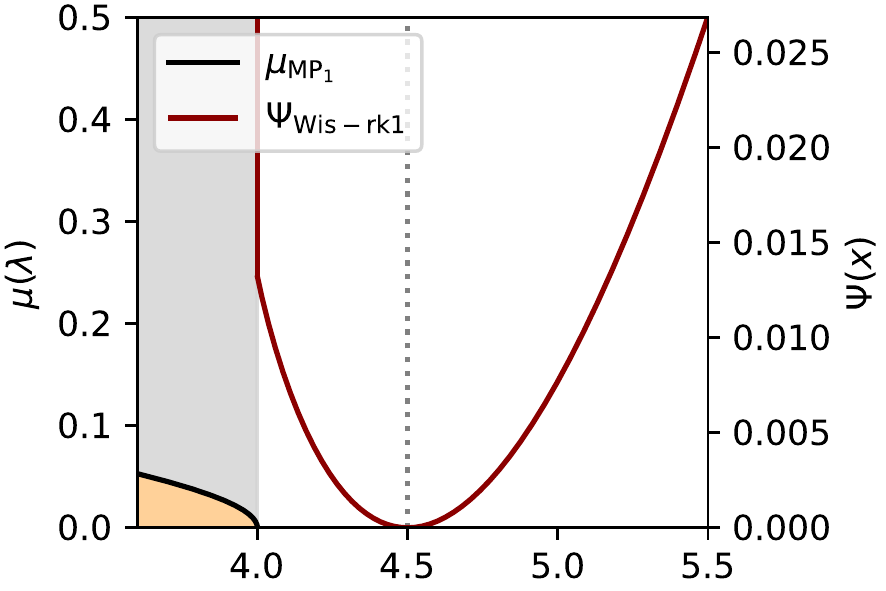}}
\caption{On the left, the Rate function (in red) of the largest eigenvalue of a spiked Wishart matrix with  $q =1$ and the value of the spike is $\gamma = 1/2$, as described by Eq.\ \eqref{eq:RateFunction_GOE_rk1.1}. This function admits a phase transition at $x_{c_1} =4.5$ represented by the vertical dotted line. On the right, the Rate function (in red) of the largest eigenvalue of the sum of a  spiked Wishart matrix with $q =1$ and a spike $\gamma = 2$, as described by Eq.\ \eqref{eq:RateFunction_GOE_rk1.2}. In this case, one is above the BBP transition and the typical position of the largest eigenvalue of this matrix is represented in dotted line.    }
    \label{fig:RateFunction_Wis_rk1}
\end{figure}

\vskip 0.3cm
 \noindent \textit{Example (spiked square Wishart):} Let's consider the case where $\mathbf{B}$ is Wishart, where in order to have simple analytical formula for the rate function, we consider the shape parameter to be equal to one, $q=1$. In this case, the density of  Eq.\ \eqref{eq:MP_dist} has a top edge at $\mathrm{b}_+ =4$.   The matrix $\mathbf{C}$ given by Eq.\ \eqref{eq:rk1_prod_deformation} is known as a spiked (square) Wishart matrix and the rate function for its largest eigenvalue is given by:
 \begin{itemize}
     \item if $\gamma \leq 1$, then there is no outlier and the rate function is given by:
     \begin{align}
     \label{eq:RateFunction_LOE_rk1.1}
         \Psi_{C}(x) =&\left\{
    \begin{array}{ll}
   \frac{\sqrt{x(x-4)}}{2} + \log \left( \frac{x-2 -\sqrt{x(x-4)} }{2}\right)  & \mbox{for } 4 \leq x  \leq 2 + \gamma + \frac{1}{\gamma}  \, ,\\
\\
   \frac{x-\gamma x +(1+\gamma) \sqrt{x(x-4)}}{4(1+\gamma)} + \frac{1}{2} \log \left( \frac{x \gamma - 2\gamma - \gamma \sqrt{x(x-4)}}{2} \right) & \mbox{for } x  \geq  2 + \gamma + \frac{1}{\gamma} \, ,
    \end{array}
\right. &&
\end{align}
One can find a plot of this function for $\gamma=1/2$ and $q=1$ in Fig.\ \ref{fig:RateFunction_Wis_rk1} (Left).
\item if $\gamma > 1$, then there is an outlier at $\lambda^* = 2+\gamma +\frac{1}{\gamma}$ and in this case the rate function is given by: \begin{align}
 \label{eq:RateFunction_LOE_rk1.2}
    \begin{array}{ll} \Psi_C(x)= \frac{1 +\frac{1}{\gamma}}{2} +  \frac{x-2-\gamma -\frac{1}{\gamma} }{2(1+\gamma)} +  \frac{\sqrt{x(x-4)} -x}{4}  + \frac{1}{2} \log \left( \gamma \frac{1-\sqrt{\frac{x-4}{x}}}{1+\sqrt{\frac{x-4}{x}}} \right)   \, . 
  \end{array} &&
\end{align} 
One can find a plot of this function for $\gamma=2$ and $q=1$ in Fig.\ \ref{fig:RateFunction_Wis_rk1} (Right).
 \end{itemize}
 
\subsection{Rectangular case}
\label{sec:RateFunction.rk1.rect}
In this case, the position of the top singular eigenvalue $s_1(\mathbf{C})$ of the matrix 
\begin{align}
    \mathbf{C} = \mathbf{A} + \gamma \mathbf{u} \mathbf{v}^{\mathsf{T}}
\end{align}
admits a phase transition given by \cite{BenaychGeorges2012}:
\begin{itemize}
    \item for $\gamma <\frac{1}{d_B(\mathrm{s}_+)}$: the top singular value sticks to the edge $\mathrm{b}_+$: 
    \begin{align}
        s_1(\mathbf{C}) &\to \mathrm{b}_+ \, ; &&
    \end{align}
    \item for $\gamma > \frac{1}{d_B(\mathrm{s}_+)}$: the top singular value pops out of the bulk and is equal to:
        \begin{align}
        s_1(\mathbf{C}) &\to  s^* := d_B^{\langle -1\rangle} \left(\frac{1}{\gamma}\right) \, ; &&
    \end{align}
\end{itemize}
and the large deviation can be obtained similarly. 

\section{Rank-one plus rank-one}
\label{sec:Ap:RateFunction.rk1rk1}
\begin{figure}
     \centering
         \includegraphics[width= 0.55\textwidth]{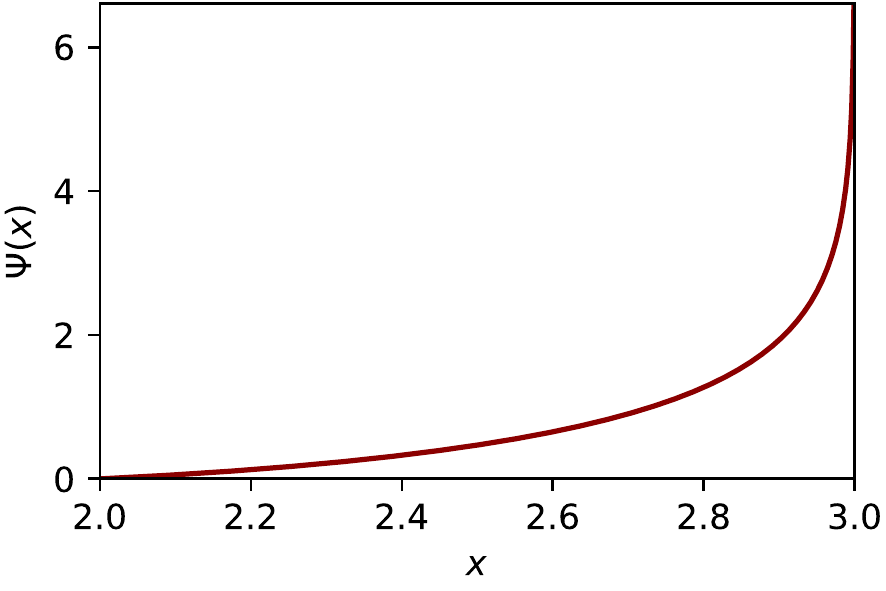}
    \caption{The rate function for the eigenvalue of the toy model of the sum of two rank-one matrices with non-zero eigenvalues being respectively given by $w_A =2$ and $w_B =1$, as described by Eq.\ \eqref{eq:Ratefunction_rk1rk1}.   } 
\label{fig:RateFunction_rk1rk1}
\end{figure}

In this section, we consider the toy model of the sum of two rank-one matrix where one is randomly rotated:
\begin{align}
\label{eq:rk1_plus_rk1}
    \mathbf{C}&:= w_A\mathbf{e}\mathbf{e}^{\mathsf{T}} + w_B \mathbf{v}\mathbf{v}^{\mathsf{T}} \, , &&
\end{align}
with $\mathbf{e}$ the unit vector in the first canonical direction and $\mathbf{v}$ a  unit vector taken uniformly on $\mathbb{S}^{N-1}$, we choose $w_A\geq w_B$ without loss of generality.
The largest eigenvalue of $\mathbf{C}$ is given by
\begin{align}
\label{eq:eigval_rk1rk1}
        \lambda_1(\mathbf{C}) &=\frac{w_A+w_B+\sqrt{(w_A-w_B)^2+4 w_A w_B |\vect{e}^{\mathsf{T}}\vect{v}|^2}}{2} \, , && 
\end{align}
Note that since $0\leq |\vect{e}^{\mathsf{T}}\vect{v}|^2 \leq 1$, we have $w_A\leq\lambda_1(\mathbf{C})\leq w_A+w_B$ as expected. In the limit  $N \to \infty$, the vector $\vect{v}$ is almost-surely orthogonal to the vector $\mathbf{e}$ so that the top eigenvalue of the matrix $\mathbf{C}$ is given by $w_A$. Note that this can be checked by taking the limit $|\vect{e}^{\mathsf{T}}\vect{v}| \to 0^+$ in Eq.\ \eqref{eq:eigval_rk1rk1}. Now at large but finite $N$, we can ask what is the probability of finding $\lambda_1(\mathbf{C})$ at a position $x$ higher than $w_A$. To do so, let's remark, that since $\mathbf{v}$ is uniform on the sphere, the square of each of its component - and hence the squared overlap -  is known to follow a \emph{Beta distribution} of parameters ($1/2$,$N/2$). Its probability density given by: 
\begin{align}
\label{eq:density_beta}
p \left( | \vect{e}^{\mathsf{T}}\vect{v}|^2 = \phi \right) &=\frac{ \phi^{-1/2}(1-\phi)^{N/2-1} }{ B(1/2,N/2) } \, , &&
\end{align}
where $B(1/2,N/2)$ is the Euler Beta function. From this we can write the exact probability density $\mathcal{P}_N$ of the law of $\lambda_1(\mathbf{C})$:
\begin{align}
\mathcal{P}_N( \lambda_1)&=\frac{2\lambda_1-w_A-w_B}{B(1/2,N/2)w_Aw_B}
\left(1-\frac{\lambda_1(w_A+w_B-\lambda_1)}{w_Aw_B}\right)^{-1/2}
\left(\frac{\lambda_1(w_A+w_B-\lambda_1)}{w_Aw_B}\right)^{N/2-1} \, . &&
\end{align}
In the large $N$ limit we obtain the following rate function for the large deviations of $\lambda_1(\mathbf{C})$
\begin{align}
\label{eq:Ratefunction_rk1rk1}
    \Psi_C(x)&=-\log\left(\frac{w_A+w_B-x}{w_B}\right)-\log\left(\frac{x}{w_A}\right) \, . &&
\end{align}
This rate function is represented in Fig.\ \ref{fig:RateFunction_rk1rk1}. This is the same rate function that we had in the third regime of Eq.\ \eqref{eq:RateFunction_Sum} with $\mathrm{c}_+=x_{c_2}=w_A$ and $g_C(z)=1/z$.

\end{document}